\documentclass[aps,prd,groupedaddress,nofootinbib]{revtex4-2}
\usepackage{graphicx}
\usepackage{slashed}
\usepackage{xcolor}

\usepackage{caption}
\usepackage{subcaption}

\usepackage{amsmath,amssymb}
\usepackage{verbatim}
\usepackage{hyperref} 
\hypersetup{
    colorlinks,
    citecolor=black,
    filecolor=black,
    linkcolor=black,
    urlcolor=black
}

\newcommand{\be}{\begin{equation}}
\newcommand{\ee}{\end{equation}}
\newcommand{\ba}{\begin{align}}
\newcommand{\ea}{\end{align}}
\newcommand{\ban}{\begin{eqnarray*}} 
\newcommand{\ean}{\end{eqnarray*}}
\newcommand \nn {\nonumber}

\def\tr{{\rm Tr}}

\def\sq{\slashed{q}}

\def\P{{\mathbf P}}
\def\b{{\mathbf b}}
\def\p{{\mathbf p}}
\def\q{{\mathbf q}}
\def\k{{\mathbf k}}
\def\x{{\mathbf x}}  
\def\y{{\mathbf y}}

\def\z{{\mathbf z}}

\def\v{{\mathbf v}}

\def\ux{{\underline{x}}}

\def\uk{{\underline{k}}}

\begin{document}

\title{Parton model contributions as next-to-eikonal corrections to the dipole factorization of DIS and SIDIS at low $x_{Bj}$}
\author{Tolga Altinoluk$^{a}$, Guillaume Beuf$\,^{a}$ and Swaleha Mulani$^{a,b,c}$}
\affiliation{ $^{a}$Theoretical Physics Division, National Centre for Nuclear Research,
Pasteura 7, Warsaw 02-093, Poland}
\affiliation{ $^{b}$Department of Physics, University of Jyv\"askyl\"a, P.O. Box 35, 40014 University of Jyv\"askyl\"a, Finland}
\affiliation{ $^{c}$Helsinki Institute of Physics, P.O. Box 64, 00014 University of Helsinki, Finland} 

\date{\today}

\begin{abstract}

We compute the next-to-eikonal (NEik) power corrections to inclusive deep inelastic scattering (DIS) and semi-inclusive deep inelastic scattering (SIDIS) at low $x$ beyond dipole factorization, which represent the eikonal result. The analysis is restricted to contributions arising from t-channel quark exchanges, thereby probing the quark background field of the target. For a transversely polarized virtual photon, the NEik corrections to inclusive DIS are expressed in terms of quark and antiquark collinear parton distribution functions (PDFs), while the corresponding corrections to SIDIS are formulated in terms of the unpolarized quark transverse-momentum–dependent distribution (TMD). In contrast, for a longitudinally polarized photon, the NEik corrections to both inclusive DIS and SIDIS vanish at lowest order in $\alpha_s$.

\end{abstract}

\maketitle


\section{Introduction}
\label{sec:intro}

Deep Inelastic Scattering (DIS) of leptons off hadronic targets provides a powerful probe of the internal partonic structure of nucleons and nuclei in  Quantum Chromodynamics (QCD). The process, mediated by the exchange of a virtual photon, is characterized by two Lorentz-invariant variables: the virtuality of the exchanged photon, $Q^2=-q^2$, and the Bjorken scaling variable $x=x_{Bj}=Q^2/(2P\cdot q)$, where $P$ and $q$ are the four-momenta of the target and the virtual photon, respectively. 

In the one-photon exchange approximation,  after integration over the azimuthal angle of the scattered lepton, the unpolarized DIS cross section can be written as  
\begin{align}
\frac{d\sigma^{l+p\rightarrow l'+X}}{dx_{Bj}\, d Q^2}
=&\, \frac{4\pi\, \alpha_{\textrm{em}}^2}{ x_{Bj}\, Q^4 }\left[\left(1\!-\!y\!+\!\frac{y^2}{2}\right) \; F_{T}(x_{Bj},Q^2)
+ \left(1\!-\!y\right)\; F_{L}(x_{Bj},Q^2) \right]
\nonumber\\
=&\, \frac{\alpha_{\textrm{em}}}{\pi\, x_{Bj}\, Q^2}\left[\left(1\!-\!y\!+\!\frac{y^2}{2}\right) \; \sigma^{\gamma^*}_{T}\!(x_{Bj},Q^2)
+ \left(1\!-\!y\right)\; \sigma^{\gamma^*}_{L}\!(x_{Bj},Q^2) \right]
\label{DIS_xsect_one_photon}
\, ,
\end{align}
where $\sigma^{\gamma^*}_{T,L}$ is interpreted as total cross section for the scattering of a transverse or longitudinal virtual photon on the target.
Moreover, the inelasticity $y$ is defined as $y=(2P\cdot q)/s$ where the Mandelstam $s$ variable of the lepton-target collision which is defined as $s=(P+l)^2$ with $l$ being the four momentum of the incoming lepton. The total DIS cross sections for the scattering of transverse or longitudinal photon on the target are related to the transverse or longitudinal structure functions via 
\begin{align}
\sigma^{\gamma^*}_{T,L}(x_{Bj},Q^2)=&\, \frac{(2\pi)^2 \alpha_{\textrm{em}}}{Q^2}\; F_{T,L}(x_{Bj},Q^2)
\, ,
\label{rel_sigmaTL_FTL}
\end{align}
with the latter being related to the $F_1$ and $F_2$ structure functions as 
\begin{align}
2\, x_{Bj}\, F_{1} =&\,  F_{T} \\
F_{2}=&\, F_{T}+F_{L}
\label{rel_FTL_F12}
\, .
\end{align}

In the parton model, formulated in the infinite-momentum frame, the nucleon is described as a collection of quasi-free partons, each carrying a longitudinal momentum fraction $x$ of the nucleon's momentum. The inclusive DIS cross section factorizes~\cite{Collins:1989gx}  into a perturbatively calculable hard scattering coefficient and universal, nonperturbative parton distribution functions (PDFs) that encode the partonic structure of the nucleon. The evolution of the PDFs with increasing $Q^2$ at fixed values of $x$ is given by the 
Dokshitzer-Gribov-Lipatov-Altarelli-Parisi (DGLAP) evolution equations \cite{Gribov:1972ri,Altarelli:1977zs,Dokshitzer:1977sg}. DGLAP evolution equations resum the leading logarithms of hard scale $Q^2$ at fixed $x$, corresponding to strongly ordered emissions in transverse momentum. At moderate to small $x$, this formalism forms the foundation of global QCD analyses.   

At low Bjorken $x$, corresponding to the high-energy limit, the gluon density in the hadron increases rapidly. The Balitsky-Fadin-Kuraev-Lipatov (BFKL) equation \cite{Lipatov:1976zz,Kuraev:1977fs,Balitsky:1978ic} resums leading logarithms in $1/x$ and predicts a power-like growth of the gluon distribution with decreasing $x$, in qualitative agreement with HERA data \cite{H1:2009pze}. However, at sufficiently small $x$, high partonic density effects become significant, leading to nonlinear QCD dynamics and eventually to the phenomena known as gluon saturation.  The Color Glass Condensate (CGC) effective theory  provides a convenient framework for describing high energy (or saturation) regime of hadronic collisions (see \cite{Gelis:2010nm,Albacete:2014fwa,Blaizot:2016qgz} for recent reviews and references therein). In the CGC picture, the small-$x$ gluons are treated as classical color fields generated by static color sources corresponding to large-$x$ partons. The transition between dilute and saturated dynamics is governed by the saturation scale $Q_s(x)$. The evolution of the system with decreasing $x$ is described by the nonlinear generalization of the linear BFKL equation which is known as Balitsky-Kovchegov/Jalilian-Marian-Iancu-McLerran-Wiegert-Leonidov-Kovner (BK-JIMWLK) equation \cite{Balitsky:1995ub,Kovchegov:1999yj,Kovchegov:1999ua,Jalilian-Marian:1996mkd,Jalilian-Marian:1997qno,Jalilian-Marian:1997jhx,Jalilian-Marian:1997ubg,Kovner:2000pt,Weigert:2000gi,Iancu:2000hn,Iancu:2001ad,Ferreiro:2001qy}. 

In the leading power at high energy or low $x$ (keeping $Q^2$ fixed), DIS can be conveniently described within the dipole factorization \cite{Bjorken:1970ah,Nikolaev:1990ja,Nikolaev:1991et}. In this formulation, the virtual photon fluctuates into a quark–antiquark dipole before interacting with the target. Then the quark-antiquark dipole scatters on the dense target.  The total cross section factorizes as a convolution of the photon light-front wave function squared, which can be computed perturbatively, and the dipole operator, which encodes the QCD dynamics of the interaction.

Moreover, within the dipole factorization framework, the leading order (LO) inclusive DIS cross sections for longitudinally and transversely polarized virtual photon are given by
\begin{align}
\sigma^{\gamma^*}_L(x_{Bj},Q^2)&=\frac{4\, \alpha_{em}Q^2N_c}{(2\pi)^2}\sum_f e_f^2 \, \int_0^1 dz \, z(1-z) \int d^2\x_1 \, d^2\x_2 \Big[1-d(\x_1,\x_2)\Big]
\, 4z(1-z)\, K_0^2(|\x_1-\x_2|\overline Q) \\
\sigma^{\gamma^*}_T(x_{Bj},Q^2)&=\frac{4\, \alpha_{em}Q^2N_c}{(2\pi)^2}\sum_f e_f^2 \, \int_0^1 dz \, z(1-z) \int d^2\x_1 \, d^2\x_2 \Big[1-d(\x_1,\x_2)\Big]
\, \big[ z^2+(1-z)^2\big]\, K_1^2(|\x_1-\x_2|\overline Q)
\end{align}
in the massless quark case, where $z$ and $(1-z)$ corresponds to the longitudinal momentum fraction carried by the integrated quark and antiquark relative to the longitudinal momentum of the incoming virtual photon. $K_{\alpha}(\cdots)$ is the modified Bessel function of the second type and ${\overline Q}^2=z(1-z)Q^2$. Finally, $d(\x_i,\x_j)$ is the aforementioned dipole operator that is defined as 
\begin{align}
d(\x_i,\x_j)=\frac{1}{N_c}\Big\langle\tr\big[ U_F(\x_i)U^\dagger_F(\x_j)\big]\Big\rangle
\label{eq:Fund_dipole_opr}
\end{align}
where $U_F(\x_i)\equiv U_F(+\infty,-\infty;\x_i)$ is the fundamental Wilson line in gluon background field ${\cal A}^-(z^+,\z)$ which reads 
\begin{align}
U_F(x^+,y^+;\z)={\cal P}_+\exp\bigg[-ig\int_{y^+}^{x^+}dz^+\, t\cdot{\cal A}^-(z^+,\z)\bigg]
\label{eq:Fund_Wilson_line}
\end{align}
with $t^a$ being the $SU(N_c)$ generators in the fundamental representation. The bracket in Eq. \eqref{eq:Fund_dipole_opr} represent an averaging over the gluon background field representing the target.

Beyond the collinear framework and inclusive structure functions, a comprehensive understanding of nucleon structure requires incorporating the transverse-momentum degrees of freedom of partons. This is achieved through Transverse-Momentum-Dependent parton distributions (TMDs) \cite{Collins:2011zzd,Boussarie:2023izj}, which depend on both the longitudinal momentum fraction $x$ and the intrinsic transverse momentum. TMDs provide a three-dimensional description of the nucleon in momentum space and are essential for processes sensitive to parton transverse motion, such as semi-inclusive DIS (SIDIS) and Drell–Yan production. When computed in the dipole factorization, the LO SIDIS cross section at small-$x$ (see for example \cite{Marquet:2009ca}) for longitudinally and transversely polarized incoming virtual photon can be written respectively as 
\begin{align}
\frac{d\sigma_L^{\gamma^*A\to q+X}}{d^2\p\, dz}&=\frac{4\pi \alpha_{em}Q^2N_c}{(2\pi)^5}\sum_fe_f^2
\, z(1-z)\int d^2\x_1 \, d^2\x_2\,  d^2\x_{1'} \, e^{i\p\cdot (\x_{1'}-\x_1)}\Big[ d(\x_1,\x_{1'})-d(\x_1,\x_2)-d(\x_{1'},\x_2)+1\Big] 
\nn \\
& \hspace{3.5cm} \times
4z(1-z)\, K_0(|\x_1-\x_2|\overline Q) K_0(|\x_{1'}-\x_2|\overline Q) 
\\
\frac{d\sigma_T^{\gamma^*A\to q+X}}{d^2\p\, dz}&=\frac{4\pi \alpha_{em}Q^2N_c}{(2\pi)^5}\sum_fe_f^2
\, z(1-z)\int d^2\x_1 \, d^2\x_2\,  d^2\x_{1'} \, e^{i\p\cdot (\x_{1'}-\x_1)}\Big[ d(\x_1,\x_{1'})-d(\x_1,\x_2)-d(\x_{1'},\x_2)+1\Big] 
\nn \\
& \hspace{3.5cm} \times
\big[z^2+(1-z)^2\big]\frac{(\x_1-\x_2)}{|\x_1-\x_2|}\cdot\frac{(\x_{1'}-\x_2)}{|\x_{1'}-\x_2|}\, K_1(|\x_1-\x_2|\overline Q) K_1(|\x_{1'}-\x_2|\overline Q) 
\Big\}
\label{eq:SIDIS_dip_fac}
\end{align}
%

At small $x$, the relation between the dipole operators (together with the kinematic coefficients) and the sea quark contribution to the quark TMD distribution have been discussed widely in literature \cite{Catani:1994sq,Ji:2005nu,Marquet:2009ca,Dominguez:2011wm,Xiao:2017yya,Hentschinski:2017ayz,Caucal:2025xxh} for different kinematic regimes. Moreover, we would like to mention that the SIDIS cross section given in Eq. \eqref{eq:SIDIS_dip_fac} is at partonic level. In order to get the hadronic cross section it should be convoluted with the TMD fragmentation function. As discussed in \cite{Marquet:2009ca}, 
 the TMD fragmentation function reduces to a collinear fragmentation function at lowest order in $\alpha_s$. In the rest of the manuscript, for the sake of simplicity, our calculation will be presented at partonic level without convoluting it with a fragmentation function.

A key simplification in this regime arises from the eikonal approximation, valid at leading power at high energies, when the interaction time is much shorter than the lifetime of the photon’s partonic fluctuation. The quark and antiquark propagate through the target along straight, light-like trajectories, undergoing color rotations described by Wilson lines in the fundamental representation given in Eq.~\eqref{eq:Fund_Wilson_line}. The multiple scattering of the dipole off the target’s color field exponentiates into these Wilson lines, leading naturally to nonlinear dynamics and gluon saturation. The dipole cross section is thus determined by the expectation value of a color dipole operator given in Eq.~\eqref{eq:Fund_dipole_opr}. 

Comprehensive DIS measurements across wide ranges in $x$ and $Q^2$ thus offer a unique opportunity to study the interplay between DGLAP evolution, BFKL dynamics, gluon saturation, and TMD physics. Future experimental programs at the Electron–Ion Collider (EIC) \cite{Accardi:2012qut} and the Large Hadron–Electron Collider (LHeC) \cite{LHeCStudyGroup:2012zhm} will explore this rich landscape with unprecedented precision, providing stringent tests of QCD factorization, universality, and nonlinear dynamics in the high-energy limit.

With the continuous improvement of experimental measurements, it has become essential to enhance the theoretical precision of QCD predictions for the corresponding observables. This can be achieved either by extending perturbative calculations to higher orders in the strong coupling constant or by relaxing the kinematical approximations employed in leading-order (LO) computations. Over the past two decades, significant progress has been made in calculating next-to-leading order (NLO) corrections to DIS structure functions \cite{Balitsky:2010ze,Balitsky:2012bs,Beuf:2011xd,Beuf:2016wdz,Beuf:2017bpd,Ducloue:2017ftk,Hanninen:2017ddy,Beuf:2020dxl,Beuf:2021qqa,Beuf:2021srj,Beuf:2022ndu,Hanninen:2022gje,Casuga:2025etc} and inclusive SIDIS processes \cite{Caucal:2024cdq,Bergabo:2024ivx,Bergabo:2022zhe,Altinoluk:2025dwd,Caucal:2024vbv}. As mentioned earlier, a complementary route to improving theoretical accuracy in the high-energy regime involves going beyond the eikonal approximation. In practice, this approximation corresponds to retaining only the leading-power contributions in energy, while neglecting subleading (energy-suppressed) terms in the calculation of observables. Systematically including these power corrections offers an important step toward a more precise description of QCD dynamics at small Bjorken $x$. 

Within the CGC framework, the high-energy limit can be realized by boosting the target along the $x^-$ direction with a Lorentz factor $\gamma_t$. In a high-energy dilute–dense scattering process, the eikonal approximation relies on three key assumptions:
{\text(i)} the highly boosted background field describing the target becomes localized in the longitudinal direction (around $x^+=0$) due to Lorentz contraction; {\text (ii)} the background field is dominated by its leading component in powers of $\gamma_t$, while subleading components are neglected; and {\text(iii)} as a result of Lorentz time dilation, the background field is assumed to be independent of the light-cone coordinate $x^-$, effectively treating the target as static and neglecting its internal dynamics. Relaxing any of these assumptions introduces corrections beyond the eikonal approximation. To systematically account for such effects, one must include next-to-eikonal (NEik) corrections, which correspond to terms suppressed by powers of $1/\gamma_t$ at the level of the boosted background field. These corrections encode finite-energy effects of the target, extending the applicability of the CGC formalism beyond the strict high-energy limit.

All three approximations discussed above are valid when the target is described solely by a gluon background field. However, an additional source of NEik corrections arises from incorporating the quark background field of the target. In this case, the projectile parton can interact with the target via a $t$-channel quark exchange, leading to subeikonal contributions beyond those associated with the gluon background field of the target. Under a boost of $\gamma_t$ along $x^-$ direction, current associated with the target scale as 
\begin{align}
J^-(x)\propto \gamma_t, \:\: J^j(x)\propto (\gamma_t)^0, \:\: J^+(x)\propto (\gamma_t)^{-1} \, .
\label{eq:Target_Current_Scaling_1}
\end{align}
The good and bad components of the quark background field $\Psi(x)$ are defined as 
\begin{align}
\Psi^{(-)}(x)&=\frac{\gamma^+\gamma^-}{2}\Psi(x) \, ,\\
\Psi^{(+)}(x)&=\frac{\gamma^-\gamma^+}{2}\Psi(x) \, .
\end{align}
The currents associated with the target are constructed as bilinears of the quark background field $\Psi(x)$ and its components satisfy 
\begin{align}
{\overline \Psi}(x)\gamma^-\Psi(x)&=\overline {\Psi^{(-)}}(x)\gamma^-\Psi^{(-)}(x) \, , \\
{\overline \Psi}(x)\gamma^j\Psi(x)&=\overline {\Psi^{(-)}}(x)\gamma^j\Psi^{(+)}(x)+\overline {\Psi^{(+)}}(x)\gamma^j\Psi^{(-)}(x) \, , \\
{\overline \Psi}(x)\gamma^+\Psi(x)&=\overline {\Psi^{(+)}}(x)\gamma^-\Psi^{(+)}(x) \, .
\end{align}
Since the currents associated with the target have to follow the same scaling behavior as introduced in Eq. \eqref{eq:Target_Current_Scaling_1}, the good and the bad components of the quark background field scale as 
\begin{align}
\Psi^{(-)}(x) \propto (\gamma_t)^{1/2} \, ,\\
\Psi^{(+)}(x) \propto (\gamma_t)^{-1/2} \, ,
\end{align}
with the boosting parameter $\gamma_t$. Thus, the quark background field does not contribute at eikonal order; the enhanced component $\Psi^{(-)}(x)$ contributes at NEik order, while the suppressed component $\Psi^{(+)}(x)$ contributes at next-to-next-to-eikonal (NNEik) order and beyond.  

Over the last decade, significant efforts have been devoted to computing NEik corrections within the CGC framework. Early studies focused on NEik corrections arising from the finite longitudinal width of the target \cite{Altinoluk:2014oxa,Altinoluk:2015gia}, which were subsequently applied to particle production and correlations in both dilute-dilute \cite{Altinoluk:2015xuy,Agostini:2019avp,Agostini:2019hkj} and dilute-dense \cite{Agostini:2022ctk,Agostini:2022oge} collisions. NEik corrections to quark and scalar propagators were computed in \cite{Altinoluk:2020oyd,Altinoluk:2021lvu,Agostini:2022oge} and applied to DIS dijet production in \cite{Altinoluk:2022jkk,Agostini:2024xqs}, while back-to-back quark-gluon dijet production in DIS, including $t$-channel quark exchange, was studied to probe quark TMDs \cite{Altinoluk:2023qfr}. Analogous studies for gluon TMDs and the interplay between NEik and kinematic twist corrections were performed in \cite{Altinoluk:2024zom}, and the gluon propagator incorporating all NEik corrections was revisited in \cite{Altinoluk:2024dba} for applications in parton-nucleus scattering. In \cite{Altinoluk:2024tyx}, NEik corrections stemming from the $t$-channel quark exchanges are computed in the back-to-back dijet production in proton-nucleus collision at forward rapidities and various quark TMDs are probed. Helicity-dependent observables, quark and gluon helicity evolutions as well as single and double spin asymmetries, have been investigated extensively in \cite{Kovchegov:2015pbl,Kovchegov:2016zex,Kovchegov:2016weo,Kovchegov:2017jxc,Kovchegov:2017lsr,Kovchegov:2018znm,Kovchegov:2018zeq,Kovchegov:2020kxg,Kovchegov:2020hgb,Adamiak:2021ppq,Kovchegov:2021lvz,Kovchegov:2021iyc,Cougoulic:2022gbk,Kovchegov:2022kyy,Borden:2023ugd,Kovchegov:2024aus,Borden:2024bxa,Adamiak:2025dpw,Kovchegov:2025gcg,Borden:2025ehe} at NEik accuracy and the helicity-dependent extensions of the CGC framework have been formulated in \cite{Cougoulic:2019aja,Cougoulic:2020tbc}. Rapidity evolution of gluon TMDs, which interpolates between moderate and low $x$, was studied in \cite{Balitsky:2015qba,Balitsky:2016dgz,Balitsky:2017flc}. A similar interpolation between the moderate and low $x$ has been also studied in the context of inclusive DIS \cite{Boussarie:2020fpb,Boussarie:2021wkn} and exclusive Compton scattering \cite{Boussarie:2023xun}. NEik corrections to both quark and gluon propagators have also been formulated within the high-energy operator product expansion (OPE) in \cite{Chirilli:2018kkw,Chirilli:2021lif}. Sub-eikonal corrections in the CGC were studied using an effective Hamiltonian approach in \cite{Li:2023tlw,Li:2024fdb,Li:2024xra}, and an alternative framework allowing longitudinal momentum exchange between projectile and target was developed in \cite{Jalilian-Marian:2017ttv,Jalilian-Marian:2018iui,Jalilian-Marian:2019kaf}. Finally, the impact of sub-eikonal corrections on orbital angular momentum has been analyzed in \cite{Hatta:2016aoc,Kovchegov:2019rrz,Boussarie:2019icw,Kovchegov:2023yzd,Kovchegov:2024wjs}.  

In this manuscript, we consider the NEik corrections which stem from $t$-channel quark exchanges and focus on inclusive DIS and SIDIS. The manuscript is organized as follows. In Section \ref{sec:sidis} and \ref{sec:dis}, we study the quark background contribution to SIDIS cross section and to DIS structure functions, respectively. In Section \ref{sec:discussion}, we provide a concise discussion of the interplay between contributions from the quark background field and those arising from dipole factorization at eikonal order, focusing on DIS structure functions and the SIDIS cross section. In Section \ref{sec:summary}, we provide a short summary of our results. Appendix \ref{App:qbar_DIS} is devoted to derivation of the antiquark contribution to the DIS structure functions.  

\section{Quark background field contribution to SIDIS}
\label{sec:sidis}

\begin{figure}
    \centering
    \includegraphics[scale=0.5]{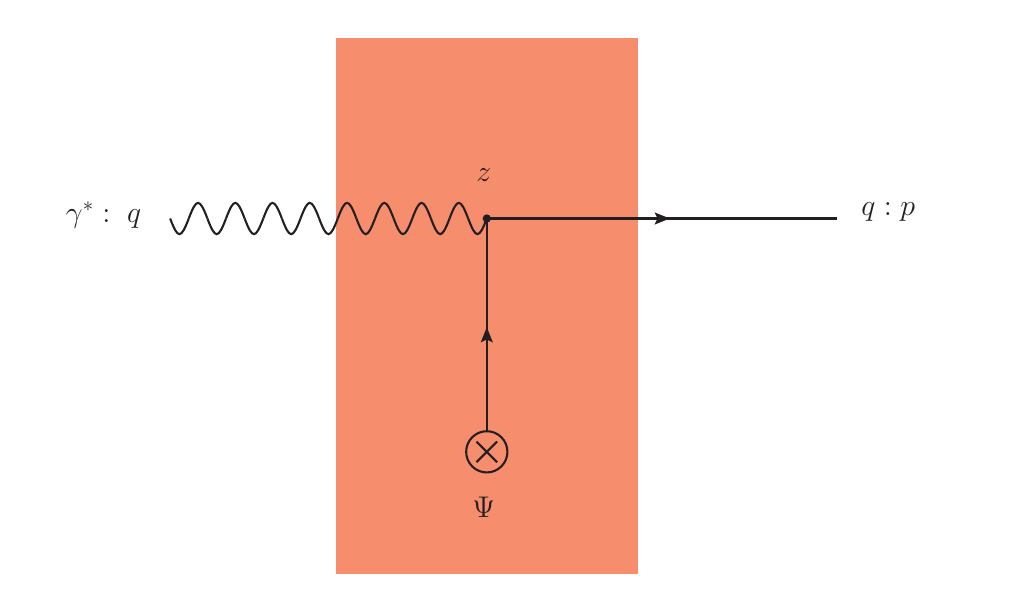}
    \caption{Semi-inclusive deep inelastic scattering (SIDIS) diagram: quark production from quark background field}
    \label{fig: SIDIS_q}
\end{figure}

For the SIDIS quark production at low $x_{Bj}$, the first contribution involving a quark background field, beyond the eikonal approximation and beyond dipole factorization, corresponds to the diagram on Fig.~\ref{fig: SIDIS_q}, with the incoming virtual photon being converted into the produced quark by coupling with the quark background field. In general, the S-matrix element associated with that diagram can be written from the Feynman rules and LSZ reduction formula as\footnote{
We use the metric signature $(+,-,-,-)$. We use $x^\mu$ for a Minkowski 4-vector. In a light-cone basis we have
$
x^\mu=(x^+,{\bf x},x^-)
$
where $x^\pm=(x^0\pm x^3)/\sqrt{2}$ and ${\bf x}$ denotes a transverse vector with components $x^i$. 
We will also use the notations 
$
\ux = (x^+,{\bf x})
$
and 
$\uk = (k^+,{\bf k})$.
For a given momentum 4-vector $k^{\mu}$, we use the notation $\check{k}^{\mu}$ for its on-shell analog. More precisely, it is defined in such a way that their $+$ and transverse components coincide, $\check{k}^{+}={k}^{+}$ and $\check{\k}=\k$, whereas the $-$ component of $\check{k}^{\mu}$ is adjusted to make it on-shell, i.e. $\check{k}^-=(\k^2+m^2)/(2k^+)$ for a massive quark or $\check{k}^-=\k^2/(2k^+)$ for a gluon.}
\begin{align}
S_{\gamma^*_{\lambda}\rightarrow q} 
=&\,
 \lim_{x^{+}\to \infty} \int \! d^{2}\x \int\! dx^{-} \ e^{i\check{p}\cdot x}  \ \overline{u}(\check{p},h) \ \gamma^{+} 
\int\! d^{4}z \ \epsilon_{\lambda}^{\mu}(q) \ e^{-i{q} \cdot z}\ S_{F}(x,z) (-iee_{f}\gamma_{\mu}) \Psi(z)
\label{SIDIS_Smat_q_1}
\, .
\end{align}

Note that the quark Feynman propagator from the QED vertex $z$ to the point $x$ in the asymptotic future is the quark propagator in a gluon background field, accounting for rescattering of the produced quark with the target remnants. As recalled in the previous section, both the high-energy limit and the power suppressed corrections beyond that limit can be understood from the behaviour under large longitudinal Lorentz boosts of the target, with a boost factor $\gamma_t$. Under such boost, the quark background field is more and more peaked around $z^+=0$, so that the integration over the position $z^+$ of the vertex effectively brings a suppression by a factor $1/\gamma_t$. And as explained earlier, the quark background field can be split into so called good and bad components, behaving as $\sqrt{\gamma_t}$ and as $1/\sqrt{\gamma_t}$ respectively under a large boost of the target. Finally, the quark propagator $S_{F}(x,z)$ from $z$ inside to $x$ after the target stays finite under large boost of the target, with a limit given by its eikonal expression \cite{Altinoluk:2022jkk,Altinoluk:2023qfr,Altinoluk:2024dba}
\begin{align}
S_{F}(x,z)\big|^{\rm IA,q}_{\rm Eik} &= \int\! \frac{d^{3}\uk}{(2\pi)^{3}} \frac{\theta(k^{+})}{2k^{+}}\ e^{-ix\cdot\check{k}} 
\ (\check{\slashed k}+m) U_{F}(x^{+},z^{+};\z) 
\ \Big[1-\frac{\gamma^{+}\gamma^{i}}{2k^{+}}i\overleftarrow{D}^{F}_{z^i}\Big] 
\ e^{iz^{-}k^{+}} \ e^{-i\z\cdot \k}
\label{q_prop_IA} 
\end{align}
where the fundamental Wilson line in gluon background field ${\cal A}^-_a(z^+,\z)$ is defined in Eq. \eqref{eq:Fund_Wilson_line}.
The covariant derivatives are defined as 
\begin{align}
\overrightarrow{D}^F_{z^\mu}&=\overrightarrow{\partial_{z^\mu}}+ig\, t\cdot{\cal A}_\mu(z)
\nn \\
\overleftarrow{D}^{F}_{z^i}&=\overleftarrow{\partial_{z^\mu}}-ig\, t\cdot{\cal A}_\mu(z)\, . 
\end{align}
Hence, in the limit of large boost, the leading contribution in Eq.~\eqref{SIDIS_Smat_q_1} scales as  $1/\sqrt{\gamma_t}$ overall. Squaring this result, one thus expects a contribution of order $1/{\gamma_t}$ at the cross section level, corresponding to NEik order.  
The leading power contribution in Eq.~\eqref{SIDIS_Smat_q_1} is then obtained by using the expression \eqref{q_prop_IA} for the propagator, and by including only the enhanced   components $ \Psi^{(-)}(z)$ of the quark background field, as
\begin{align}
S_{\gamma^*_{\lambda}\rightarrow q} 
=&\,
\lim_{x^{+}\to \infty} \int \! d^{2}\x \int\! dx^{-} \ e^{i\check{p}\cdot x} \ \overline{u}(\check{p},h) \ \gamma^{+}  \int\! d^{4}z \ 
\int\! \frac{d^{3}\uk}{(2\pi)^{3}} \frac{\theta(k^{+})}{2k^{+}}\ e^{-ix\cdot\check{k}} 
\nn\\
&
\times\, (\check{\not k}+m) U_{F}(x^{+},z^{+};\z)
\ \Big[1-\frac{\gamma^{+}\gamma^{i}}{2k^{+}}i\overleftarrow{D}^{F}_{z^i}\Big] 
\ e^{iz^{-}k^{+}} \ e^{-i\z\cdot \k}
(-iee_{f}\gamma_{\mu})\Psi^{(-)}(z)\, \epsilon_{\lambda}^{\mu}(q) \ e^{-i{q} \cdot z}
+O({\gamma_t}^{-\frac{3}{2}})
\nn\\
=&\,
-iee_{f}\int\! d^{4}z\, \frac{1}{2p^+}\, \overline{u}(\check{p},h) \ \gamma^{+} (\check{\not p}+m) 
\ U_{F}(+\infty,z^{+};\z)\ 
 \Big[1-\frac{\gamma^{+}\gamma^{i}}{2p^{+}}i\overleftarrow{D}^{F}_{z^i}\Big]
\nn\\
&
\times\,
e^{iz^{-}(p^{+}-q^+)} \ e^{-i\z\cdot(\p-\q)}\, 
\slashed{\epsilon}_{\lambda}(q)
 \frac{\gamma^{+}\gamma^{-}}{2}\Psi(z^+,\z)
 +O({\gamma_t}^{-\frac{3}{2}})
\label{SIDIS_Smat_q_2}
\, .
\end{align}
In the last step, we have neglected the phase factor $e^{-i{q}^- z^+}$, which would contribute only to subleading corrections beyond the shockwave approximation, and the dependence of $\Psi(z)$ on $z^-$, which would contribute only to subleading corrections beyond the static approximation. In this calculation, we indeed want to keep the insertion of the quark background field as the only effect beyond the eikonal limit.
Choosing the light cone gauge for QED, one has $\epsilon_{\lambda}^{+}(q)=0$ for any photon polarization vector, so that
\begin{align}
\gamma^{+}\gamma^{i} \slashed{\epsilon}_{\lambda}(q)\gamma^{+} &=\, \gamma^{i}\gamma^{+} \gamma^{+}\slashed{\epsilon}_{\lambda}(q) =0
\, ,
\end{align}
remembering that $\gamma^+\gamma^+ =g^{++}=0$, and thus the term with the covariant derivative in Eq.~\eqref{SIDIS_Smat_q_2} vanishes. Moreover, one has
\begin{align}
\overline{u}(\check{p},h) \ \gamma^{+} (\check{\not p}+m)
&=\, 
\overline{u}(\check{p},h) \Big( \{\gamma^{+},\check{\not p}\} -  (\check{\not p}-m)\gamma^{+}
\Big)
= (2p^+) \overline{u}(\check{p},h)
\, , 
\end{align}
where the second term vanishes from the definition of the spinor ${u}(\check{p},h) $ as a solution of the free Dirac equation.
Hence, the expression  \eqref{SIDIS_Smat_q_2} becomes
\begin{align}
S_{\gamma^*_{\lambda}\rightarrow q} 
 =&\,
2\pi \delta(p^{+}\!-\!q^+)\, (-i)ee_{f}
\int dz^+\!\int\! d^{2}\z\,  e^{-i\z\cdot(\p-\q)}\,
 \overline{u}(\check{p},h) \,  
 \slashed{\epsilon}_{\lambda}(q)\,
 \frac{\gamma^{+}\gamma^{-}}{2}\,
 U_{F}(+\infty,z^{+};\z) \Psi(z^+,\z)
 +O({\gamma_t}^{-\frac{3}{2}})
\label{SIDIS_Smat_q_3}
\, .
\end{align}

In order to calculate the transverse and longitudinal photon contributions to the SIDIS cross section, we are calculating a formal S-matrix element for scattering of an off-shell photon with the target, in which this off-shell photon can have not only one of the physical transverse polarizations, but also the unphysical longitudinal polarization. For that purpose, we can use the following expression for the longitudinal photon polarization vector in light-cone gauge:  
\begin{align}
\epsilon_{L}^{\mu}(q)
=&\,
\frac{Q}{q^{+}}g^{+\mu}
\label{pol_vect_L}
\, ,
\end{align}
then,
\begin{align}
 \slashed{\epsilon}_{L}(q)\,\gamma^{+}
 =&\,
  \frac{Q}{q^{+}}\,\gamma^{+}\gamma^{+} =0
\, ,
\label{vanishing_L_contrib}
\end{align}
so that in the longitudinal photon case, the diagram \ref{fig: SIDIS_q} vanishes at the considered accuracy,
\begin{align}
S_{\gamma^*_{L}\rightarrow q}  = 0 +O({\gamma_t}^{-\frac{3}{2}})
\, .
\end{align}

In the rest of this section, we will thus focus on the case in which the polarization $\lambda$ of the incoming virtual photon is one of the two transverse polarizations. In the light-cone gauge, the transverse polarization four-vectors can be written  as
\begin{align}
\epsilon^{\mu}_{\lambda}(q) 
=&\,
\Big[-g^{\mu j} + g^{\mu +}\, \frac{\q^j}{q^+}\Big] \varepsilon_{\lambda}^j
\label{4_to_2_trans_pol_vect}
\end{align}
in terms of corresponding two dimensional polarization vectors $\varepsilon_{\lambda}^j$, which are independent of the photon momentum, and form a basis of the transverse plane. For any choice of basis of these polarization vectors, one has the completeness relation 
\begin{align}
\sum_{\lambda}  {\varepsilon_{\lambda}^i}^*  \varepsilon_{\lambda}^j
 = \delta^{ij}
 \, .
 \label{completeness_pol_vect}
\end{align}
Inserting the relation \eqref{4_to_2_trans_pol_vect} into the expression \eqref{SIDIS_Smat_q_3}, one finds
\begin{align}
S_{\gamma^*_{\lambda}\rightarrow q} 
 =&\,
2\pi \delta(p^{+}\!-\!q^+)\, i ee_{f}\, \varepsilon_{\lambda}^j
\int dz^+\!\int\! d^{2}\z\,  e^{-i\z\cdot(\p-\q)}\,
 \overline{u}(\check{p},h) \,  
 \frac{\gamma^{j}\gamma^{+}\gamma^{-}}{2}\,
 U_{F}(+\infty,z^{+};\z) \Psi(z^+,\z)
 +O({\gamma_t}^{-\frac{3}{2}})
\label{SIDIS_Smat_q_4}
\, .
\end{align}

For such a process of scattering of one particle on background fields which are independent of the coordinate $z^-$, it is convenient to define the scattering amplitude from the S-matrix element as
\begin{align}
S_{\gamma^*_{\lambda}\rightarrow q}  
=&\,
 2q^{+}(2\pi) \delta(p^{+}\!-\!q^{+})\, i \mathcal{M}_{\gamma^*_{\lambda}\rightarrow q}(q,\p)
\label{ref_Smat_ampl}
\, .
\end{align}
Hence, from Eq.~\eqref{SIDIS_Smat_q_4}, one finds the amplitude
\begin{align}
 i \mathcal{M}_{\gamma^*_{\lambda}\rightarrow q}(q,\p)
 =&\,
 \frac{i ee_{f}}{2q^{+}}\, \varepsilon_{\lambda}^j
\int dz^+\!\int\! d^{2}\z\,  e^{-i\z\cdot(\p-\q)}\,
 \overline{u}(\check{p},h) \,  
 \frac{\gamma^{j}\gamma^{+}\gamma^{-}}{2}\,
 U_{F}(+\infty,z^{+};\z) \Psi(z^+,\z)
 +O({\gamma_t}^{-\frac{3}{2}})
\label{SIDIS_ampl_q_1}
\, .
\end{align}

The cross section is then obtained from the amplitude as
\begin{align}
2p^+ (2\pi)^3\frac{d^{3}\sigma^{\gamma_{T}^*\rightarrow q}}{dp^+ d^{2}\p} 
=&\,
2q^{+}(2\pi) \delta(p^{+}\!-\!q^{+})\,
   \frac{1}{2} \sum_{h, \lambda}\sum_{\alpha} 
   \Big\langle\mathcal{M}_{\gamma^*_{\lambda}\rightarrow q}(q,\p)^{\dagger} \  \mathcal{M}_{\gamma^*_{\lambda}\rightarrow q}(q,\p) \Big\rangle
\label{cross_sec_def_SIDIS_q}
\, ,
\end{align}
where $\alpha$ is the fundamental color index for the produced quark, the factor $1/2$ comes from the averaging over the two transverse polarizations of the incoming photon, and the bracket represent an averaging over the gluon and quark background fields representing the target.
Inserting the amplitude \eqref{SIDIS_ampl_q_1} in the formula \eqref{cross_sec_def_SIDIS_q}, one gets the quark background field contribution to the transverse photon to quark cross section
\begin{align}
2p^+ (2\pi)^3&\, 
\frac{d^{3}\sigma^{\gamma_{T}^*\rightarrow q}}{dp^+ d^{2}\p} \bigg|_{\Psi}
=
2q^{+}(2\pi) \delta(p^{+}\!-\!q^{+})\,
   \frac{1}{2} \sum_{h, \lambda}
   \frac{e^2\,  e_{f}^2}{(2q^{+})^2}\,  {\varepsilon_{\lambda}^i}^*  \varepsilon_{\lambda}^j   
   \int dz^+dz'^+\!\int\! d^{2}\z\, d^{2}\z'\,  e^{i(\z'-\z)\cdot(\p-\q)}\,
   \nn\\
  &\times\, 
 \Big\langle\overline{\Psi}(z'^+,\z')  U_{F}(+\infty,z'^{+};\z')^{\dag} 
 \frac{\gamma^{-}\gamma^{+}\gamma^{i}}{2}\,   {u}(\check{p},h) \,  
   \overline{u}(\check{p},h) \,  
 \frac{\gamma^{j}\gamma^{+}\gamma^{-}}{2}\,
 U_{F}(+\infty,z^{+};\z) \Psi(z^+,\z)\Big\rangle
 +O({\gamma_t}^{-2})
\label{cross_sec_SIDIS_q_1}
\, .
\end{align}
Using the completness relations \eqref{completeness_pol_vect} and
\begin{align}
\sum_{h}  {u}(\check{p},h) \,  
   \overline{u}(\check{p},h)
 = &\,
  (\check{\not p}+m) 
 \, ,
 \label{completeness_u_spinors}
\end{align}
the Dirac structure of the cross section can be simplified as
\begin{align}
\sum_{\lambda} {\epsilon^{i}_{\lambda}}^{*} \epsilon^{j}_{\lambda} &\,\sum_{h}\frac{\gamma^{-}\gamma^{+}\gamma^{i}}{2}  u(p,h) \ \overline{u}(p,h) \ \frac{\gamma^{j}\gamma^{+}\gamma^{-}}{2}
=
\delta^{ij} \frac{\gamma^{-}\gamma^{i}\gamma^{+}}{2}  \,  (\check{\not p}+ m) \frac{\gamma^{+}\gamma^{j}\gamma^{-}}{2}
=(2p^+) \delta^{ij}  \frac{\gamma^{-}\gamma^{i}\gamma^{+}}{2}  \, \frac{\gamma^{j}\gamma^{-}}{2}
\nn\\
=&\, 
(2p^+) \delta^{ij} (-1) \frac{\gamma^{-}\gamma^{+}}{2}  \, \frac{\gamma^{-}}{2}\, \frac{\{\gamma^{i}, \gamma^{j}\}}{2}
=
(2p^+) \delta^{ij} (-1)   \, \frac{\gamma^{-}}{2}\, \big(-\delta^{ij}\big)
=(2p^+)\gamma^{-}
\, ,
\end{align}
so that
\begin{align}
p^+ 
\frac{d^{3}\sigma^{\gamma_{T}^*\rightarrow q}}{dp^+ d^{2}\p} \bigg|_{\Psi}
=&\, 
q^{+} \delta(p^{+}\!-\!q^{+})\,
   \frac{\alpha_{\textrm{em}}\, e_{f}^2}{2q^{+}(2\pi)} \,    
   \int dz^+dz'^+\!\int\! d^{2}\z\, d^{2}\z'\,  e^{i(\z'-\z)\cdot(\p-\q)}\,
   \nn\\
  &\times\, 
 \Big\langle\overline{\Psi}(z'^+,\z')\gamma^{-}  U_{F}(+\infty,z'^{+};\z')^{\dag} 
 U_{F}(+\infty,z^{+};\z) \Psi(z^+,\z)\Big\rangle
 +O({\gamma_t}^{-2})
\label{cross_sec_SIDIS_q_2}
\, .
\end{align}

The next step is to specify how to take the average over the background fields. Since quark background fields are included, this case is beyond the scope the McLerran-Venugopalan model \cite{McLerran:1993ni,McLerran:1993ka,McLerran:1994vd} frequently used in the high-energy QCD literature to evaluate such target average. Instead, we reinterpret the target average of such quantity ${\cal O}$ depending on background fields in terms of the quantum expectation of the corresponding operator $\hat{\cal O}$ in the state of the target, with momentum $P$, 
(see for example Refs. \cite{Belitsky:2002sm,Dominguez:2011wm,Marquet:2016cgx,Altinoluk:2019wyu,Altinoluk:2023qfr,Altinoluk:2024tyx})
\begin{align}
\langle{\cal O} \rangle =&\, \lim_{P'\rightarrow P}
\frac{\langle P'|\hat{\cal O}| P\rangle}{\langle P'| P\rangle}
\, .
\label{def_average}
\end{align}
We choose target states normalized as
\begin{align}
\langle P'| P\rangle =&\, 2P^-\, (2\pi)^3 \delta(P^{'-}\!-\!P^-)\, \delta^{(2)}(\P'\!-\!\P)
\, .
\label{norm_states}
\end{align}
Then, following the derivation presented in Ref.~\cite{Altinoluk:2023qfr}, it can be shown that
\begin{align}
& \hspace{-2.5cm}
 \int dz^+dz'^+\!\int\! d^{2}\z\, d^{2}\z'\,  e^{i(\z'-\z)\cdot(\p-\q)}\,
 \Big\langle\overline{\Psi}(z'^+,\z')\gamma^{-}  U_{F}(+\infty,z'^{+};\z')^{\dag} 
 U_{F}(+\infty,z^{+};\z) \Psi(z^+,\z)\Big\rangle
  \nn\\
 =&\,
 \frac{1}{2P^-}
 \int d^2\b\, e^{i\b\cdot(\p-\q)}\int db^+\, \langle P|\overline{\Psi}(b^+,\b)\gamma^-\, U_F(+\infty,b^+; \b)^\dagger U_F(+\infty, 0; {\bf 0})\Psi(0,{\bf 0})|P\rangle
 \label{CGC_average_to_quantum_SIDIS}
 \, .
\end{align}

However, the unpolarized  quark TMD distribution is defined (up to UV and rapidity regularization and renormalization) as 
\begin{align}
f_1^q({\rm x},\k)=\frac{1}{(2\pi)^3}\int d^2\b\, e^{i\k\cdot\b}\int db^+\,e^{-i{\rm x}P^-b^+}\langle P|\overline{\Psi}(b^+,\b)\frac{\gamma^-}{2}U_F(+\infty,b^+; \b)^\dagger U_F(+\infty, 0; {\bf 0})\Psi(0,{\bf 0})|P\rangle,
\label{q_TMD_def}
\end{align}
where we have neglected the transverse gauge link at infinity, which which do not play a major role for a left-moving target in the light-cone gauge $A^+=0$ used for the derivation of the propagator \eqref{q_prop_IA}. 
Comparing Eqs.~\eqref{CGC_average_to_quantum_SIDIS} and \eqref{q_TMD_def}, one finds that the contribution \eqref{cross_sec_SIDIS_q_2} to the photon to quark cross section  can be written in terms of the unpolarized quark TMD as 
\begin{align}
p^+ 
\frac{d^{3}\sigma^{\gamma_{T}^*\rightarrow q}}{dp^+ d^{2}\p} \bigg|_{\Psi}
 =&\, 
q^{+} \delta(p^{+}\!-\!q^{+})\,
   (2\pi)^2\, \frac{\alpha_{\textrm{em}}\, e_{f}^2}{W^2} \,    
   f_1^q({\rm x}=0,\p\!-\!\q)
 +\textrm{NNEik}
\label{cross_sec_SIDIS_q_2}
\, ,
\end{align}
where we have used that $W^2=(q+P)^2\sim 2q^+ P^-$ up to subleading power corrections at high energy $W$. Now, the explicit $1/W^2$ factor present in Eq.~\eqref{cross_sec_SIDIS_q_2} is a confirmation that the calculated term is a NEik contribution. As a reminder, due to the relation \eqref{vanishing_L_contrib}, one also has
\begin{align}
p^+ 
\frac{d^{3}\sigma^{\gamma_{L}^*\rightarrow q}}{dp^+ d^{2}\p} \bigg|_{\Psi}
 =&\, 
0
 +\textrm{NNEik}
\label{cross_sec_SIDIS_q_L}
\, .
\end{align}
%

\begin{figure}
    \centering
    \includegraphics[scale=0.5]{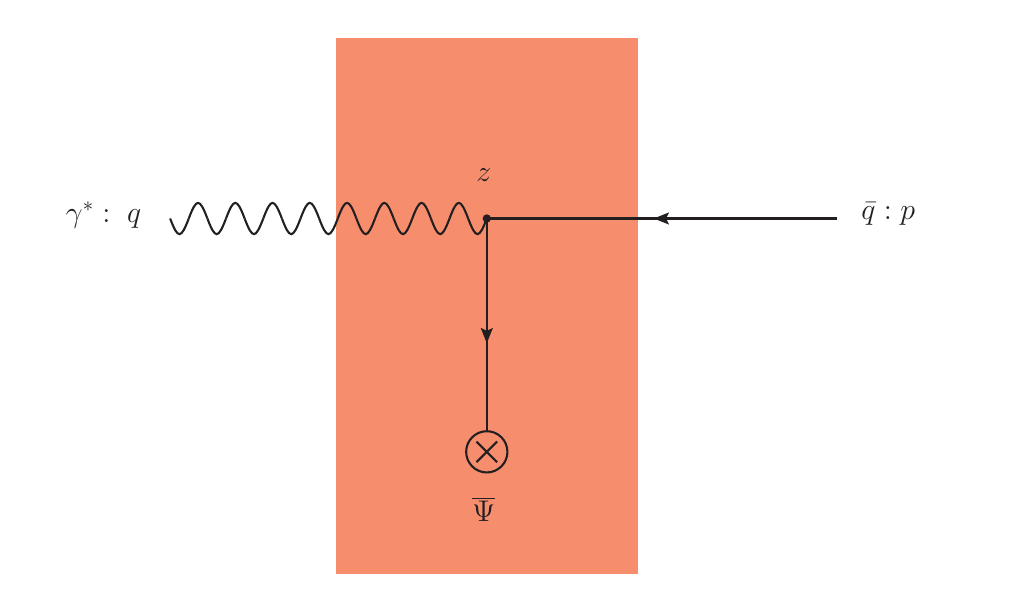}
    \caption{Semi-inclusive deep inelastic scattering (SIDIS) diagram: antiquark production from quark background field}
    \label{fig: SIDIS_qbar}
\end{figure}
Similarly, one can consider antiquark production in SIDIS at partonic level, focusing on the NEik contribution from the quark background field of the target, which corresponds to the diagram represented on Fig.~\eqref{fig: SIDIS_qbar}. Its S-matrix element can be written as
\begin{align}
S_{\gamma^*_{\lambda}\rightarrow \bar{q}} 
=&\,
\lim_{x^{+}\to \infty} \int \! d^{2}\x \int\! dx^{-} \ e^{i\check{p}\cdot x}
 \int\! d^{4}z \ \epsilon_{\lambda}^{\mu}(q) \ e^{-i{q} \cdot z}\
   \ \overline{\Psi}(z)(-iee_{f}\gamma_{\mu}) 
   S_{F}(z,x) 
   \gamma^{+} {v}(\check{p},h)
\label{SIDIS_Smat_qbar_1}
\, .
\end{align}
In Eq.~\eqref{SIDIS_Smat_qbar_1}, for the propagator $S_{F}(z,x) $  with the point $z$ inside the target and the point $x$ in the asymptotic future, one can use the eikonal expression~\cite{Altinoluk:2022jkk,Altinoluk:2024dba} 
\begin{align}
S_{F}(x,y)
\big|^{\rm IA,\bar{q}}_{\rm Eik} &= \int\! \frac{d^{3}\underline{q}}{(2\pi)^{3}} (-1)\frac{\theta(-q^{+})}{2q^{+}}\ e^{iy\check{q}} \
e^{-ix^{-}q^{+}} \ e^{i\x\cdot\q}\  
\bigg[1-\frac{\gamma^{+}\gamma^{i}}{2q^{+}}i\overrightarrow{D}^{F}_{\x^i}\bigg] \ 
      \ U_{F}^{\dagger}(y^{+},x^{+},\x)
       (\check{\sq}+m)
\end{align}
The calculation then follows closely the one for the quark production presented above, and its final result is
\begin{align}
p^+ 
\frac{d^{3}\sigma^{\gamma_{T}^*\rightarrow \bar{q}}}{dp^+ d^{2}\p} \bigg|_{\Psi}
 =&\, 
q^{+} \delta(p^{+}\!-\!q^{+})\,
   (2\pi)^2\, \frac{\alpha_{\textrm{em}}\, e_{f}^2}{W^2} \,    
   f_1^{\bar{q}}({\rm x}=0,\p\!-\!\q)
 +\textrm{NNEik}
\label{cross_sec_SIDIS_qbar_T}
\\
p^+ 
\frac{d^{3}\sigma^{\gamma_{L}^*\rightarrow \bar{q}}}{dp^+ d^{2}\p} \bigg|_{\Psi}
 =&\, 
0
 +\textrm{NNEik}
\label{cross_sec_SIDIS_qbar_L}
\, ,
\end{align}
with the unpolarized  antiquark TMD distribution defined, up to renormalization issues, as 
\begin{align}
f_1^{\bar{q}}({\rm x},\k)=\frac{1}{(2\pi)^3}\int d^2\b\, e^{i\k\cdot\b}\int db^+\,e^{-i{\rm x}P^-b^+}
\tr_{D,F}\left[\langle P|
U_F(+\infty,b^+; \b)\Psi(b^+,\b)
\overline{\Psi}(0,{\bf 0})\frac{\gamma^-}{2}U_F(+\infty, 0; {\bf 0})^\dagger
 |P\rangle
 \right]\, ,
\label{qbar_TMD_def}
\end{align}
where the trace is taken both over the Dirac indices and over the fundamental color indices.

\section{Quark background field contribution to DIS structure functions}
\label{sec:dis}

The inclusive DIS cross section in the single photon exchange approximation can be written as a linear combination of total photon-target cross sections for a virtual photon of either longitudinal or transverse polarization, see Eq.~\eqref{DIS_xsect_one_photon}. 
In turn, these total photon-target cross section can be calculated in two different ways. Either one can calculate the differential cross section for the production of each possible final state separately, at the considered accuracy, and then sum over the final states. Or one can use an appropriate version of the optical theorem, in order to relate the total cross section to the forward elastic scattering amplitude based thanks to unitarity. 

Let us consider the elastic scattering of a photon on the target represented by static background fields, with an incoming photon of momentum $q$ and polarization $\lambda_1$, and an outgoing photon of momentum $p$ and polarization $\lambda_2$, where both photons are virtual, and can be longitudinal or transverse.
In that case, the amplitude can be defined from the S-matrix as
\begin{align}
   S_{\gamma^*_{\lambda_1}\rightarrow \gamma^*_{\lambda_2}}
   =& \,  (2\pi)^4 \delta^{(4)}(p\!-\!q)\, \delta_{\lambda_1, \lambda_2}
   + 2q^{+}(2\pi) \delta(p^{+}\!-\!q^{+})\, i \mathcal{M}_{\gamma^*_{\lambda_1}\rightarrow \gamma^*_{\lambda_2}}\!(q,p)
   \, ,
\label{Smat_DIS_def_ampl}
\end{align}
with the first term corresponding to the possibility of no scattering of the photon with the target. The forward elastic scattering amplitude is then the amplitude with same initial and final states: $p=q$ and $\lambda_2=\lambda_1$.
Using the optical theorem, the total cross section for longitudinal photon is then obtained as
\begin{align}
 \sigma^{\gamma*}_{L} =&\,  2 \text{Im}\,  \langle\mathcal{M}_{\gamma^*_{L}\rightarrow \gamma^*_{L}}\!(q,q) \rangle
 = 2\text{Re}\, \langle (-i)\mathcal{M}_{\gamma^*_{L}\rightarrow \gamma^*_{L}}\!(q,q) \rangle
 \label{optical_th_L}
\end{align}
Similarly, for the transverse photon one has
\begin{align}
 \sigma^{\gamma*}_{T} =&\, \frac{1}{2}\sum_{\lambda}\, 2 \text{Im}\,  \langle\mathcal{M}_{\gamma^*_{\lambda}\rightarrow \gamma^*_{\lambda}}\!(q,q) \rangle
 = \frac{1}{2}\sum_{\lambda}\, 2\text{Re}\, \langle (-i)\mathcal{M}_{\gamma^*_{\lambda}\rightarrow \gamma^*_{\lambda}}\!(q,q) \rangle
 \, , \label{optical_th_T}
\end{align}
averaging over the two transverse polarizations $\lambda$.
 The goal of this section is then to calculate, using Eqs.~\eqref{optical_th_L} and \eqref{optical_th_T}, the NEik contributions to inclusive DIS due to the quark background field of the target. 
We will also cross check the obtained results using the other approach to inclusive DIS, integrating the produced quark or antiquark in the SIDIS result of the previous section.
\begin{figure}[ht]
\subfloat[Diagram with photon absorption before emission along the quark line\label{Fig:DIS_q}]{%
       \includegraphics[width=0.45\textwidth]{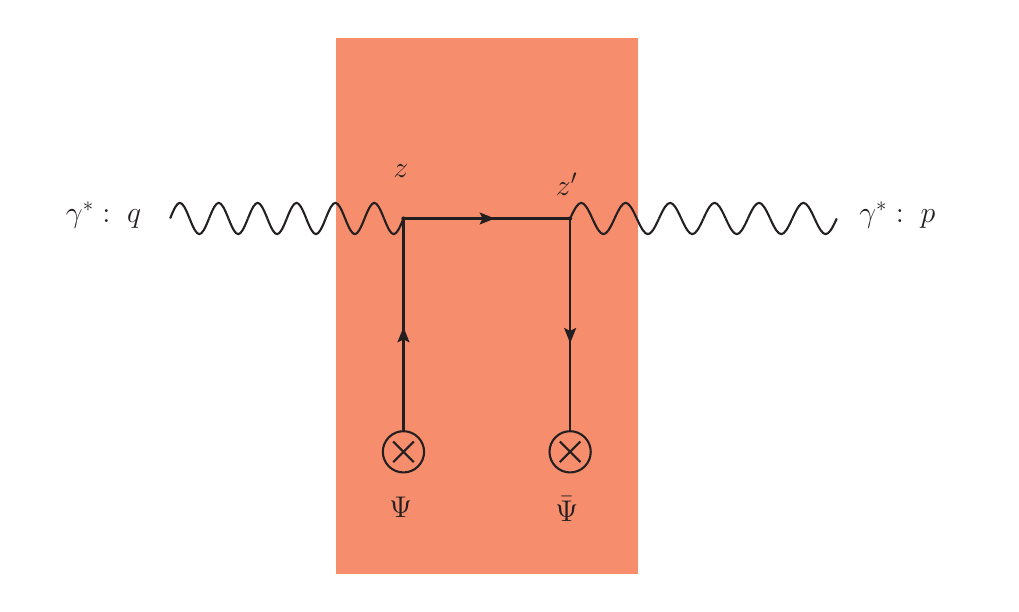}
     }
\hfill
\subfloat[Diagram with photon absorption after emission along the quark line\label{Fig:DIS_qbar}]{%
       \includegraphics[width=0.45\textwidth]{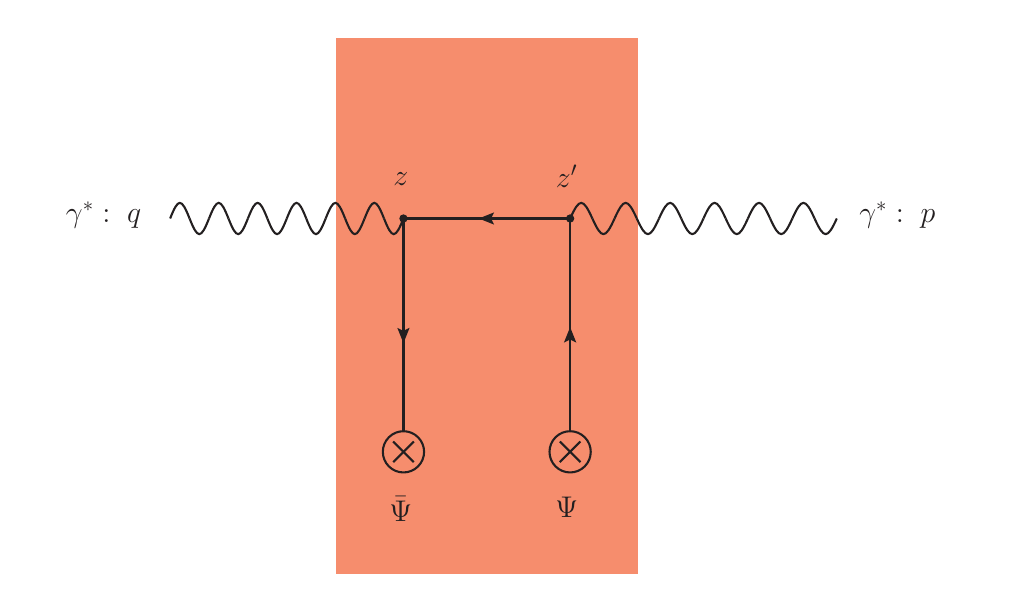}
     }
\caption{\label{Fig:DIS_background_diags}  Quark background field contributions to the photon-target elastic scattering amplitude, and to inclusive DIS via the optical theorem}
\end{figure}

The first two diagrams for photon-target scattering which include the quark background of the target are shown on Fig.~\ref{Fig:DIS_background_diags}. They differ only by the ordering of the vertices along the quark line.
The diagram \ref{Fig:DIS_q} corresponds to a contribution
\begin{align}
   S_{\gamma^*_{\lambda_1}\rightarrow \gamma^*_{\lambda_2}}\bigg|_{\textrm{Diag.}~\ref{Fig:DIS_q}} 
   =& \, 
   \int\! d^{4}z \int\! d^{4}z' \ {\epsilon_{\lambda_2}^{\mu}(p)}^{*} \ e^{ip \cdot z'} \ {\cal T}\, \overline{\Psi}(z') (-iee_{f}\gamma_{\mu}) \ S_{F}(z',z)\ (-iee_{f}\gamma_{\nu}) \ \Psi(z) 
    \epsilon_{\lambda_{1}}^{\nu}(q) \ e^{-iq \cdot z}
\label{Smat_DIS_q_1}
\end{align}
to the S-matrix element, whereas the diagram  \ref{Fig:DIS_qbar} corresponds to a contribution
\begin{align}
   S_{\gamma^*_{\lambda_1}\rightarrow \gamma^*_{\lambda_2}}\bigg|_{\textrm{Diag.}~\ref{Fig:DIS_qbar}} 
=& \,    
   \int\! d^{4}z \int\! d^{4}z' \ {\epsilon_{\lambda_2}^{\mu}(p)}^{*} \ e^{ip \cdot z'} \ {\cal T}\, \overline{\Psi}(z) (-iee_{f}\gamma_{\nu}) \ S_{F}(z,z')\ (-iee_{f}\gamma_{\mu}) \ \Psi(z') 
   \epsilon_{\lambda_{1}}^{\nu}(q) \ e^{-iq \cdot z}
   \label{Smat_DIS_qbar_1}
   \, .
\end{align}
In this section the detailed calculation of the diagram~\ref{Fig:DIS_q} following Eq.~\eqref{Smat_DIS_q_1} is presented, whereas the calculation of the diagram~\ref{Fig:DIS_qbar}, based on Eq.~\eqref{Smat_DIS_qbar_1}, is outlined in the Appendix \ref{App:qbar_DIS}.  

In Eqs.~\eqref{Smat_DIS_q_1} and \eqref{Smat_DIS_qbar_1}, ${\cal T}$ indicates the time ordering of the background fields as quantum operators, and the intermediate quark propagator is the propagator in gluon background field. Under a large longitudinal boost of the target with a parameter $\gamma_t$, the quark background field is non negligeable only over a smaller and smaller range in $z^+$ around  $z^+=0$. Hence, the integrations over $z^+$ and  $z'^+$ each brings a priori a power suppression as $1/\gamma_t$. The intermediate propagator in Eqs.~\eqref{Smat_DIS_q_1} and \eqref{Smat_DIS_qbar_1} thus corresponds to an inside-inside propagator, calculated at Eikonal accuracy in Ref.~\cite{Altinoluk:2024dba} for example, as
\begin{align}
 \label{eq:inside-inside_quark_full}
 S_{F}(x,y)
 \big|_{\rm Eik}^{\rm II} = &\, \int \! \frac{dk^+}{2\pi}   \frac{1}{2k^+} \ e^{-ik^+(x^- - y^-)}  \bigg\{  
i\gamma^+ \delta(x^{+}\!-\!y^{+}) \delta^{2}(\x \!-\! \y)
 \nn \\ 
 & \hspace{-1cm}
 +
 \bigg[ k^+ \gamma^- + m + i \gamma^i \overrightarrow{D}_{\x^i}^{F} \bigg] \, \frac{\gamma^+}{2k^+}
 \int \! d^{2}\z \
\delta^2(\x - \z) \ \delta^2(\z- \y) \ 
 \nn \\ 
 & \hspace{-1cm}
 \times 
\bigg[\theta(x^+\!-\!y^+)\theta(k^+) U_{F}(x^+,y^+;\z)
-\theta(y^+\!-\!x^+)\theta(-k^+)
U_{F}(y^+,x^+;\z)^{\dagger}
\bigg]  
\ \bigg[k^+ \gamma^- + m -i \gamma^j \overleftarrow{D}_{\y^j}^{F}\bigg] \bigg\}
\, .
\end{align}
Note that in the expression \eqref{eq:inside-inside_quark_full}, the first term, proportional to $\gamma^+\delta(x^{+}\!-\!y^{+})$, is formally power enhanced in the high-energy limit compared to the other terms, since it would eliminate one of the integrations along the $+$ direction inside the target, and thus remove one $1/\gamma_t$ suppression factor.
Inserting that term for the propagator in Eq.~\eqref{Smat_DIS_q_1}, one obtains the Dirac structure
\begin{align}
 \overline{\Psi}(z') \slashed{\epsilon}_{\lambda_2}(p)^* \ \gamma^+\ \slashed{\epsilon}_{\lambda_1}(q) \ \Psi(z) 
 =&\,
  \overline{\Psi^{(+)}}(z') \slashed{\epsilon}_{\lambda_2}(p)^* \ \gamma^+\ \slashed{\epsilon}_{\lambda_1}(q) \ \Psi^{+}(z) 
\, .
\end{align}
Indeed, thanks to ${\epsilon_{\lambda_2}^{+}(p)}^{*}=\epsilon_{\lambda_{1}}^{+}(q)=0$ in light-cone gauge, the $\gamma^+$ is projecting out the enhanced components of the background field, and one is left with the insertions of the suppressed components $\overline{\Psi^{(+)}}(z')$ and $\Psi^{+}(z) $, each bringing a suppression by a factor $1/\sqrt{\gamma_t}$ under a large boost of the target. Hence, the instantaneous term in the inside-inside propagator \eqref{eq:inside-inside_quark_full} provides a contribution overall of order $1/(\gamma_t)^2$ to Eqs.~\eqref{Smat_DIS_q_1} and \eqref{Smat_DIS_qbar_1}, corresponding to NNEik order, which is beyond the scope of this study.

By contrast, taking the non-instantaneous contributions in the propagator \eqref{eq:inside-inside_quark_full}, the good components of the quark background field of the target survive. In that case, each quark field insertion brings a $\sqrt{\gamma_t}$ enhancement  factor, and each of the two integrations along the $+$ direction brings a  $1/\gamma_t$ suppression factor, so that the leading contributions to Eqs.~\eqref{Smat_DIS_q_1} and \eqref{Smat_DIS_qbar_1} are overall suppressed as  $1/\gamma_t$, corresponding to NEik order.  This is the contribution that we will focus on in this section.

Hence, neglecting NNEik contributions in Eq.\eqref{Smat_DIS_q_1}, one finds
\begin{align}
   S_{\gamma^*_{\lambda_1}\rightarrow \gamma^*_{\lambda_2}}\bigg|_{\textrm{Diag.}~\ref{Fig:DIS_q}} 
   =& \, 
   -e^2\, e_{f}^2
   \int\! d^{4}z \int\! d^{4}z' \  \ e^{ip^+  z'^-} e^{-i\p \cdot \z'} \int \! \frac{dk^+}{2\pi}   \frac{1}{(2k^+)^2} \ e^{-ik^+(z'^- - z^-)} \int \! d^{2}\v \,
   {\cal T}_+\, \overline{\Psi}(z'^+, \z') \frac{\gamma^- \gamma^+}{2} \slashed{\epsilon}_{\lambda_2}(p)^*
   \nn\\
   &\, \times\,
   \bigg[ k^+ \gamma^- + m + i \gamma^i \overrightarrow{D}_{\z'^i}^{F} \bigg] \, \gamma^+\delta^2(\z' - \v) \ \delta^2(\v- \z) 
   \bigg[\theta(z'^+\!-\!z^+)\theta(k^+) U_{F}(z'^+,z^+;\v)
  \nn\\
   &\,
-\theta(z^+\!-\!z'^+)\theta(-k^+)
U_{F}(z^+,z'^+;\v)^{\dagger}
\bigg]\,
\bigg[k^+ \gamma^- + m -i \gamma^j \overleftarrow{D}_{\z^j}^{F}\bigg]
 \slashed{\epsilon}_{\lambda_1}(q) \, \frac{\gamma^+ \gamma^-}{2}\, \Psi(z^+,\z) 
  \nn\\
   &\, \times\,
        \ e^{-iq^+ z^-}e^{i\q \cdot \z}
        + \text{NNEik}
        \, .
\label{Smat_DIS_q_2}
\end{align}
In particular, in Eq.~\eqref{Smat_DIS_q_2}, we have neglected the phases dependent on $z^+$ or $z'^+$, following the shockwave approximation, and taken all the background field at $z^-=0$, following the static approximation. We have also explicitly written, for clarity, the projection matrices on the good components of the quark field of the target. Then, the unique effect beyond the eikonal approximation in Eq.~\eqref{Smat_DIS_q_2} is the insertion of the quark background fields. Since all the background field insertions are now at $z^-=0$, the time ordering is now equivalent to ordering along the light-cone time $x^+$, denoted by ${\cal T}_+$. Integrating over $z^-$ and $z'^-$, and then over $k^+$, one obtains
\begin{align}
   S_{\gamma^*_{\lambda_1}\rightarrow \gamma^*_{\lambda_2}}\bigg|_{\textrm{Diag.}~\ref{Fig:DIS_q}} 
   =& \, 
   2\pi \delta(p^+\!-\!q^+)
   (-1)\frac{e^2\, e_{f}^2}{(2q^+)^2}
   \int\! dz^+ \int\! d^{2}\z \int\! dz'^+ \int\! d^{2}\z' \  \  e^{-i\p \cdot \z'}  \int \! d^{2}\v \,
   {\cal T}_+\, \overline{\Psi}(z'^+, \z')\frac{\gamma^- \gamma^+}{2} \slashed{\epsilon}_{\lambda_2}(p)^*
   \nn\\
   &\, \times\,
    \bigg[ q^+ \gamma^- + m + i \gamma^i \overrightarrow{D}_{\z'^i}^{F} \bigg] \, \gamma^+\delta^2(\z' - \v) \ \delta^2(\v- \z) 
   \bigg[\theta(z'^+\!-\!z^+)\theta(q^+) U_{F}(z'^+,z^+;\v)
  \nn\\
   &\,
-\theta(z^+\!-\!z'^+)\theta(-q^+)
U_{F}(z^+,z'^+;\v)^{\dagger}
\bigg]\,
\bigg[q^+ \gamma^- + m -i \gamma^j \overleftarrow{D}_{\z^j}^{F}\bigg]
 \slashed{\epsilon}_{\lambda_1}(q) \, \frac{\gamma^+ \gamma^-}{2}\, \Psi(z^+,\z) 
        \ e^{i\q \cdot \z}
        + \text{NNEik}
        \, .
\label{Smat_DIS_q_3}
\end{align}
Since $q^+>0$ for the incoming photon, the contribution with $\theta(-q^+)$ vanishes.
Using ${\epsilon_{\lambda_2}^{+}(p)}^{*}=\epsilon_{\lambda_{1}}^{+}(q)=0$ in light-cone gauge, and $\gamma^+\gamma^+=0$, one finds that the terms containing covariant derivatives or the quark mass vanish, so that
\begin{align}
   S_{\gamma^*_{\lambda_1}\rightarrow \gamma^*_{\lambda_2}}\bigg|_{\textrm{Diag.}~\ref{Fig:DIS_q}} 
   =& \, 
   2\pi \delta(p^+\!-\!q^+)
   (-1)\frac{e^2\, e_{f}^2}{(2q^+)^2}
   \int\! dz^+ \int\! d^{2}\z \int\! dz'^+ \int\! d^{2}\z' \  \  e^{-i\p \cdot \z'}  \int \! d^{2}\v \,
   {\cal T}_+\, \overline{\Psi}(z'^+, \z')\frac{\gamma^- \gamma^+}{2} \slashed{\epsilon}_{\lambda_2}(p)^*
   \nn\\
   &\, \times\,
    [ q^+ \gamma^-  ] \, \gamma^+\delta^2(\z' - \v) \ \delta^2(\v- \z) 
   \theta(z'^+\!-\!z^+) U_{F}(z'^+,z^+;\v)
  \nn\\
   &\, \times\,
[q^+ \gamma^- ]
 \slashed{\epsilon}_{\lambda_1}(q) \, \frac{\gamma^+ \gamma^-}{2}\, \Psi(z^+,\z) 
        \ e^{i\q \cdot \z}
        + \text{NNEik}
\nn\\
   =& \, 
   2\pi \delta(p^+\!-\!q^+)
   (-1)\frac{e^2\, e_{f}^2}{2}
   \int\! dz^+\int\! dz'^+\,   \theta(z'^+\!-\!z^+)     \int\! d^{2}\z   \  e^{-i\z \cdot (\p\!-\!\q)}   \,
   \nn\\
   &\, \times\,
    \overline{\Psi}(z'^+, \z)\frac{\gamma^- \gamma^+}{2} \slashed{\epsilon}_{\lambda_2}(p)^*
      \gamma^-   \, 
    U_{F}(z'^+,z^+;\z)
 \slashed{\epsilon}_{\lambda_1}(q) \, \frac{\gamma^+ \gamma^-}{2}\, \Psi(z^+,\z) 
        + \text{NNEik}        
        \, .
\label{Smat_DIS_q_4}
\end{align}
Note that the light-cone time ordering ${\cal T}_+$ is now redundant is this contribution due to the constraint $z'^+>z^+$, and has thus been dropped.

If one takes either the incoming or the outgoing photon (or both), to have longitudinal polarization, one should take the longitudinal polarization vector defined in Eq.~\eqref{pol_vect_L}.  In either case, the expression \eqref{Smat_DIS_q_4} would then vanish (see Eq.~\eqref{vanishing_L_contrib}). 
Hence, the diagram \ref{Fig:DIS_q} does not contribute to the longitudinal photon - target total cross section, via the optical theorem \eqref{optical_th_L},
\begin{align}
 \sigma^{\gamma*}_{L}\bigg|_{\textrm{Diag.}~\ref{Fig:DIS_q}}  =&\,  0 + \textrm{NNEik}
 \, .
 \label{sigma_tot_L_res}
\end{align}

In order to get a non-zero contribution from Eq.~\eqref{Smat_DIS_q_4}, both the incoming and the outgoing photon have to be transverse, with polarization vectors given in Eq.~\eqref{4_to_2_trans_pol_vect}. Focusing on that case from now on, one finds
\begin{align}
   S_{\gamma^*_{\lambda_1}\rightarrow \gamma^*_{\lambda_2}}\bigg|_{\textrm{Diag.}~\ref{Fig:DIS_q}} 
   =& \, 
   2\pi \delta(p^+\!-\!q^+)
   (-1)\frac{e^2\, e_{f}^2}{2}
   \int\! dz^+\int\! dz'^+\,   \theta(z'^+\!-\!z^+)     \int\! d^{2}\z   \  e^{-i\z \cdot (\p\!-\!\q)}   \,
   \nn\\
   &\, \times\,
    \overline{\Psi}(z'^+, \z)\frac{\gamma^- \gamma^+}{2} (-{\varepsilon_{\lambda_2}^i}^*\, \gamma^i) 
      \gamma^-   \, 
    U_{F}(z'^+,z^+;\z)
 (-{\varepsilon_{\lambda_1}^j} \gamma^j) 
  \frac{\gamma^+ \gamma^-}{2}\, \Psi(z^+,\z) 
        + \text{NNEik}        
 \nn\\
    =& \, 
   2\pi \delta(p^+\!-\!q^+)
  \frac{e^2\, e_{f}^2}{2}\, {\varepsilon_{\lambda_2}^i}^*\,  {\varepsilon_{\lambda_1}^j}
   \int\! dz^+\int\! dz'^+\,   \theta(z'^+\!-\!z^+)     \int\! d^{2}\z   \  e^{-i\z \cdot (\p\!-\!\q)}   \,
   \nn\\
   &\, \times\,
    \overline{\Psi}(z'^+, \z) 
      \gamma^-  \gamma^i  \gamma^j\, 
    U_{F}(z'^+,z^+;\z)
  \Psi(z^+,\z) 
        + \text{NNEik}           
        \, .
\label{Smat_DIS_q_5}
\end{align}
Note that this contribution, from the diagram  \ref{Fig:DIS_q}, vanishes if the background fields are set to zero. Hence, it does not contribute to the first term in Eq.~\eqref{Smat_DIS_def_ampl}, but only to the second term. Hence, the forward scattering amplitude is obtained from the expression \eqref{Smat_DIS_q_5} as
\begin{align}
   i \mathcal{M}_{\gamma^*_{\lambda}\rightarrow \gamma^*_{\lambda}}\!(q,q)\bigg|_{\textrm{Diag.}~\ref{Fig:DIS_q}} 
    =& \, 
  \frac{e^2\, e_{f}^2}{2(2q^+)}\, {\varepsilon_{\lambda}^i}^*\,  {\varepsilon_{\lambda}^j}
   \int\! dz^+\int\! dz'^+\,   \theta(z'^+\!-\!z^+)     \int\! d^{2}\z     \,
   \nn\\
   &\, \times\,
    \overline{\Psi}(z'^+, \z) 
      \gamma^-  \gamma^i  \gamma^j\, 
    U_{F}(z'^+,z^+;\z)
  \Psi(z^+,\z) 
        + \text{NNEik}           
        \, .
\label{ampl_DIS_q_1}
\end{align}
Performing the average over the transverse photon polarizations as 
\begin{align}
\frac{1}{2}\sum_{\lambda}\,  {\varepsilon_{\lambda}^i}^*\,  {\varepsilon_{\lambda}^j}\;   \gamma^i  \gamma^j\, 
=&\, 
\frac{1}{2}\, \delta^{ij}\,  \frac{\{\gamma^i,\gamma^j\}}{2}
=
-1\,
\end{align}
one finds the total cross section for transverse photon from the optical theorem \eqref{optical_th_T} as
\begin{align}
  \sigma^{\gamma*}_{T}\bigg|_{\textrm{Diag.}~\ref{Fig:DIS_q}} 
    =& \, 
  \frac{e^2\, e_{f}^2}{4q^+}\, 
   \int\! dz^+\int\! dz'^+\,   \theta(z'^+\!-\!z^+)     \int\! d^{2}\z     \,
    2\text{Re}\, \big\langle \overline{\Psi}(z'^+, \z) 
      \gamma^-  
    U_{F}(z'^+,z^+;\z)
  \Psi(z^+,\z) \big\rangle
        + \text{NNEik}   
  \nn\\      
    =& \,  
   \frac{\pi\, \alpha_{\textrm{em}}\, e_{f}^2}{q^+}\,      \int\! dz^+\int\! dz'^+\,   \theta(z'^+\!-\!z^+)     \int\! d^{2}\z     \,
   \big\langle \overline{\Psi}(z'^+, \z) \gamma^-  
    U_{F}(+\infty,z'^+;\z)^{\dag}\,  U_{F}(+\infty,z^+;\z)   \Psi(z^+,\z)
  \nn\\  
    &\, \;\;
    +       \overline{\Psi}(z^+, \z) \gamma^-  
    U_{F}(+\infty,z^+;\z)^{\dag}\,  U_{F}(+\infty,z'^+;\z)   \Psi(z'^+,\z)  \big\rangle
        + \text{NNEik}    
          \nn\\      
    =& \,  
      \frac{\pi\, \alpha_{\textrm{em}}\, e_{f}^2}{q^+}\,      \int\! dz^+\int\! dz'^+     \int\! d^{2}\z     \,
   \big\langle       \overline{\Psi}(z'^+, \z) \gamma^-  
    U_{F}(+\infty,z'^+;\z)^{\dag}\,  U_{F}(+\infty,z^+;\z)   \Psi(z^+,\z)  \big\rangle
        + \text{NNEik}       
        \, .
\label{xsec_DIS_q_1}
\end{align}

As a remark, let us note that starting from the contribution \eqref{cross_sec_SIDIS_q_2} to the quark production cross section, one recovers the result \eqref{xsec_DIS_q_1} by integration over the produced quark momentum, as
\begin{align}
 \int dp^+\! \int d^{2}\p\;
\frac{d^{2}\sigma^{\gamma_{T}^*\rightarrow q}}{dp^+ d^{2}\p} \bigg|_{\Psi}
=&\, 
   \frac{\alpha_{\textrm{em}}\, e_{f}^2}{2q^{+}(2\pi)} \,    
   \int dz^+dz'^+\!\int\! d^{2}\z\, d^{2}\z'\,  (2\pi)^2 \delta^{(2)}(\z'-\z)\,
   \nn\\
  &\times\, 
 \Big\langle\overline{\Psi}(z'^+,\z')\gamma^{-}  U_{F}(+\infty,z'^{+};\z')^{\dag} 
 U_{F}(+\infty,z^{+};\z) \Psi(z^+,\z)\Big\rangle
 + \text{NNEik}  
   \nn\\
=&\, 
    \sigma^{\gamma*}_{T}\bigg|_{\textrm{Diag.}~\ref{Fig:DIS_q}} 
\label{xsec_DIS_q_unitarity_rel}
\, .
\end{align}
Hence, one finds that the imaginary part of the amplitude associated with the diagram \ref{Fig:DIS_q} and the modulus square of the diagram \ref{fig: SIDIS_q} are directly related through the optical theorem, which provides a crosscheck of the result \eqref{xsec_DIS_q_1} in our formalism. 

Integrating the identity~\eqref{CGC_average_to_quantum_SIDIS} over $\p$ leads to the relation 
\begin{align}
  &
  \hspace{-3.5cm}
  \int dz^+dz'^+\!\int\! d^{2}\z\, 
 \Big\langle\overline{\Psi}(z'^+,\z)\gamma^{-}  U_{F}(+\infty,z'^{+};\z)^{\dag} 
 U_{F}(+\infty,z^{+};\z) \Psi(z^+,\z)\Big\rangle
  \nn\\
 =&\,
 \frac{1}{2P^-}
 \int db^+\, \langle P|\overline{\Psi}(b^+,{\bf 0})\gamma^-U_F^\dagger(+\infty,b^+; {\bf 0})U_F(+\infty, 0; {\bf 0})\Psi(0,{\bf 0})|P\rangle
 \label{CGC_average_to_quantum_DIS}
 \, , 
\end{align}
so that
\begin{align}
  \sigma^{\gamma*}_{T}\bigg|_{\textrm{Diag.}~\ref{Fig:DIS_q}}    
    =& \,  
      \frac{\pi\, \alpha_{\textrm{em}}\, e_{f}^2}{(2P^- q^+)}\,      \int db^+\, \langle P|\overline{\Psi}(b^+,{\bf 0})\gamma^- U_F^\dagger(+\infty,b^+; {\bf 0})U_F(+\infty, 0; {\bf 0})\Psi(0,{\bf 0})|P\rangle
        + \text{NNEik}       
        \, .
\label{xsec_DIS_q_2}
\end{align}
Since the quark collinear pdf is defined (up to UV renormalization) as  
\begin{align}
q_f({\rm x})=\int \frac{db^+}{2\pi}\, e^{-ib^+{\rm x}P^-}\langle P|\overline{\Psi}(b^+,{\bf 0})\frac{\gamma^-}{2}U_F^\dagger(+\infty,b^+; {\bf 0})U_F(+\infty, 0; {\bf 0})\Psi(0,{\bf 0})|P\rangle
\label{pdf_def}
\, ,
\end{align}
one has
\begin{align}
  \sigma^{\gamma*}_{T}\bigg|_{\textrm{Diag.}~\ref{Fig:DIS_q}}    
    =& \,  
      \frac{(2\pi)^2\, \alpha_{\textrm{em}}\, e_{f}^2}{(2P^- q^+)}\,     
    q_f({\rm x}=0) 
        + \text{NNEik}       
        \, ,
\label{xsec_DIS_q_3}
\end{align}
and, using the relation \eqref{rel_sigmaTL_FTL}, 
\begin{align}
F_{T}(x_{Bj},Q^2)\bigg|_{\textrm{Diag.}~\ref{Fig:DIS_q}}
=&\, 
e_{f}^2\;  \frac{Q^2}{(2P^- q^+)}\,     
    q_f({\rm x}=0) 
        + \text{NNEik}  
=
e_{f}^2\;  x_{Bj}\,     
    q_f({\rm x}=0) 
        + \text{NNEik}  
        \label{FT_q_1}
\\
  F_{L}(x_{Bj},Q^2)\bigg|_{\textrm{Diag.}~\ref{Fig:DIS_q}}
=&\, 
0
        + \text{NNEik}        
\, .
\label{FL_q_1}
\end{align}

In Ref.~\cite{Altinoluk:2024zom}, DIS dijet production is studied including Eikonal and NEik contributions in pure gluon background field. At the leading power in the back-to-back jets regime, it is found that the Eikonal contributions can be written as gluon TMD distributions with momentum fraction ${\rm x}=0$, whereas NEik contributions correspond to the first order correction in the Taylor expansion of the gluon TMDs around ${\rm x}=0$. These NEik corrections then provide an estimate of the physical value of ${\rm x}$ for the gluon TMDs in that process, that would be obtained from a partial resummation of all order power corrections beyond the eikonal approximation. That estimate for the value of ${\rm x}$ is indeed consistent with the value obtained from the TMD factorization formalism for back-to-back dijet production in DIS, not relying on the eikonal approximation.

By comparison, it is reasonable to expect a similar situation here. One should obtain the correct value of ${\rm x}$ for the quark pdf in Eq.~\eqref{FT_q_1} away from the high-energy regime by calculating NNEik contributions to DIS structure functions, more precisely terms which are power suppressed at high energy once by insertions of the quark background field, and once by going beyond the shockwave approximation. That calculation is clearly beyond the scope of the present study, but it might be possible to perform it in the future, thus providing an important confirmation of our results. Nevertheless, by comparison with the collinear factorization formalism for DIS, the expected value for the momentum fraction ${\rm x}$ in the PDFs is $x_{Bj}$. In the high-energy regime,
\begin{align}
x_{Bj} \sim &\,  \frac{Q^2}{(2P^- q^+)}  \sim  \frac{Q^2}{W^2} \ll 1
  \end{align}
so that the difference between $x_{Bj} q_f({\rm x}=0) $ and $x_{Bj} q_f(x_{Bj}) $ is of order $x_{Bj}^2$, and thus NNEik indeed. Hence, at the order of our calculation, we can equivalently write 
\begin{align}
F_{T}(x_{Bj},Q^2)\bigg|_{\textrm{Diag.}~\ref{Fig:DIS_q}}
=&\, 
e_{f}^2\;  x_{Bj}\,     
    q_f(x_{Bj}) 
        + \text{NNEik}  
        \label{FT_q_2}
\, .
\end{align}

The calculation of the Diagram~\ref{Fig:DIS_qbar} is explained in Appendix~\ref{App:qbar_DIS}, starting from the expression \eqref{Smat_DIS_qbar_1}. It leads to
\begin{align}
  \sigma^{\gamma*}_{T}\bigg|_{\textrm{Diag.}~\ref{Fig:DIS_qbar}}    
    =& \,  
      \frac{(2\pi)^2\, \alpha_{\textrm{em}}\, e_{f}^2}{(2P^- q^+)}\,     
    \bar{q}_f({\rm x}=0) 
        + \text{NNEik}  
        \label{xsec_DIS_qbar_3}
 \\
  \sigma^{\gamma*}_{L}\bigg|_{\textrm{Diag.}~\ref{Fig:DIS_qbar}}    
    =& \,  
0       + \text{NNEik}              
        \, ,
\end{align}
with the antiquark collinear pdf 
\begin{align}
\bar{q}_f({\rm x})=\int \frac{db^+}{2\pi}\, e^{-ib^+{\rm x}P^-}
\tr_{D,F}\left[\langle P|
U_F(+\infty,b^+; {\bf 0})\Psi(b^+,{\bf 0})
\overline{\Psi}(0,{\bf 0})\frac{\gamma^-}{2}U_F(+\infty, 0; {\bf 0})^\dagger
 |P\rangle
 \right]
 \, .
\label{qbar_pdf_def}
\end{align}
Hence, at the level of the structure functions, one gets
\begin{align}
F_{T}(x_{Bj},Q^2)\bigg|_{\textrm{Diag.}~\ref{Fig:DIS_qbar}}
=&\, 
e_{f}^2\;  x_{Bj}\,     
    \bar{q}_f({\rm x}=0) 
        + \text{NNEik}  
=
e_{f}^2\;  x_{Bj}\,     
    \bar{q}_f(x_{Bj}) 
        + \text{NNEik} 
        \label{FT_qbar_1} 
\\
  F_{L}(x_{Bj},Q^2)\bigg|_{\textrm{Diag.}~\ref{Fig:DIS_qbar}}
=&\, 
0
        + \text{NNEik}        
\, .
\label{FL_qbar_1}
\end{align}

All in all, combining the contributions from both diagrams, and summing over quark flavors, one arrives at the total NEik contribution to DIS structure functions induced by the quark background field of the target
\begin{align}
F_{T}(x_{Bj},Q^2)\bigg|_{q\textrm{ Backgd.}}
=&\, 
 F_{2}(x_{Bj},Q^2)\bigg|_{q\textrm{ Backgd.}}
=2x_{Bj}\, F_{1}(x_{Bj},Q^2)\bigg|_{q\textrm{ Backgd.}}
 \nn\\       
=&\,
\sum_f
e_{f}^2\;  x_{Bj}\,     
    \Big[{q}_f(x_{Bj})+\bar{q}_f(x_{Bj})\Big] 
        + \text{NNEik} 
        \label{FT_q_back_tot} 
\\
  F_{L}(x_{Bj},Q^2)\bigg|_{q\textrm{ Backgd.}}
=&\, 
0
        + \text{NNEik}        
\, ,
\label{FL_q_back_tot}
\end{align}
which are the same expressions as in the naive parton model.

\section{Discussion: dipole versus quark background field contributions to DIS and SIDIS}
\label{sec:discussion}

It is now in order to discuss the interplay between the quark background field contribution to DIS and SIDIS that we have derived in this article, with the dipole factorization contribution usually considered in the low $x$ literature. Most of the discussion will focus on the inclusive DIS case for definiteness, but the same remarks apply to the SIDIS case.


First of all, let us remind that the approach we follow, generalizing the CGC formalism, is based on a double expansion for dense-dilute scattering processes at high energy. The two small parameters are the QCD coupling $g$, and the ratio of any momentum, mass or virtuality scale over the energy of the collision (or equivalently the inverse of the Lorentz boost factor of the target). Moreover, the target is considered dense in the sense that its associated gluon background field is large, in the nonlinear regime of Yang-Mills theories: ${\cal A}=O(1/g) $, and thus  $g{\cal A}=O(g^0) $ and for the background field strength $g{\cal F}=O(g^0) $ in terms of the small coupling $g$ expansion.    
With this power counting for the gluon background field, the dipole factorization to DIS observables corresponds to the Leading Order term, of order $O(g^0)$, within the Eikonal contribution (leading power at high energy) of order $O(\gamma_t^0)$, with usually the resummation of the high-energy logarithms appearing at higher orders in $g^2$. In particular, the photon always splits into a quark-antiquark before the target in that approximation, and both the quark and antiquark cross the whole target, and interact with it in a fully coherent way described via Wilson lines.  
NLO corrections to the dipole factorization, of order $O(g^2)$, have been calculated \cite{Balitsky:2010ze,Balitsky:2012bs,Beuf:2011xd,Beuf:2016wdz,Beuf:2017bpd,Ducloue:2017ftk,Hanninen:2017ddy,Beuf:2021qqa,Beuf:2021srj,Beuf:2022ndu}, staying within the Eikonal approximation.

Some of the NEik corrections to DIS and SIDIS are obtained by relaxing the Eikonal approximation for the quark propagators through the target in the dipole factorization expression. Such NEik contributions can still be written in a dipole factorization form, but now with a dipole operator decorated by gluon field strength insertions $g{\cal F}$ typically. Such NEik corrections will start at order  $O(g^0)$ in the dense regime for the target.

By contrast, the NEik contributions derived in the two previous sections are of a different form, with the photon vertex located inside the target, and a single quark or antiquark propagating through part of the target only. At the cross section level for inclusive DIS or SIDIS,  as can be seen from our results (Eqs. \eqref{cross_sec_SIDIS_qbar_T} and \eqref{cross_sec_SIDIS_qbar_L} together with Eqs. \eqref{q_TMD_def}  and \eqref{qbar_TMD_def} for SIDIS; Eqs. \eqref{FT_q_back_tot} and  \eqref{FL_q_back_tot} together with \eqref{pdf_def} and \eqref{qbar_pdf_def} for the inclusive DIS), these contributions involve bilinear operators in the quark background field, but without extra power of the coupling $g^2$. The power counting in $g$ for these contributions then depends on the power of $g$ assigned to the quark background field. The Yang-Mills equation for the gluon background field 
\begin{align}  
\partial_{\mu}{\cal F}^{\mu \nu}_a  -g f^{abc} {\cal A}_{\mu}^b  {\cal F}^{\mu \nu}_c
=& \, g\,  \overline{\Psi} \gamma^{\nu} t^a  \Psi
\label{EOM_back}
\end{align}
 suggests that, in the dense regime for the target in which ${\cal A}=O(1/g) $ and ${\cal F}=O(1/g) $, one should take as well $\Psi =O(1/g)$ for consistency. This power counting convention for the quark background field is indeed widely used in the literature for the calculation of NEik corrections \cite{Kovchegov:2015pbl,Kovchegov:2016zex,Kovchegov:2016weo,Kovchegov:2017jxc,Kovchegov:2017lsr,Kovchegov:2018znm,Kovchegov:2018zeq,Kovchegov:2020kxg,Kovchegov:2020hgb,Adamiak:2021ppq,Kovchegov:2021lvz,Kovchegov:2021iyc,Cougoulic:2022gbk,Kovchegov:2022kyy,Borden:2023ugd,Kovchegov:2024aus,Borden:2024bxa,Adamiak:2025dpw,Kovchegov:2025gcg,Borden:2025ehe,Chirilli:2018kkw,Chirilli:2021lif,Altinoluk:2023qfr,Altinoluk:2024dba,Altinoluk:2024tyx}. In this case, the quark background field NEik contributions calculated in this paper are overall of order
 $O(1/g^2)$, and thus perturbatively enhanced. In particular, they are suppressed in terms of powers of energy but enhanced in terms of powers of $g$ compared to the  Eikonal   LO dipole factorization expressions. Moreover, the NEik corrections involving $g{\cal F}$  decorations into the dipole factorization are perturbatively suppressed compared to our results, Eqs. \eqref{cross_sec_SIDIS_qbar_T} and \eqref{FT_q_back_tot}\footnote{There is a third type of NEik contributions to DIS and SIDIS. They consist in having the quark (or the antiquark) from the dipole factorization converting into a gluon and then back into a quark by two successive insertions of the quark background field. Since these conversions occur via QCD vertices, each insertion of the quark background field is accompanied by a factor $g$. Hence, that type of NEik correction is overall of order $O(g^0)$ in the dense target case, like the ones with $g{\cal F}$ insertions. }.

 Hence, the parton-model like contributions from quark background field calculated in the previous sections are the lowest order contributions in $g$ among the NEik corrections to DIS and SIDIS. They are thus the first contributions which should be included in order to complement the dipole factorization results beyond the Eikonal approximation at low  $x_{Bj}$.

As a remark, the fact that the parton-model-like contributions are perturbatively enhanced compared to the dipole factorization contributions is not so surprising: moving away from the low $x_{Bj}$ limit (and instead considering the large $Q^2$ limit), the power suppression of NEik terms disappears, and the parton-model-like terms would become the dominant terms, in agreement with the well known results from collinear factorization physics. However, an in-depth study of such matching between the low $x_{Bj}$ formalism including NEik corrections and the collinear factorization is far beyond the scope of the present study.

As we have discussed, the parton-model-like contributions dominate among the NEik corrections, by power counting in $g$. However, these contributions represent a different channel than the dipole factorization contributions, and involve different operators. Hence, it is not fully guaranteed that the enhancement in terms of $g$ translates into a quantitative enhancement. In order to perform a numerical evaluation of the parton-model-like contributions, one would need to build a model for it, for example by extending the McLerran-Venugopalan model \cite{McLerran:1993ni,McLerran:1993ka,McLerran:1994vd} to include quark background fields. However, even then, the relative size of the contributions would ultimately depend on the model parameters. For that reason, only a fit to the data could confirm the perturbative enhancement of the parton-model-like contributions among the NEik corrections. This would nevertheless be particularly challenging, and would most likely require to consider multiple observables at that accuracy, since the NEik corrections should first be disentangled from the Eikonal contributions.

At the partonic level, a clear difference between the dipole factorization and the parton model-like expressions calculated in this study concern the flow of light-cone momentum. In the parton-model-like expressions (and at lowest order in $g$), the produced quark or antiquark carries the entire light-cone momentum $q^+$ of the incoming photon. These correspond to the so-called aligned jet configurations. 
By contrast, in the dipole factorization, the $q^+$ of the photon is shared between the quark and antiquark in the dipole: $zq^+$ for the quark and $(1\!-\!z)q^+$ for the antiquark, with $0<z<1$. 
Still, even within the dipole factorization, the aligned jet configurations $z\simeq 1$ and $z\simeq 0$ give crucial contributions to inclusive DIS \cite{Mantysaari:2018nng} and SIDIS \cite{Marquet:2009ca,Iancu:2020jch,Altinoluk:2024vgg}. 
Hence, in order to include the parton-model-like contributions that we have calculated as subleading power corrections at high energy to the dipole factorization, further studies are needed to properly treat the aligned jet region  
and avoid double counting of quark modes with small light-cone momentum. This would represent an important step towards generalizing the CGC formalism into an effective theory systematically improvable beyond leading power in the high-energy limit.

\section{Summary}
\label{sec:summary} 

In this work, we investigate the inclusive DIS and SIDIS processes at NEik accuracy, focusing exclusively on NEik corrections arising from $t$-channel quark exchanges. These corrections probe the quark background field of the target at NEik order. First, we computed the quark background field contribution to the SIDIS and found that the result vanishes for  longitudinal photon at NEik order, up to higher order corrections in $\alpha_s$. The contribution for  transverse photon polarization  is given in Eq. \eqref{cross_sec_SIDIS_q_2}, written in terms of unpolarized quark TMD at ${\rm x}=0$ and with an explicit $1/W^2$ factor confirming the power suppression of the NEik correction. Similar results are also obtained for the antiquark production in SIDIS. 

We have also computed the quark background field contributions to the DIS structure functions. The computations are performed in two ways: either using the optical theorem or integrating out the quark production phase space in the SIDIS cross section obtained earlier, and an agreement between the results is obtained as expected. We have shown that quark background contribution to the longitudinal structure function vanishes at NEik, up to higher order corrections in $\alpha_s$.
On the other hand, the quark background contribution to the transverse structure function is given in Eq. \eqref{FT_q_back_tot} and it is written as a combination of quark and antiquark collinear PDF. 

The contributions obtained in such a way for SIDIS and DIS,  induced by the quark background field, coincide with the low $x$ expansion of the parton model results.  
We have then discussed power counting for these quark background field induced corrections in the dilute-dense limit and argued that such parton-model–like contributions 
 represent the lowest-order terms in the coupling $g$ among the NEik corrections to DIS and SIDIS. These contributions therefore constitute the first NEik power corrections at low $x_{Bj}$ that must be included beyond the dipole factorization results, derived in the eikonal approximation.

\acknowledgements{
TA is supported in part by the National Science Centre (Poland) under the research Grant No. 2023/50/E/ST2/00133 (SONATA BIS 13). GB and SM are supported in part by the National Science Centre (Poland) under the research Grant No. 2020/38/E/ST2/00122 (SONATA BIS 10). 
SM is also supported by the Research Council of Finland, the Centre of Excellence in Quark Matter, and by the European Research Council (ERC, grant agreements No. ERC-2023-101123801 GlueSatLight and ERC-2018-ADG-835105 YoctoLHC).
}
\appendix



\section{Calculation of the antiquark contribution to DIS structure functions.}
\label{App:qbar_DIS} 

In this appendix, the main steps of the calculation of the Diagram~\ref{Fig:DIS_qbar} are described, highlighting its main differences compared to the calculation of the Diagram~\ref{Fig:DIS_q} presented in section \ref{sec:dis}. Starting from the expression \eqref{Smat_DIS_qbar_1}, on can perform similar approximation as in the case of the Diagram~\ref{Fig:DIS_q}, neglecting effects which would matter only at NNEik accuracy onwards. In particular, the expression \eqref{eq:inside-inside_quark_full} can be used for the quark propagator, excluding the instantaneous term. In such a way, one finds
\begin{align}
   S_{\gamma^*_{\lambda_1}\rightarrow \gamma^*_{\lambda_2}}\bigg|_{\textrm{Diag.}~\ref{Fig:DIS_qbar}} 
   =& \, 
   -e^2\, e_{f}^2
   \int\! d^{4}z \int\! d^{4}z' \  \ e^{ip^+  z'^-} e^{-i\p \cdot \z'} \int \! \frac{dk^+}{2\pi}   \frac{1}{(2k^+)^2} \ e^{-ik^+(z^- - z'^-)} \int \! d^{2}\v \,
   {\cal T}_+\, \overline{\Psi}(z^+, \z) \frac{\gamma^- \gamma^+}{2} 
   \slashed{\epsilon}_{\lambda_1}(q)
   \nn\\
   &\, \times\,
   \bigg[ k^+ \gamma^- + m + i \gamma^i \overrightarrow{D}_{\z^i}^{F} \bigg] \, \gamma^+\delta^2(\z - \v) \ \delta^2(\v- \z') 
   \bigg[\theta(z^+\!-\!z'^+)\theta(k^+) U_{F}(z^+,z'^+;\v)
  \nn\\
   &\,
-\theta(z'^+\!-\!z^+)\theta(-k^+)
U_{F}(z'^+,z^+;\v)^{\dagger}
\bigg]\,
\bigg[k^+ \gamma^- + m -i \gamma^j \overleftarrow{D}_{\z'^j}^{F}\bigg]
 \slashed{\epsilon}_{\lambda_2}(p)^* \, 
 \frac{\gamma^+ \gamma^-}{2}\, \Psi(z'^+,\z') 
  \nn\\
   &\, \times\,
        \ e^{-iq^+ z^-}e^{i\q \cdot \z}
        + \text{NNEik}
\nn\\        
   =& \, 
   2\pi \delta(p^+\!-\!q^+)
   (-1)\frac{e^2\, e_{f}^2}{(-2q^+)^2}
   \int\! dz^+ \int\! d^{2}\z \int\! dz'^+ \int\! d^{2}\z' \  \  e^{-i\p \cdot \z'}  \int \! d^{2}\v \,
   {\cal T}_+\, \overline{\Psi}(z^+, \z)\frac{\gamma^- \gamma^+}{2} 
    \slashed{\epsilon}_{\lambda_1}(q)
   \nn\\
   &\, \times\,
    \bigg[ -q^+ \gamma^- + m + i \gamma^i \overrightarrow{D}_{\z^i}^{F} \bigg] \, \gamma^+\delta^2(\z - \v) \ \delta^2(\v- \z') 
   \bigg[\theta(z^+\!-\!z'^+)\theta(-q^+) U_{F}(z^+,z'^+;\v)
  \nn\\
   &\,
-\theta(z'^+\!-\!z^+)\theta(q^+)
U_{F}(z'^+,z^+;\v)^{\dagger}
\bigg]\,
\bigg[-q^+ \gamma^- + m -i \gamma^j \overleftarrow{D}_{\z'^j}^{F}\bigg]
\slashed{\epsilon}_{\lambda_2}(p)^*
  \, \frac{\gamma^+ \gamma^-}{2}\, \Psi(z'^+,\z') 
        \ e^{i\q \cdot \z}
        + \text{NNEik}
        \, ,
\label{Smat_DIS_qbar_2}
\end{align}
because the integrations over $z^-$ and $z'^-$ force $k^+=-q^+=-p^+$. Then, since $q^+>0$, it is now the term with $U_{F}^{\dagger}$ from the propagator which is selected this time, bringing a $(-1)$ factor which was not there in the case of the Diagram~\ref{Fig:DIS_q}. Once again, the terms with covariant derivatives or quark mass drop, and one arrives at 
\begin{align}
   S_{\gamma^*_{\lambda_1}\rightarrow \gamma^*_{\lambda_2}}\bigg|_{\textrm{Diag.}~\ref{Fig:DIS_qbar}}       
   =& \, 
   2\pi \delta(p^+\!-\!q^+)
   (-1)\frac{e^2\, e_{f}^2}{2}
   \int\! dz^+  \int\! dz'^+  \, (-1) \theta(z'^+\!-\!z^+) \int\! d^{2}\z\  e^{-i\z \cdot (\p\!-\!\q)}  
   {\cal T}_+\, \overline{\Psi}(z^+, \z)\frac{\gamma^- \gamma^+}{2}
  \slashed{\epsilon}_{\lambda_1}(q)   
   \nn\\
   &\, \times\,
    \gamma^-  
U_{F}(z'^+,z^+;\z)^{\dagger}
 \slashed{\epsilon}_{\lambda_2}(p)^*
 \, \frac{\gamma^+ \gamma^-}{2}\, \Psi(z'^+,\z) 
        + \text{NNEik}
        \, .
\label{Smat_DIS_qbar_3}
\end{align}
Like in the case of Diagram~\ref{Fig:DIS_q}, one obtains a non-zero contribution only if both the incoming and the outgoing photons have a transverse polarization. In that case, one finds
\begin{align}
   S_{\gamma^*_{\lambda_1}\rightarrow \gamma^*_{\lambda_2}}\bigg|_{\textrm{Diag.}~\ref{Fig:DIS_qbar}} 
    =& \, 
   2\pi \delta(p^+\!-\!q^+)
  \frac{e^2\, e_{f}^2}{2}\, {\varepsilon_{\lambda_2}^i}^*\,  {\varepsilon_{\lambda_1}^j}
   \int\! dz^+\int\! dz'^+\,   \theta(z'^+\!-\!z^+)     \int\! d^{2}\z   \  e^{-i\z \cdot (\p\!-\!\q)}   \,
   \nn\\
   &\, \times\,
   (-1){\cal T}_+\, \overline{\Psi}(z^+, \z) 
      \gamma^-  \gamma^j  \gamma^i\, 
    U_{F}(z'^+,z^+;\z)^{\dagger}
  \Psi(z'^+,\z) 
        + \text{NNEik}           
        \, .
\label{Smat_DIS_qbar_4}
\end{align}
Hence, the corresponding contribution to the forward scattering amplitude is
\begin{align}
   i \mathcal{M}_{\gamma^*_{\lambda}\rightarrow \gamma^*_{\lambda}}\!(q,q)\bigg|_{\textrm{Diag.}~\ref{Fig:DIS_qbar}} 
    =& \, 
  \frac{e^2\, e_{f}^2}{2(2q^+)}\, {\varepsilon_{\lambda}^i}^*\,  {\varepsilon_{\lambda}^j}
   \int\! dz^+\int\! dz'^+\,   \theta(z'^+\!-\!z^+)     \int\! d^{2}\z     \,
   \nn\\
   &\, \times\,
    (-1){\cal T}_+\, \overline{\Psi}(z^+, \z) 
      \gamma^-  \gamma^j  \gamma^i\, 
    U_{F}(z'^+,z^+;\z)^{\dagger}
  \Psi(z'^+,\z) 
        + \text{NNEik}
\nn\\
=& \,  
   \frac{e^2\, e_{f}^2}{2(2q^+)}\, {\varepsilon_{\lambda}^i}^*\,  {\varepsilon_{\lambda}^j}
   \int\! dz^+\int\! dz'^+\,   \theta(z'^+\!-\!z^+)     \int\! d^{2}\z     \,
   \nn\\
   &\, \times\,
   \tr_{D,F}\left[
      U_{F}(+\infty,z'^+;\z)
  \Psi(z'^+,\z)\,      
   \, \overline{\Psi}(z^+, \z) 
      \gamma^-  \gamma^j  \gamma^i\, 
    U_{F}(+\infty,z^+;\z)^{\dagger}    
  \right]  
  + \text{NNEik}
        \, .
\label{ampl_DIS_qbar_1}
\end{align}
Using the constraint $z'^+>z^+$ to explicitly place the quark background fields in the $z^+$ ordered way, one generates an extra factor $(-1)$, which compensate the one appearing earlier, so that both Eq.~\eqref{ampl_DIS_qbar_1} and Eq.~\eqref{ampl_DIS_q_1} have the same overall sign. 
Then, following the same steps as in the case of the Diagram~\ref{Fig:DIS_q}, one obtains
\begin{align}
  \sigma^{\gamma*}_{T}\bigg|_{\textrm{Diag.}~\ref{Fig:DIS_qbar}} 
    =& \, 
  \frac{e^2\, e_{f}^2}{4q^+}\, 
   \int\! dz^+\int\! dz'^+\,   \theta(z'^+\!-\!z^+)     \int\! d^{2}\z     \,
   \nn\\
   &\, \times\,
    2\text{Re}\, \big\langle
    \tr_{D,F}\left[
      U_{F}(+\infty,z'^+;\z)
  \Psi(z'^+,\z)\,      
   \, \overline{\Psi}(z^+, \z) 
      \gamma^-  
    U_{F}(+\infty,z^+;\z)^{\dagger}    
  \right]  
  \big\rangle
        + \text{NNEik}   
  \nn\\      
    =& \,  
      \frac{\pi\, \alpha_{\textrm{em}}\, e_{f}^2}{q^+}      \int\! dz^+\!\int\! dz'^+ \!\!    \int\! d^{2}\z     \,
   \big\langle
    \tr_{D,F}\left[
      U_{F}(+\infty,z'^+;\z)
  \Psi(z'^+,\z)\,      
   \, \overline{\Psi}(z^+, \z) 
      \gamma^-  
    U_{F}(+\infty,z^+;\z)^{\dagger}    
  \right]  
  \big\rangle
        + \text{NNEik}       
        \, , 
\label{xsec_DIS_qbar_1}
\end{align}
and then the relation \eqref{xsec_DIS_qbar_3}.



\bibliography{mybib_New}

\begin{thebibliography}{120}%
\makeatletter
\providecommand \@ifxundefined [1]{%
 \@ifx{#1\undefined}
}%
\providecommand \@ifnum [1]{%
 \ifnum #1\expandafter \@firstoftwo
 \else \expandafter \@secondoftwo
 \fi
}%
\providecommand \@ifx [1]{%
 \ifx #1\expandafter \@firstoftwo
 \else \expandafter \@secondoftwo
 \fi
}%
\providecommand \natexlab [1]{#1}%
\providecommand \enquote  [1]{``#1''}%
\providecommand \bibnamefont  [1]{#1}%
\providecommand \bibfnamefont [1]{#1}%
\providecommand \citenamefont [1]{#1}%
\providecommand \href@noop [0]{\@secondoftwo}%
\providecommand \href [0]{\begingroup \@sanitize@url \@href}%
\providecommand \@href[1]{\@@startlink{#1}\@@href}%
\providecommand \@@href[1]{\endgroup#1\@@endlink}%
\providecommand \@sanitize@url [0]{\catcode `\\12\catcode `\$12\catcode
  `\&12\catcode `\#12\catcode `\^12\catcode `\_12\catcode `\%12\relax}%
\providecommand \@@startlink[1]{}%
\providecommand \@@endlink[0]{}%
\providecommand \url  [0]{\begingroup\@sanitize@url \@url }%
\providecommand \@url [1]{\endgroup\@href {#1}{\urlprefix }}%
\providecommand \urlprefix  [0]{URL }%
\providecommand \Eprint [0]{\href }%
\providecommand \doibase [0]{https://doi.org/}%
\providecommand \selectlanguage [0]{\@gobble}%
\providecommand \bibinfo  [0]{\@secondoftwo}%
\providecommand \bibfield  [0]{\@secondoftwo}%
\providecommand \translation [1]{[#1]}%
\providecommand \BibitemOpen [0]{}%
\providecommand \bibitemStop [0]{}%
\providecommand \bibitemNoStop [0]{.\EOS\space}%
\providecommand \EOS [0]{\spacefactor3000\relax}%
\providecommand \BibitemShut  [1]{\csname bibitem#1\endcsname}%
\let\auto@bib@innerbib\@empty
\bibitem [{\citenamefont {Collins}\ \emph {et~al.}(1989)\citenamefont
  {Collins}, \citenamefont {Soper},\ and\ \citenamefont
  {Sterman}}]{Collins:1989gx}%
  \BibitemOpen
  \bibfield  {author} {\bibinfo {author} {\bibfnamefont {J.~C.}\ \bibnamefont
  {Collins}}, \bibinfo {author} {\bibfnamefont {D.~E.}\ \bibnamefont {Soper}},\
  and\ \bibinfo {author} {\bibfnamefont {G.~F.}\ \bibnamefont {Sterman}},\
  }\href {https://doi.org/10.1142/9789814503266_0001} {\bibfield  {journal}
  {\bibinfo  {journal} {Adv. Ser. Direct. High Energy Phys.}\ }\textbf
  {\bibinfo {volume} {5}},\ \bibinfo {pages} {1} (\bibinfo {year} {1989})},\
  \Eprint {https://arxiv.org/abs/hep-ph/0409313} {arXiv:hep-ph/0409313}
  \BibitemShut {NoStop}%
\bibitem [{\citenamefont {Gribov}\ and\ \citenamefont
  {Lipatov}(1972)}]{Gribov:1972ri}%
  \BibitemOpen
  \bibfield  {author} {\bibinfo {author} {\bibfnamefont {V.~N.}\ \bibnamefont
  {Gribov}}\ and\ \bibinfo {author} {\bibfnamefont {L.~N.}\ \bibnamefont
  {Lipatov}},\ }\href@noop {} {\bibfield  {journal} {\bibinfo  {journal} {Sov.
  J. Nucl. Phys.}\ }\textbf {\bibinfo {volume} {15}},\ \bibinfo {pages} {438}
  (\bibinfo {year} {1972})}\BibitemShut {NoStop}%
\bibitem [{\citenamefont {Altarelli}\ and\ \citenamefont
  {Parisi}(1977)}]{Altarelli:1977zs}%
  \BibitemOpen
  \bibfield  {author} {\bibinfo {author} {\bibfnamefont {G.}~\bibnamefont
  {Altarelli}}\ and\ \bibinfo {author} {\bibfnamefont {G.}~\bibnamefont
  {Parisi}},\ }\href {https://doi.org/10.1016/0550-3213(77)90384-4} {\bibfield
  {journal} {\bibinfo  {journal} {Nucl. Phys. B}\ }\textbf {\bibinfo {volume}
  {126}},\ \bibinfo {pages} {298} (\bibinfo {year} {1977})}\BibitemShut
  {NoStop}%
\bibitem [{\citenamefont {Dokshitzer}(1977)}]{Dokshitzer:1977sg}%
  \BibitemOpen
  \bibfield  {author} {\bibinfo {author} {\bibfnamefont {Y.~L.}\ \bibnamefont
  {Dokshitzer}},\ }\href@noop {} {\bibfield  {journal} {\bibinfo  {journal}
  {Sov. Phys. JETP}\ }\textbf {\bibinfo {volume} {46}},\ \bibinfo {pages} {641}
  (\bibinfo {year} {1977})}\BibitemShut {NoStop}%
\bibitem [{\citenamefont {Lipatov}(1976)}]{Lipatov:1976zz}%
  \BibitemOpen
  \bibfield  {author} {\bibinfo {author} {\bibfnamefont {L.~N.}\ \bibnamefont
  {Lipatov}},\ }\href@noop {} {\bibfield  {journal} {\bibinfo  {journal} {Sov.
  J. Nucl. Phys.}\ }\textbf {\bibinfo {volume} {23}},\ \bibinfo {pages} {338}
  (\bibinfo {year} {1976})}\BibitemShut {NoStop}%
\bibitem [{\citenamefont {Kuraev}\ \emph {et~al.}(1977)\citenamefont {Kuraev},
  \citenamefont {Lipatov},\ and\ \citenamefont {Fadin}}]{Kuraev:1977fs}%
  \BibitemOpen
  \bibfield  {author} {\bibinfo {author} {\bibfnamefont {E.~A.}\ \bibnamefont
  {Kuraev}}, \bibinfo {author} {\bibfnamefont {L.~N.}\ \bibnamefont
  {Lipatov}},\ and\ \bibinfo {author} {\bibfnamefont {V.~S.}\ \bibnamefont
  {Fadin}},\ }\href@noop {} {\bibfield  {journal} {\bibinfo  {journal} {Sov.
  Phys. JETP}\ }\textbf {\bibinfo {volume} {45}},\ \bibinfo {pages} {199}
  (\bibinfo {year} {1977})}\BibitemShut {NoStop}%
\bibitem [{\citenamefont {Balitsky}\ and\ \citenamefont
  {Lipatov}(1978)}]{Balitsky:1978ic}%
  \BibitemOpen
  \bibfield  {author} {\bibinfo {author} {\bibfnamefont {I.~I.}\ \bibnamefont
  {Balitsky}}\ and\ \bibinfo {author} {\bibfnamefont {L.~N.}\ \bibnamefont
  {Lipatov}},\ }\href@noop {} {\bibfield  {journal} {\bibinfo  {journal} {Sov.
  J. Nucl. Phys.}\ }\textbf {\bibinfo {volume} {28}},\ \bibinfo {pages} {822}
  (\bibinfo {year} {1978})}\BibitemShut {NoStop}%
\bibitem [{\citenamefont {Aaron}\ \emph {et~al.}(2010)\citenamefont {Aaron}
  \emph {et~al.}}]{H1:2009pze}%
  \BibitemOpen
  \bibfield  {author} {\bibinfo {author} {\bibfnamefont {F.~D.}\ \bibnamefont
  {Aaron}} \emph {et~al.} (\bibinfo {collaboration} {H1, ZEUS}),\ }\href
  {https://doi.org/10.1007/JHEP01(2010)109} {\bibfield  {journal} {\bibinfo
  {journal} {JHEP}\ }\textbf {\bibinfo {volume} {01}},\ \bibinfo {pages}
  {109}},\ \Eprint {https://arxiv.org/abs/0911.0884} {arXiv:0911.0884 [hep-ex]}
  \BibitemShut {NoStop}%
\bibitem [{\citenamefont {Gelis}\ \emph {et~al.}(2010)\citenamefont {Gelis},
  \citenamefont {Iancu}, \citenamefont {Jalilian-Marian},\ and\ \citenamefont
  {Venugopalan}}]{Gelis:2010nm}%
  \BibitemOpen
  \bibfield  {author} {\bibinfo {author} {\bibfnamefont {F.}~\bibnamefont
  {Gelis}}, \bibinfo {author} {\bibfnamefont {E.}~\bibnamefont {Iancu}},
  \bibinfo {author} {\bibfnamefont {J.}~\bibnamefont {Jalilian-Marian}},\ and\
  \bibinfo {author} {\bibfnamefont {R.}~\bibnamefont {Venugopalan}},\ }\href
  {https://doi.org/10.1146/annurev.nucl.010909.083629} {\bibfield  {journal}
  {\bibinfo  {journal} {Ann. Rev. Nucl. Part. Sci.}\ }\textbf {\bibinfo
  {volume} {60}},\ \bibinfo {pages} {463} (\bibinfo {year} {2010})},\ \Eprint
  {https://arxiv.org/abs/1002.0333} {arXiv:1002.0333 [hep-ph]} \BibitemShut
  {NoStop}%
\bibitem [{\citenamefont {Albacete}\ and\ \citenamefont
  {Marquet}(2014)}]{Albacete:2014fwa}%
  \BibitemOpen
  \bibfield  {author} {\bibinfo {author} {\bibfnamefont {J.~L.}\ \bibnamefont
  {Albacete}}\ and\ \bibinfo {author} {\bibfnamefont {C.}~\bibnamefont
  {Marquet}},\ }\href {https://doi.org/10.1016/j.ppnp.2014.01.004} {\bibfield
  {journal} {\bibinfo  {journal} {Prog. Part. Nucl. Phys.}\ }\textbf {\bibinfo
  {volume} {76}},\ \bibinfo {pages} {1} (\bibinfo {year} {2014})},\ \Eprint
  {https://arxiv.org/abs/1401.4866} {arXiv:1401.4866 [hep-ph]} \BibitemShut
  {NoStop}%
\bibitem [{\citenamefont {Blaizot}(2017)}]{Blaizot:2016qgz}%
  \BibitemOpen
  \bibfield  {author} {\bibinfo {author} {\bibfnamefont {J.-P.}\ \bibnamefont
  {Blaizot}},\ }\href {https://doi.org/10.1088/1361-6633/aa5435} {\bibfield
  {journal} {\bibinfo  {journal} {Rept. Prog. Phys.}\ }\textbf {\bibinfo
  {volume} {80}},\ \bibinfo {pages} {032301} (\bibinfo {year} {2017})},\
  \Eprint {https://arxiv.org/abs/1607.04448} {arXiv:1607.04448 [hep-ph]}
  \BibitemShut {NoStop}%
\bibitem [{\citenamefont {Balitsky}(1996)}]{Balitsky:1995ub}%
  \BibitemOpen
  \bibfield  {author} {\bibinfo {author} {\bibfnamefont {I.}~\bibnamefont
  {Balitsky}},\ }\href {https://doi.org/10.1016/0550-3213(95)00638-9}
  {\bibfield  {journal} {\bibinfo  {journal} {Nucl. Phys. B}\ }\textbf
  {\bibinfo {volume} {463}},\ \bibinfo {pages} {99} (\bibinfo {year} {1996})},\
  \Eprint {https://arxiv.org/abs/hep-ph/9509348} {arXiv:hep-ph/9509348}
  \BibitemShut {NoStop}%
\bibitem [{\citenamefont {Kovchegov}(1999)}]{Kovchegov:1999yj}%
  \BibitemOpen
  \bibfield  {author} {\bibinfo {author} {\bibfnamefont {Y.~V.}\ \bibnamefont
  {Kovchegov}},\ }\href {https://doi.org/10.1103/PhysRevD.60.034008} {\bibfield
   {journal} {\bibinfo  {journal} {Phys. Rev. D}\ }\textbf {\bibinfo {volume}
  {60}},\ \bibinfo {pages} {034008} (\bibinfo {year} {1999})},\ \Eprint
  {https://arxiv.org/abs/hep-ph/9901281} {arXiv:hep-ph/9901281} \BibitemShut
  {NoStop}%
\bibitem [{\citenamefont {Kovchegov}(2000)}]{Kovchegov:1999ua}%
  \BibitemOpen
  \bibfield  {author} {\bibinfo {author} {\bibfnamefont {Y.~V.}\ \bibnamefont
  {Kovchegov}},\ }\href {https://doi.org/10.1103/PhysRevD.61.074018} {\bibfield
   {journal} {\bibinfo  {journal} {Phys. Rev. D}\ }\textbf {\bibinfo {volume}
  {61}},\ \bibinfo {pages} {074018} (\bibinfo {year} {2000})},\ \Eprint
  {https://arxiv.org/abs/hep-ph/9905214} {arXiv:hep-ph/9905214} \BibitemShut
  {NoStop}%
\bibitem [{\citenamefont {Jalilian-Marian}\ \emph
  {et~al.}(1997{\natexlab{a}})\citenamefont {Jalilian-Marian}, \citenamefont
  {Kovner}, \citenamefont {McLerran},\ and\ \citenamefont
  {Weigert}}]{Jalilian-Marian:1996mkd}%
  \BibitemOpen
  \bibfield  {author} {\bibinfo {author} {\bibfnamefont {J.}~\bibnamefont
  {Jalilian-Marian}}, \bibinfo {author} {\bibfnamefont {A.}~\bibnamefont
  {Kovner}}, \bibinfo {author} {\bibfnamefont {L.~D.}\ \bibnamefont
  {McLerran}},\ and\ \bibinfo {author} {\bibfnamefont {H.}~\bibnamefont
  {Weigert}},\ }\href {https://doi.org/10.1103/PhysRevD.55.5414} {\bibfield
  {journal} {\bibinfo  {journal} {Phys. Rev. D}\ }\textbf {\bibinfo {volume}
  {55}},\ \bibinfo {pages} {5414} (\bibinfo {year} {1997}{\natexlab{a}})},\
  \Eprint {https://arxiv.org/abs/hep-ph/9606337} {arXiv:hep-ph/9606337}
  \BibitemShut {NoStop}%
\bibitem [{\citenamefont {Jalilian-Marian}\ \emph
  {et~al.}(1997{\natexlab{b}})\citenamefont {Jalilian-Marian}, \citenamefont
  {Kovner}, \citenamefont {Leonidov},\ and\ \citenamefont
  {Weigert}}]{Jalilian-Marian:1997qno}%
  \BibitemOpen
  \bibfield  {author} {\bibinfo {author} {\bibfnamefont {J.}~\bibnamefont
  {Jalilian-Marian}}, \bibinfo {author} {\bibfnamefont {A.}~\bibnamefont
  {Kovner}}, \bibinfo {author} {\bibfnamefont {A.}~\bibnamefont {Leonidov}},\
  and\ \bibinfo {author} {\bibfnamefont {H.}~\bibnamefont {Weigert}},\ }\href
  {https://doi.org/10.1016/S0550-3213(97)00440-9} {\bibfield  {journal}
  {\bibinfo  {journal} {Nucl. Phys. B}\ }\textbf {\bibinfo {volume} {504}},\
  \bibinfo {pages} {415} (\bibinfo {year} {1997}{\natexlab{b}})},\ \Eprint
  {https://arxiv.org/abs/hep-ph/9701284} {arXiv:hep-ph/9701284} \BibitemShut
  {NoStop}%
\bibitem [{\citenamefont {Jalilian-Marian}\ \emph
  {et~al.}(1998{\natexlab{a}})\citenamefont {Jalilian-Marian}, \citenamefont
  {Kovner}, \citenamefont {Leonidov},\ and\ \citenamefont
  {Weigert}}]{Jalilian-Marian:1997jhx}%
  \BibitemOpen
  \bibfield  {author} {\bibinfo {author} {\bibfnamefont {J.}~\bibnamefont
  {Jalilian-Marian}}, \bibinfo {author} {\bibfnamefont {A.}~\bibnamefont
  {Kovner}}, \bibinfo {author} {\bibfnamefont {A.}~\bibnamefont {Leonidov}},\
  and\ \bibinfo {author} {\bibfnamefont {H.}~\bibnamefont {Weigert}},\ }\href
  {https://doi.org/10.1103/PhysRevD.59.014014} {\bibfield  {journal} {\bibinfo
  {journal} {Phys. Rev. D}\ }\textbf {\bibinfo {volume} {59}},\ \bibinfo
  {pages} {014014} (\bibinfo {year} {1998}{\natexlab{a}})},\ \Eprint
  {https://arxiv.org/abs/hep-ph/9706377} {arXiv:hep-ph/9706377} \BibitemShut
  {NoStop}%
\bibitem [{\citenamefont {Jalilian-Marian}\ \emph
  {et~al.}(1998{\natexlab{b}})\citenamefont {Jalilian-Marian}, \citenamefont
  {Kovner},\ and\ \citenamefont {Weigert}}]{Jalilian-Marian:1997ubg}%
  \BibitemOpen
  \bibfield  {author} {\bibinfo {author} {\bibfnamefont {J.}~\bibnamefont
  {Jalilian-Marian}}, \bibinfo {author} {\bibfnamefont {A.}~\bibnamefont
  {Kovner}},\ and\ \bibinfo {author} {\bibfnamefont {H.}~\bibnamefont
  {Weigert}},\ }\href {https://doi.org/10.1103/PhysRevD.59.014015} {\bibfield
  {journal} {\bibinfo  {journal} {Phys. Rev. D}\ }\textbf {\bibinfo {volume}
  {59}},\ \bibinfo {pages} {014015} (\bibinfo {year} {1998}{\natexlab{b}})},\
  \Eprint {https://arxiv.org/abs/hep-ph/9709432} {arXiv:hep-ph/9709432}
  \BibitemShut {NoStop}%
\bibitem [{\citenamefont {Kovner}\ \emph {et~al.}(2000)\citenamefont {Kovner},
  \citenamefont {Milhano},\ and\ \citenamefont {Weigert}}]{Kovner:2000pt}%
  \BibitemOpen
  \bibfield  {author} {\bibinfo {author} {\bibfnamefont {A.}~\bibnamefont
  {Kovner}}, \bibinfo {author} {\bibfnamefont {J.~G.}\ \bibnamefont
  {Milhano}},\ and\ \bibinfo {author} {\bibfnamefont {H.}~\bibnamefont
  {Weigert}},\ }\href {https://doi.org/10.1103/PhysRevD.62.114005} {\bibfield
  {journal} {\bibinfo  {journal} {Phys. Rev. D}\ }\textbf {\bibinfo {volume}
  {62}},\ \bibinfo {pages} {114005} (\bibinfo {year} {2000})},\ \Eprint
  {https://arxiv.org/abs/hep-ph/0004014} {arXiv:hep-ph/0004014} \BibitemShut
  {NoStop}%
\bibitem [{\citenamefont {Weigert}(2002)}]{Weigert:2000gi}%
  \BibitemOpen
  \bibfield  {author} {\bibinfo {author} {\bibfnamefont {H.}~\bibnamefont
  {Weigert}},\ }\href {https://doi.org/10.1016/S0375-9474(01)01668-2}
  {\bibfield  {journal} {\bibinfo  {journal} {Nucl. Phys. A}\ }\textbf
  {\bibinfo {volume} {703}},\ \bibinfo {pages} {823} (\bibinfo {year}
  {2002})},\ \Eprint {https://arxiv.org/abs/hep-ph/0004044}
  {arXiv:hep-ph/0004044} \BibitemShut {NoStop}%
\bibitem [{\citenamefont {Iancu}\ \emph
  {et~al.}(2001{\natexlab{a}})\citenamefont {Iancu}, \citenamefont {Leonidov},\
  and\ \citenamefont {McLerran}}]{Iancu:2000hn}%
  \BibitemOpen
  \bibfield  {author} {\bibinfo {author} {\bibfnamefont {E.}~\bibnamefont
  {Iancu}}, \bibinfo {author} {\bibfnamefont {A.}~\bibnamefont {Leonidov}},\
  and\ \bibinfo {author} {\bibfnamefont {L.~D.}\ \bibnamefont {McLerran}},\
  }\href {https://doi.org/10.1016/S0375-9474(01)00642-X} {\bibfield  {journal}
  {\bibinfo  {journal} {Nucl. Phys. A}\ }\textbf {\bibinfo {volume} {692}},\
  \bibinfo {pages} {583} (\bibinfo {year} {2001}{\natexlab{a}})},\ \Eprint
  {https://arxiv.org/abs/hep-ph/0011241} {arXiv:hep-ph/0011241} \BibitemShut
  {NoStop}%
\bibitem [{\citenamefont {Iancu}\ \emph
  {et~al.}(2001{\natexlab{b}})\citenamefont {Iancu}, \citenamefont {Leonidov},\
  and\ \citenamefont {McLerran}}]{Iancu:2001ad}%
  \BibitemOpen
  \bibfield  {author} {\bibinfo {author} {\bibfnamefont {E.}~\bibnamefont
  {Iancu}}, \bibinfo {author} {\bibfnamefont {A.}~\bibnamefont {Leonidov}},\
  and\ \bibinfo {author} {\bibfnamefont {L.~D.}\ \bibnamefont {McLerran}},\
  }\href {https://doi.org/10.1016/S0370-2693(01)00524-X} {\bibfield  {journal}
  {\bibinfo  {journal} {Phys. Lett. B}\ }\textbf {\bibinfo {volume} {510}},\
  \bibinfo {pages} {133} (\bibinfo {year} {2001}{\natexlab{b}})},\ \Eprint
  {https://arxiv.org/abs/hep-ph/0102009} {arXiv:hep-ph/0102009} \BibitemShut
  {NoStop}%
\bibitem [{\citenamefont {Ferreiro}\ \emph {et~al.}(2002)\citenamefont
  {Ferreiro}, \citenamefont {Iancu}, \citenamefont {Leonidov},\ and\
  \citenamefont {McLerran}}]{Ferreiro:2001qy}%
  \BibitemOpen
  \bibfield  {author} {\bibinfo {author} {\bibfnamefont {E.}~\bibnamefont
  {Ferreiro}}, \bibinfo {author} {\bibfnamefont {E.}~\bibnamefont {Iancu}},
  \bibinfo {author} {\bibfnamefont {A.}~\bibnamefont {Leonidov}},\ and\
  \bibinfo {author} {\bibfnamefont {L.}~\bibnamefont {McLerran}},\ }\href
  {https://doi.org/10.1016/S0375-9474(01)01329-X} {\bibfield  {journal}
  {\bibinfo  {journal} {Nucl. Phys. A}\ }\textbf {\bibinfo {volume} {703}},\
  \bibinfo {pages} {489} (\bibinfo {year} {2002})},\ \Eprint
  {https://arxiv.org/abs/hep-ph/0109115} {arXiv:hep-ph/0109115} \BibitemShut
  {NoStop}%
\bibitem [{\citenamefont {Bjorken}\ \emph {et~al.}(1971)\citenamefont
  {Bjorken}, \citenamefont {Kogut},\ and\ \citenamefont
  {Soper}}]{Bjorken:1970ah}%
  \BibitemOpen
  \bibfield  {author} {\bibinfo {author} {\bibfnamefont {J.~D.}\ \bibnamefont
  {Bjorken}}, \bibinfo {author} {\bibfnamefont {J.~B.}\ \bibnamefont {Kogut}},\
  and\ \bibinfo {author} {\bibfnamefont {D.~E.}\ \bibnamefont {Soper}},\ }\href
  {https://doi.org/10.1103/PhysRevD.3.1382} {\bibfield  {journal} {\bibinfo
  {journal} {Phys. Rev. D}\ }\textbf {\bibinfo {volume} {3}},\ \bibinfo {pages}
  {1382} (\bibinfo {year} {1971})}\BibitemShut {NoStop}%
\bibitem [{\citenamefont {Nikolaev}\ and\ \citenamefont
  {Zakharov}(1991)}]{Nikolaev:1990ja}%
  \BibitemOpen
  \bibfield  {author} {\bibinfo {author} {\bibfnamefont {N.~N.}\ \bibnamefont
  {Nikolaev}}\ and\ \bibinfo {author} {\bibfnamefont {B.~G.}\ \bibnamefont
  {Zakharov}},\ }\href {https://doi.org/10.1007/BF01483577} {\bibfield
  {journal} {\bibinfo  {journal} {Z. Phys. C}\ }\textbf {\bibinfo {volume}
  {49}},\ \bibinfo {pages} {607} (\bibinfo {year} {1991})}\BibitemShut
  {NoStop}%
\bibitem [{\citenamefont {Nikolaev}\ and\ \citenamefont
  {Zakharov}(1992)}]{Nikolaev:1991et}%
  \BibitemOpen
  \bibfield  {author} {\bibinfo {author} {\bibfnamefont {N.}~\bibnamefont
  {Nikolaev}}\ and\ \bibinfo {author} {\bibfnamefont {B.~G.}\ \bibnamefont
  {Zakharov}},\ }\href {https://doi.org/10.1007/BF01597573} {\bibfield
  {journal} {\bibinfo  {journal} {Z. Phys. C}\ }\textbf {\bibinfo {volume}
  {53}},\ \bibinfo {pages} {331} (\bibinfo {year} {1992})}\BibitemShut
  {NoStop}%
\bibitem [{\citenamefont {Collins}(2023)}]{Collins:2011zzd}%
  \BibitemOpen
  \bibfield  {author} {\bibinfo {author} {\bibfnamefont {J.}~\bibnamefont
  {Collins}},\ }\href {https://doi.org/10.1017/9781009401845} {\emph {\bibinfo
  {title} {{Foundations of Perturbative QCD}}}},\ \bibinfo {series} {Cambridge
  Monographs on Particle Physics, Nuclear Physics and Cosmology}, Vol.~\bibinfo
  {volume} {32}\ (\bibinfo  {publisher} {Cambridge University Press},\ \bibinfo
  {year} {2023})\BibitemShut {NoStop}%
\bibitem [{\citenamefont {Boussarie}\ \emph {et~al.}(2023)\citenamefont
  {Boussarie} \emph {et~al.}}]{Boussarie:2023izj}%
  \BibitemOpen
  \bibfield  {author} {\bibinfo {author} {\bibfnamefont {R.}~\bibnamefont
  {Boussarie}} \emph {et~al.},\ }\href@noop {} {\  (\bibinfo {year} {2023})},\
  \Eprint {https://arxiv.org/abs/2304.03302} {arXiv:2304.03302 [hep-ph]}
  \BibitemShut {NoStop}%
\bibitem [{\citenamefont {Marquet}\ \emph {et~al.}(2009)\citenamefont
  {Marquet}, \citenamefont {Xiao},\ and\ \citenamefont
  {Yuan}}]{Marquet:2009ca}%
  \BibitemOpen
  \bibfield  {author} {\bibinfo {author} {\bibfnamefont {C.}~\bibnamefont
  {Marquet}}, \bibinfo {author} {\bibfnamefont {B.-W.}\ \bibnamefont {Xiao}},\
  and\ \bibinfo {author} {\bibfnamefont {F.}~\bibnamefont {Yuan}},\ }\href
  {https://doi.org/10.1016/j.physletb.2009.10.099} {\bibfield  {journal}
  {\bibinfo  {journal} {Phys. Lett. B}\ }\textbf {\bibinfo {volume} {682}},\
  \bibinfo {pages} {207} (\bibinfo {year} {2009})},\ \Eprint
  {https://arxiv.org/abs/0906.1454} {arXiv:0906.1454 [hep-ph]} \BibitemShut
  {NoStop}%
\bibitem [{\citenamefont {Catani}\ and\ \citenamefont
  {Hautmann}(1994)}]{Catani:1994sq}%
  \BibitemOpen
  \bibfield  {author} {\bibinfo {author} {\bibfnamefont {S.}~\bibnamefont
  {Catani}}\ and\ \bibinfo {author} {\bibfnamefont {F.}~\bibnamefont
  {Hautmann}},\ }\href {https://doi.org/10.1016/0550-3213(94)90636-X}
  {\bibfield  {journal} {\bibinfo  {journal} {Nucl. Phys. B}\ }\textbf
  {\bibinfo {volume} {427}},\ \bibinfo {pages} {475} (\bibinfo {year}
  {1994})},\ \Eprint {https://arxiv.org/abs/hep-ph/9405388}
  {arXiv:hep-ph/9405388} \BibitemShut {NoStop}%
\bibitem [{\citenamefont {Ji}\ \emph {et~al.}(2005)\citenamefont {Ji},
  \citenamefont {Ma},\ and\ \citenamefont {Yuan}}]{Ji:2005nu}%
  \BibitemOpen
  \bibfield  {author} {\bibinfo {author} {\bibfnamefont {X.-d.}\ \bibnamefont
  {Ji}}, \bibinfo {author} {\bibfnamefont {J.-P.}\ \bibnamefont {Ma}},\ and\
  \bibinfo {author} {\bibfnamefont {F.}~\bibnamefont {Yuan}},\ }\href
  {https://doi.org/10.1088/1126-6708/2005/07/020} {\bibfield  {journal}
  {\bibinfo  {journal} {JHEP}\ }\textbf {\bibinfo {volume} {07}},\ \bibinfo
  {pages} {020}},\ \Eprint {https://arxiv.org/abs/hep-ph/0503015}
  {arXiv:hep-ph/0503015} \BibitemShut {NoStop}%
\bibitem [{\citenamefont {Dominguez}\ \emph {et~al.}(2011)\citenamefont
  {Dominguez}, \citenamefont {Marquet}, \citenamefont {Xiao},\ and\
  \citenamefont {Yuan}}]{Dominguez:2011wm}%
  \BibitemOpen
  \bibfield  {author} {\bibinfo {author} {\bibfnamefont {F.}~\bibnamefont
  {Dominguez}}, \bibinfo {author} {\bibfnamefont {C.}~\bibnamefont {Marquet}},
  \bibinfo {author} {\bibfnamefont {B.-W.}\ \bibnamefont {Xiao}},\ and\
  \bibinfo {author} {\bibfnamefont {F.}~\bibnamefont {Yuan}},\ }\href
  {https://doi.org/10.1103/PhysRevD.83.105005} {\bibfield  {journal} {\bibinfo
  {journal} {Phys. Rev. D}\ }\textbf {\bibinfo {volume} {83}},\ \bibinfo
  {pages} {105005} (\bibinfo {year} {2011})},\ \Eprint
  {https://arxiv.org/abs/1101.0715} {arXiv:1101.0715 [hep-ph]} \BibitemShut
  {NoStop}%
\bibitem [{\citenamefont {Xiao}\ \emph {et~al.}(2017)\citenamefont {Xiao},
  \citenamefont {Yuan},\ and\ \citenamefont {Zhou}}]{Xiao:2017yya}%
  \BibitemOpen
  \bibfield  {author} {\bibinfo {author} {\bibfnamefont {B.-W.}\ \bibnamefont
  {Xiao}}, \bibinfo {author} {\bibfnamefont {F.}~\bibnamefont {Yuan}},\ and\
  \bibinfo {author} {\bibfnamefont {J.}~\bibnamefont {Zhou}},\ }\href
  {https://doi.org/10.1016/j.nuclphysb.2017.05.012} {\bibfield  {journal}
  {\bibinfo  {journal} {Nucl. Phys. B}\ }\textbf {\bibinfo {volume} {921}},\
  \bibinfo {pages} {104} (\bibinfo {year} {2017})},\ \Eprint
  {https://arxiv.org/abs/1703.06163} {arXiv:1703.06163 [hep-ph]} \BibitemShut
  {NoStop}%
\bibitem [{\citenamefont {Hentschinski}\ \emph {et~al.}(2018)\citenamefont
  {Hentschinski}, \citenamefont {Kusina}, \citenamefont {Kutak},\ and\
  \citenamefont {Serino}}]{Hentschinski:2017ayz}%
  \BibitemOpen
  \bibfield  {author} {\bibinfo {author} {\bibfnamefont {M.}~\bibnamefont
  {Hentschinski}}, \bibinfo {author} {\bibfnamefont {A.}~\bibnamefont
  {Kusina}}, \bibinfo {author} {\bibfnamefont {K.}~\bibnamefont {Kutak}},\ and\
  \bibinfo {author} {\bibfnamefont {M.}~\bibnamefont {Serino}},\ }\href
  {https://doi.org/10.1140/epjc/s10052-018-5634-2} {\bibfield  {journal}
  {\bibinfo  {journal} {Eur. Phys. J. C}\ }\textbf {\bibinfo {volume} {78}},\
  \bibinfo {pages} {174} (\bibinfo {year} {2018})},\ \Eprint
  {https://arxiv.org/abs/1711.04587} {arXiv:1711.04587 [hep-ph]} \BibitemShut
  {NoStop}%
\bibitem [{\citenamefont {Caucal}\ \emph
  {et~al.}(2025{\natexlab{a}})\citenamefont {Caucal}, \citenamefont {Morales},
  \citenamefont {Iancu}, \citenamefont {Salazar},\ and\ \citenamefont
  {Yuan}}]{Caucal:2025xxh}%
  \BibitemOpen
  \bibfield  {author} {\bibinfo {author} {\bibfnamefont {P.}~\bibnamefont
  {Caucal}}, \bibinfo {author} {\bibfnamefont {M.~G.}\ \bibnamefont {Morales}},
  \bibinfo {author} {\bibfnamefont {E.}~\bibnamefont {Iancu}}, \bibinfo
  {author} {\bibfnamefont {F.}~\bibnamefont {Salazar}},\ and\ \bibinfo {author}
  {\bibfnamefont {F.}~\bibnamefont {Yuan}},\ }\href@noop {} {\  (\bibinfo
  {year} {2025}{\natexlab{a}})},\ \Eprint {https://arxiv.org/abs/2503.16162}
  {arXiv:2503.16162 [hep-ph]} \BibitemShut {NoStop}%
\bibitem [{\citenamefont {Accardi}\ \emph {et~al.}(2016)\citenamefont {Accardi}
  \emph {et~al.}}]{Accardi:2012qut}%
  \BibitemOpen
  \bibfield  {author} {\bibinfo {author} {\bibfnamefont {A.}~\bibnamefont
  {Accardi}} \emph {et~al.},\ }\href
  {https://doi.org/10.1140/epja/i2016-16268-9} {\bibfield  {journal} {\bibinfo
  {journal} {Eur. Phys. J. A}\ }\textbf {\bibinfo {volume} {52}},\ \bibinfo
  {pages} {268} (\bibinfo {year} {2016})},\ \Eprint
  {https://arxiv.org/abs/1212.1701} {arXiv:1212.1701 [nucl-ex]} \BibitemShut
  {NoStop}%
\bibitem [{\citenamefont {Abelleira~Fernandez}\ \emph
  {et~al.}(2012)\citenamefont {Abelleira~Fernandez} \emph
  {et~al.}}]{LHeCStudyGroup:2012zhm}%
  \BibitemOpen
  \bibfield  {author} {\bibinfo {author} {\bibfnamefont {J.~L.}\ \bibnamefont
  {Abelleira~Fernandez}} \emph {et~al.} (\bibinfo {collaboration} {LHeC Study
  Group}),\ }\href {https://doi.org/10.1088/0954-3899/39/7/075001} {\bibfield
  {journal} {\bibinfo  {journal} {J. Phys. G}\ }\textbf {\bibinfo {volume}
  {39}},\ \bibinfo {pages} {075001} (\bibinfo {year} {2012})},\ \Eprint
  {https://arxiv.org/abs/1206.2913} {arXiv:1206.2913 [physics.acc-ph]}
  \BibitemShut {NoStop}%
\bibitem [{\citenamefont {Balitsky}\ and\ \citenamefont
  {Chirilli}(2011)}]{Balitsky:2010ze}%
  \BibitemOpen
  \bibfield  {author} {\bibinfo {author} {\bibfnamefont {I.}~\bibnamefont
  {Balitsky}}\ and\ \bibinfo {author} {\bibfnamefont {G.~A.}\ \bibnamefont
  {Chirilli}},\ }\href {https://doi.org/10.1103/PhysRevD.83.031502} {\bibfield
  {journal} {\bibinfo  {journal} {Phys. Rev. D}\ }\textbf {\bibinfo {volume}
  {83}},\ \bibinfo {pages} {031502} (\bibinfo {year} {2011})},\ \Eprint
  {https://arxiv.org/abs/1009.4729} {arXiv:1009.4729 [hep-ph]} \BibitemShut
  {NoStop}%
\bibitem [{\citenamefont {Balitsky}\ and\ \citenamefont
  {Chirilli}(2013)}]{Balitsky:2012bs}%
  \BibitemOpen
  \bibfield  {author} {\bibinfo {author} {\bibfnamefont {I.}~\bibnamefont
  {Balitsky}}\ and\ \bibinfo {author} {\bibfnamefont {G.~A.}\ \bibnamefont
  {Chirilli}},\ }\href {https://doi.org/10.1103/PhysRevD.87.014013} {\bibfield
  {journal} {\bibinfo  {journal} {Phys. Rev. D}\ }\textbf {\bibinfo {volume}
  {87}},\ \bibinfo {pages} {014013} (\bibinfo {year} {2013})},\ \Eprint
  {https://arxiv.org/abs/1207.3844} {arXiv:1207.3844 [hep-ph]} \BibitemShut
  {NoStop}%
\bibitem [{\citenamefont {Beuf}(2012)}]{Beuf:2011xd}%
  \BibitemOpen
  \bibfield  {author} {\bibinfo {author} {\bibfnamefont {G.}~\bibnamefont
  {Beuf}},\ }\href {https://doi.org/10.1103/PhysRevD.85.034039} {\bibfield
  {journal} {\bibinfo  {journal} {Phys. Rev. D}\ }\textbf {\bibinfo {volume}
  {85}},\ \bibinfo {pages} {034039} (\bibinfo {year} {2012})},\ \Eprint
  {https://arxiv.org/abs/1112.4501} {arXiv:1112.4501 [hep-ph]} \BibitemShut
  {NoStop}%
\bibitem [{\citenamefont {Beuf}(2016)}]{Beuf:2016wdz}%
  \BibitemOpen
  \bibfield  {author} {\bibinfo {author} {\bibfnamefont {G.}~\bibnamefont
  {Beuf}},\ }\href {https://doi.org/10.1103/PhysRevD.94.054016} {\bibfield
  {journal} {\bibinfo  {journal} {Phys. Rev. D}\ }\textbf {\bibinfo {volume}
  {94}},\ \bibinfo {pages} {054016} (\bibinfo {year} {2016})},\ \Eprint
  {https://arxiv.org/abs/1606.00777} {arXiv:1606.00777 [hep-ph]} \BibitemShut
  {NoStop}%
\bibitem [{\citenamefont {Beuf}(2017)}]{Beuf:2017bpd}%
  \BibitemOpen
  \bibfield  {author} {\bibinfo {author} {\bibfnamefont {G.}~\bibnamefont
  {Beuf}},\ }\href {https://doi.org/10.1103/PhysRevD.96.074033} {\bibfield
  {journal} {\bibinfo  {journal} {Phys. Rev. D}\ }\textbf {\bibinfo {volume}
  {96}},\ \bibinfo {pages} {074033} (\bibinfo {year} {2017})},\ \Eprint
  {https://arxiv.org/abs/1708.06557} {arXiv:1708.06557 [hep-ph]} \BibitemShut
  {NoStop}%
\bibitem [{\citenamefont {Duclou\'e}\ \emph {et~al.}(2017)\citenamefont
  {Duclou\'e}, \citenamefont {H\"anninen}, \citenamefont {Lappi},\ and\
  \citenamefont {Zhu}}]{Ducloue:2017ftk}%
  \BibitemOpen
  \bibfield  {author} {\bibinfo {author} {\bibfnamefont {B.}~\bibnamefont
  {Duclou\'e}}, \bibinfo {author} {\bibfnamefont {H.}~\bibnamefont
  {H\"anninen}}, \bibinfo {author} {\bibfnamefont {T.}~\bibnamefont {Lappi}},\
  and\ \bibinfo {author} {\bibfnamefont {Y.}~\bibnamefont {Zhu}},\ }\href
  {https://doi.org/10.1103/PhysRevD.96.094017} {\bibfield  {journal} {\bibinfo
  {journal} {Phys. Rev. D}\ }\textbf {\bibinfo {volume} {96}},\ \bibinfo
  {pages} {094017} (\bibinfo {year} {2017})},\ \Eprint
  {https://arxiv.org/abs/1708.07328} {arXiv:1708.07328 [hep-ph]} \BibitemShut
  {NoStop}%
\bibitem [{\citenamefont {H\"anninen}\ \emph {et~al.}(2018)\citenamefont
  {H\"anninen}, \citenamefont {Lappi},\ and\ \citenamefont
  {Paatelainen}}]{Hanninen:2017ddy}%
  \BibitemOpen
  \bibfield  {author} {\bibinfo {author} {\bibfnamefont {H.}~\bibnamefont
  {H\"anninen}}, \bibinfo {author} {\bibfnamefont {T.}~\bibnamefont {Lappi}},\
  and\ \bibinfo {author} {\bibfnamefont {R.}~\bibnamefont {Paatelainen}},\
  }\href {https://doi.org/10.1016/j.aop.2018.04.015} {\bibfield  {journal}
  {\bibinfo  {journal} {Annals Phys.}\ }\textbf {\bibinfo {volume} {393}},\
  \bibinfo {pages} {358} (\bibinfo {year} {2018})},\ \Eprint
  {https://arxiv.org/abs/1711.08207} {arXiv:1711.08207 [hep-ph]} \BibitemShut
  {NoStop}%
\bibitem [{\citenamefont {Beuf}\ \emph {et~al.}(2020)\citenamefont {Beuf},
  \citenamefont {H\"anninen}, \citenamefont {Lappi},\ and\ \citenamefont
  {M\"antysaari}}]{Beuf:2020dxl}%
  \BibitemOpen
  \bibfield  {author} {\bibinfo {author} {\bibfnamefont {G.}~\bibnamefont
  {Beuf}}, \bibinfo {author} {\bibfnamefont {H.}~\bibnamefont {H\"anninen}},
  \bibinfo {author} {\bibfnamefont {T.}~\bibnamefont {Lappi}},\ and\ \bibinfo
  {author} {\bibfnamefont {H.}~\bibnamefont {M\"antysaari}},\ }\href
  {https://doi.org/10.1103/PhysRevD.102.074028} {\bibfield  {journal} {\bibinfo
   {journal} {Phys. Rev. D}\ }\textbf {\bibinfo {volume} {102}},\ \bibinfo
  {pages} {074028} (\bibinfo {year} {2020})},\ \Eprint
  {https://arxiv.org/abs/2007.01645} {arXiv:2007.01645 [hep-ph]} \BibitemShut
  {NoStop}%
\bibitem [{\citenamefont {Beuf}\ \emph {et~al.}(2021)\citenamefont {Beuf},
  \citenamefont {Lappi},\ and\ \citenamefont {Paatelainen}}]{Beuf:2021qqa}%
  \BibitemOpen
  \bibfield  {author} {\bibinfo {author} {\bibfnamefont {G.}~\bibnamefont
  {Beuf}}, \bibinfo {author} {\bibfnamefont {T.}~\bibnamefont {Lappi}},\ and\
  \bibinfo {author} {\bibfnamefont {R.}~\bibnamefont {Paatelainen}},\ }\href
  {https://doi.org/10.1103/PhysRevD.104.056032} {\bibfield  {journal} {\bibinfo
   {journal} {Phys. Rev. D}\ }\textbf {\bibinfo {volume} {104}},\ \bibinfo
  {pages} {056032} (\bibinfo {year} {2021})},\ \Eprint
  {https://arxiv.org/abs/2103.14549} {arXiv:2103.14549 [hep-ph]} \BibitemShut
  {NoStop}%
\bibitem [{\citenamefont {Beuf}\ \emph
  {et~al.}(2022{\natexlab{a}})\citenamefont {Beuf}, \citenamefont {Lappi},\
  and\ \citenamefont {Paatelainen}}]{Beuf:2021srj}%
  \BibitemOpen
  \bibfield  {author} {\bibinfo {author} {\bibfnamefont {G.}~\bibnamefont
  {Beuf}}, \bibinfo {author} {\bibfnamefont {T.}~\bibnamefont {Lappi}},\ and\
  \bibinfo {author} {\bibfnamefont {R.}~\bibnamefont {Paatelainen}},\ }\href
  {https://doi.org/10.1103/PhysRevLett.129.072001} {\bibfield  {journal}
  {\bibinfo  {journal} {Phys. Rev. Lett.}\ }\textbf {\bibinfo {volume} {129}},\
  \bibinfo {pages} {072001} (\bibinfo {year} {2022}{\natexlab{a}})},\ \Eprint
  {https://arxiv.org/abs/2112.03158} {arXiv:2112.03158 [hep-ph]} \BibitemShut
  {NoStop}%
\bibitem [{\citenamefont {Beuf}\ \emph
  {et~al.}(2022{\natexlab{b}})\citenamefont {Beuf}, \citenamefont {Lappi},\
  and\ \citenamefont {Paatelainen}}]{Beuf:2022ndu}%
  \BibitemOpen
  \bibfield  {author} {\bibinfo {author} {\bibfnamefont {G.}~\bibnamefont
  {Beuf}}, \bibinfo {author} {\bibfnamefont {T.}~\bibnamefont {Lappi}},\ and\
  \bibinfo {author} {\bibfnamefont {R.}~\bibnamefont {Paatelainen}},\ }\href
  {https://doi.org/10.1103/PhysRevD.106.034013} {\bibfield  {journal} {\bibinfo
   {journal} {Phys. Rev. D}\ }\textbf {\bibinfo {volume} {106}},\ \bibinfo
  {pages} {034013} (\bibinfo {year} {2022}{\natexlab{b}})},\ \Eprint
  {https://arxiv.org/abs/2204.02486} {arXiv:2204.02486 [hep-ph]} \BibitemShut
  {NoStop}%
\bibitem [{\citenamefont {H\"anninen}\ \emph {et~al.}(2023)\citenamefont
  {H\"anninen}, \citenamefont {M\"antysaari}, \citenamefont {Paatelainen},\
  and\ \citenamefont {Penttala}}]{Hanninen:2022gje}%
  \BibitemOpen
  \bibfield  {author} {\bibinfo {author} {\bibfnamefont {H.}~\bibnamefont
  {H\"anninen}}, \bibinfo {author} {\bibfnamefont {H.}~\bibnamefont
  {M\"antysaari}}, \bibinfo {author} {\bibfnamefont {R.}~\bibnamefont
  {Paatelainen}},\ and\ \bibinfo {author} {\bibfnamefont {J.}~\bibnamefont
  {Penttala}},\ }\href {https://doi.org/10.1103/PhysRevLett.130.192301}
  {\bibfield  {journal} {\bibinfo  {journal} {Phys. Rev. Lett.}\ }\textbf
  {\bibinfo {volume} {130}},\ \bibinfo {pages} {192301} (\bibinfo {year}
  {2023})},\ \Eprint {https://arxiv.org/abs/2211.03504} {arXiv:2211.03504
  [hep-ph]} \BibitemShut {NoStop}%
\bibitem [{\citenamefont {Casuga}\ \emph {et~al.}(2025)\citenamefont {Casuga},
  \citenamefont {H{\"a}nninen},\ and\ \citenamefont
  {M{\"a}ntysaari}}]{Casuga:2025etc}%
  \BibitemOpen
  \bibfield  {author} {\bibinfo {author} {\bibfnamefont {C.}~\bibnamefont
  {Casuga}}, \bibinfo {author} {\bibfnamefont {H.}~\bibnamefont
  {H{\"a}nninen}},\ and\ \bibinfo {author} {\bibfnamefont {H.}~\bibnamefont
  {M{\"a}ntysaari}},\ }\href {https://doi.org/10.1103/54zd-hyvg} {\bibfield
  {journal} {\bibinfo  {journal} {Phys. Rev. D}\ }\textbf {\bibinfo {volume}
  {112}},\ \bibinfo {pages} {034003} (\bibinfo {year} {2025})},\ \Eprint
  {https://arxiv.org/abs/2506.00487} {arXiv:2506.00487 [hep-ph]} \BibitemShut
  {NoStop}%
\bibitem [{\citenamefont {Caucal}\ \emph {et~al.}(2024)\citenamefont {Caucal},
  \citenamefont {Ferrand},\ and\ \citenamefont {Salazar}}]{Caucal:2024cdq}%
  \BibitemOpen
  \bibfield  {author} {\bibinfo {author} {\bibfnamefont {P.}~\bibnamefont
  {Caucal}}, \bibinfo {author} {\bibfnamefont {E.}~\bibnamefont {Ferrand}},\
  and\ \bibinfo {author} {\bibfnamefont {F.}~\bibnamefont {Salazar}},\ }\href
  {https://doi.org/10.1007/JHEP05(2024)110} {\bibfield  {journal} {\bibinfo
  {journal} {JHEP}\ }\textbf {\bibinfo {volume} {05}},\ \bibinfo {pages}
  {110}},\ \Eprint {https://arxiv.org/abs/2401.01934} {arXiv:2401.01934
  [hep-ph]} \BibitemShut {NoStop}%
\bibitem [{\citenamefont {Bergabo}\ and\ \citenamefont
  {Jalilian-Marian}(2024)}]{Bergabo:2024ivx}%
  \BibitemOpen
  \bibfield  {author} {\bibinfo {author} {\bibfnamefont {F.}~\bibnamefont
  {Bergabo}}\ and\ \bibinfo {author} {\bibfnamefont {J.}~\bibnamefont
  {Jalilian-Marian}},\ }\href {https://doi.org/10.1103/PhysRevD.109.074011}
  {\bibfield  {journal} {\bibinfo  {journal} {Phys. Rev. D}\ }\textbf {\bibinfo
  {volume} {109}},\ \bibinfo {pages} {074011} (\bibinfo {year} {2024})},\
  \Eprint {https://arxiv.org/abs/2401.06259} {arXiv:2401.06259 [hep-ph]}
  \BibitemShut {NoStop}%
\bibitem [{\citenamefont {Bergabo}\ and\ \citenamefont
  {Jalilian-Marian}(2023)}]{Bergabo:2022zhe}%
  \BibitemOpen
  \bibfield  {author} {\bibinfo {author} {\bibfnamefont {F.}~\bibnamefont
  {Bergabo}}\ and\ \bibinfo {author} {\bibfnamefont {J.}~\bibnamefont
  {Jalilian-Marian}},\ }\href {https://doi.org/10.1007/JHEP01(2023)095}
  {\bibfield  {journal} {\bibinfo  {journal} {JHEP}\ }\textbf {\bibinfo
  {volume} {01}},\ \bibinfo {pages} {095}},\ \Eprint
  {https://arxiv.org/abs/2210.03208} {arXiv:2210.03208 [hep-ph]} \BibitemShut
  {NoStop}%
\bibitem [{\citenamefont {Altinoluk}\ \emph
  {et~al.}(2025{\natexlab{a}})\citenamefont {Altinoluk}, \citenamefont
  {Bergabo}, \citenamefont {Jalilian-Marian}, \citenamefont {Marquet},\ and\
  \citenamefont {Shi}}]{Altinoluk:2025dwd}%
  \BibitemOpen
  \bibfield  {author} {\bibinfo {author} {\bibfnamefont {T.}~\bibnamefont
  {Altinoluk}}, \bibinfo {author} {\bibfnamefont {F.}~\bibnamefont {Bergabo}},
  \bibinfo {author} {\bibfnamefont {J.}~\bibnamefont {Jalilian-Marian}},
  \bibinfo {author} {\bibfnamefont {C.}~\bibnamefont {Marquet}},\ and\ \bibinfo
  {author} {\bibfnamefont {Y.}~\bibnamefont {Shi}},\ }\href
  {https://doi.org/10.1103/s561-vqh8} {\bibfield  {journal} {\bibinfo
  {journal} {Phys. Rev. D}\ }\textbf {\bibinfo {volume} {112}},\ \bibinfo
  {pages} {054020} (\bibinfo {year} {2025}{\natexlab{a}})},\ \Eprint
  {https://arxiv.org/abs/2505.04557} {arXiv:2505.04557 [hep-ph]} \BibitemShut
  {NoStop}%
\bibitem [{\citenamefont {Caucal}\ \emph
  {et~al.}(2025{\natexlab{b}})\citenamefont {Caucal}, \citenamefont {Iancu},
  \citenamefont {Mueller},\ and\ \citenamefont {Yuan}}]{Caucal:2024vbv}%
  \BibitemOpen
  \bibfield  {author} {\bibinfo {author} {\bibfnamefont {P.}~\bibnamefont
  {Caucal}}, \bibinfo {author} {\bibfnamefont {E.}~\bibnamefont {Iancu}},
  \bibinfo {author} {\bibfnamefont {A.~H.}\ \bibnamefont {Mueller}},\ and\
  \bibinfo {author} {\bibfnamefont {F.}~\bibnamefont {Yuan}},\ }\href
  {https://doi.org/10.1103/PhysRevLett.134.061903} {\bibfield  {journal}
  {\bibinfo  {journal} {Phys. Rev. Lett.}\ }\textbf {\bibinfo {volume} {134}},\
  \bibinfo {pages} {061903} (\bibinfo {year} {2025}{\natexlab{b}})},\ \Eprint
  {https://arxiv.org/abs/2408.03129} {arXiv:2408.03129 [hep-ph]} \BibitemShut
  {NoStop}%
\bibitem [{\citenamefont {Altinoluk}\ \emph {et~al.}(2014)\citenamefont
  {Altinoluk}, \citenamefont {Armesto}, \citenamefont {Beuf}, \citenamefont
  {Mart\'\i{}nez},\ and\ \citenamefont {Salgado}}]{Altinoluk:2014oxa}%
  \BibitemOpen
  \bibfield  {author} {\bibinfo {author} {\bibfnamefont {T.}~\bibnamefont
  {Altinoluk}}, \bibinfo {author} {\bibfnamefont {N.}~\bibnamefont {Armesto}},
  \bibinfo {author} {\bibfnamefont {G.}~\bibnamefont {Beuf}}, \bibinfo {author}
  {\bibfnamefont {M.}~\bibnamefont {Mart\'\i{}nez}},\ and\ \bibinfo {author}
  {\bibfnamefont {C.~A.}\ \bibnamefont {Salgado}},\ }\href
  {https://doi.org/10.1007/JHEP07(2014)068} {\bibfield  {journal} {\bibinfo
  {journal} {JHEP}\ }\textbf {\bibinfo {volume} {07}},\ \bibinfo {pages}
  {068}},\ \Eprint {https://arxiv.org/abs/1404.2219} {arXiv:1404.2219 [hep-ph]}
  \BibitemShut {NoStop}%
\bibitem [{\citenamefont {Altinoluk}\ \emph {et~al.}(2016)\citenamefont
  {Altinoluk}, \citenamefont {Armesto}, \citenamefont {Beuf},\ and\
  \citenamefont {Moscoso}}]{Altinoluk:2015gia}%
  \BibitemOpen
  \bibfield  {author} {\bibinfo {author} {\bibfnamefont {T.}~\bibnamefont
  {Altinoluk}}, \bibinfo {author} {\bibfnamefont {N.}~\bibnamefont {Armesto}},
  \bibinfo {author} {\bibfnamefont {G.}~\bibnamefont {Beuf}},\ and\ \bibinfo
  {author} {\bibfnamefont {A.}~\bibnamefont {Moscoso}},\ }\href
  {https://doi.org/10.1007/JHEP01(2016)114} {\bibfield  {journal} {\bibinfo
  {journal} {JHEP}\ }\textbf {\bibinfo {volume} {01}},\ \bibinfo {pages}
  {114}},\ \Eprint {https://arxiv.org/abs/1505.01400} {arXiv:1505.01400
  [hep-ph]} \BibitemShut {NoStop}%
\bibitem [{\citenamefont {Altinoluk}\ and\ \citenamefont
  {Dumitru}(2016)}]{Altinoluk:2015xuy}%
  \BibitemOpen
  \bibfield  {author} {\bibinfo {author} {\bibfnamefont {T.}~\bibnamefont
  {Altinoluk}}\ and\ \bibinfo {author} {\bibfnamefont {A.}~\bibnamefont
  {Dumitru}},\ }\href {https://doi.org/10.1103/PhysRevD.94.074032} {\bibfield
  {journal} {\bibinfo  {journal} {Phys. Rev. D}\ }\textbf {\bibinfo {volume}
  {94}},\ \bibinfo {pages} {074032} (\bibinfo {year} {2016})},\ \Eprint
  {https://arxiv.org/abs/1512.00279} {arXiv:1512.00279 [hep-ph]} \BibitemShut
  {NoStop}%
\bibitem [{\citenamefont {Agostini}\ \emph
  {et~al.}(2019{\natexlab{a}})\citenamefont {Agostini}, \citenamefont
  {Altinoluk},\ and\ \citenamefont {Armesto}}]{Agostini:2019avp}%
  \BibitemOpen
  \bibfield  {author} {\bibinfo {author} {\bibfnamefont {P.}~\bibnamefont
  {Agostini}}, \bibinfo {author} {\bibfnamefont {T.}~\bibnamefont
  {Altinoluk}},\ and\ \bibinfo {author} {\bibfnamefont {N.}~\bibnamefont
  {Armesto}},\ }\href {https://doi.org/10.1140/epjc/s10052-019-7097-5}
  {\bibfield  {journal} {\bibinfo  {journal} {Eur. Phys. J. C}\ }\textbf
  {\bibinfo {volume} {79}},\ \bibinfo {pages} {600} (\bibinfo {year}
  {2019}{\natexlab{a}})},\ \Eprint {https://arxiv.org/abs/1902.04483}
  {arXiv:1902.04483 [hep-ph]} \BibitemShut {NoStop}%
\bibitem [{\citenamefont {Agostini}\ \emph
  {et~al.}(2019{\natexlab{b}})\citenamefont {Agostini}, \citenamefont
  {Altinoluk},\ and\ \citenamefont {Armesto}}]{Agostini:2019hkj}%
  \BibitemOpen
  \bibfield  {author} {\bibinfo {author} {\bibfnamefont {P.}~\bibnamefont
  {Agostini}}, \bibinfo {author} {\bibfnamefont {T.}~\bibnamefont
  {Altinoluk}},\ and\ \bibinfo {author} {\bibfnamefont {N.}~\bibnamefont
  {Armesto}},\ }\href {https://doi.org/10.1140/epjc/s10052-019-7315-1}
  {\bibfield  {journal} {\bibinfo  {journal} {Eur. Phys. J. C}\ }\textbf
  {\bibinfo {volume} {79}},\ \bibinfo {pages} {790} (\bibinfo {year}
  {2019}{\natexlab{b}})},\ \Eprint {https://arxiv.org/abs/1907.03668}
  {arXiv:1907.03668 [hep-ph]} \BibitemShut {NoStop}%
\bibitem [{\citenamefont {Agostini}\ \emph {et~al.}(2022)\citenamefont
  {Agostini}, \citenamefont {Altinoluk}, \citenamefont {Armesto}, \citenamefont
  {Dominguez},\ and\ \citenamefont {Milhano}}]{Agostini:2022ctk}%
  \BibitemOpen
  \bibfield  {author} {\bibinfo {author} {\bibfnamefont {P.}~\bibnamefont
  {Agostini}}, \bibinfo {author} {\bibfnamefont {T.}~\bibnamefont {Altinoluk}},
  \bibinfo {author} {\bibfnamefont {N.}~\bibnamefont {Armesto}}, \bibinfo
  {author} {\bibfnamefont {F.}~\bibnamefont {Dominguez}},\ and\ \bibinfo
  {author} {\bibfnamefont {J.~G.}\ \bibnamefont {Milhano}},\ }\href
  {https://doi.org/10.1140/epjc/s10052-022-10962-1} {\bibfield  {journal}
  {\bibinfo  {journal} {Eur. Phys. J. C}\ }\textbf {\bibinfo {volume} {82}},\
  \bibinfo {pages} {1001} (\bibinfo {year} {2022})},\ \Eprint
  {https://arxiv.org/abs/2207.10472} {arXiv:2207.10472 [hep-ph]} \BibitemShut
  {NoStop}%
\bibitem [{\citenamefont {Agostini}\ \emph {et~al.}(2023)\citenamefont
  {Agostini}, \citenamefont {Altinoluk},\ and\ \citenamefont
  {Armesto}}]{Agostini:2022oge}%
  \BibitemOpen
  \bibfield  {author} {\bibinfo {author} {\bibfnamefont {P.}~\bibnamefont
  {Agostini}}, \bibinfo {author} {\bibfnamefont {T.}~\bibnamefont
  {Altinoluk}},\ and\ \bibinfo {author} {\bibfnamefont {N.}~\bibnamefont
  {Armesto}},\ }\href {https://doi.org/10.1016/j.physletb.2023.137892}
  {\bibfield  {journal} {\bibinfo  {journal} {Phys. Lett. B}\ }\textbf
  {\bibinfo {volume} {840}},\ \bibinfo {pages} {137892} (\bibinfo {year}
  {2023})},\ \Eprint {https://arxiv.org/abs/2212.03633} {arXiv:2212.03633
  [hep-ph]} \BibitemShut {NoStop}%
\bibitem [{\citenamefont {Altinoluk}\ \emph {et~al.}(2021)\citenamefont
  {Altinoluk}, \citenamefont {Beuf}, \citenamefont {Czajka},\ and\
  \citenamefont {Tymowska}}]{Altinoluk:2020oyd}%
  \BibitemOpen
  \bibfield  {author} {\bibinfo {author} {\bibfnamefont {T.}~\bibnamefont
  {Altinoluk}}, \bibinfo {author} {\bibfnamefont {G.}~\bibnamefont {Beuf}},
  \bibinfo {author} {\bibfnamefont {A.}~\bibnamefont {Czajka}},\ and\ \bibinfo
  {author} {\bibfnamefont {A.}~\bibnamefont {Tymowska}},\ }\href
  {https://doi.org/10.1103/PhysRevD.104.014019} {\bibfield  {journal} {\bibinfo
   {journal} {Phys. Rev. D}\ }\textbf {\bibinfo {volume} {104}},\ \bibinfo
  {pages} {014019} (\bibinfo {year} {2021})},\ \Eprint
  {https://arxiv.org/abs/2012.03886} {arXiv:2012.03886 [hep-ph]} \BibitemShut
  {NoStop}%
\bibitem [{\citenamefont {Altinoluk}\ and\ \citenamefont
  {Beuf}(2022)}]{Altinoluk:2021lvu}%
  \BibitemOpen
  \bibfield  {author} {\bibinfo {author} {\bibfnamefont {T.}~\bibnamefont
  {Altinoluk}}\ and\ \bibinfo {author} {\bibfnamefont {G.}~\bibnamefont
  {Beuf}},\ }\href {https://doi.org/10.1103/PhysRevD.105.074026} {\bibfield
  {journal} {\bibinfo  {journal} {Phys. Rev. D}\ }\textbf {\bibinfo {volume}
  {105}},\ \bibinfo {pages} {074026} (\bibinfo {year} {2022})},\ \Eprint
  {https://arxiv.org/abs/2109.01620} {arXiv:2109.01620 [hep-ph]} \BibitemShut
  {NoStop}%
\bibitem [{\citenamefont {Altinoluk}\ \emph
  {et~al.}(2023{\natexlab{a}})\citenamefont {Altinoluk}, \citenamefont {Beuf},
  \citenamefont {Czajka},\ and\ \citenamefont {Tymowska}}]{Altinoluk:2022jkk}%
  \BibitemOpen
  \bibfield  {author} {\bibinfo {author} {\bibfnamefont {T.}~\bibnamefont
  {Altinoluk}}, \bibinfo {author} {\bibfnamefont {G.}~\bibnamefont {Beuf}},
  \bibinfo {author} {\bibfnamefont {A.}~\bibnamefont {Czajka}},\ and\ \bibinfo
  {author} {\bibfnamefont {A.}~\bibnamefont {Tymowska}},\ }\href
  {https://doi.org/10.1103/PhysRevD.107.074016} {\bibfield  {journal} {\bibinfo
   {journal} {Phys. Rev. D}\ }\textbf {\bibinfo {volume} {107}},\ \bibinfo
  {pages} {074016} (\bibinfo {year} {2023}{\natexlab{a}})},\ \Eprint
  {https://arxiv.org/abs/2212.10484} {arXiv:2212.10484 [hep-ph]} \BibitemShut
  {NoStop}%
\bibitem [{\citenamefont {Agostini}\ \emph {et~al.}(2024)\citenamefont
  {Agostini}, \citenamefont {Altinoluk},\ and\ \citenamefont
  {Armesto}}]{Agostini:2024xqs}%
  \BibitemOpen
  \bibfield  {author} {\bibinfo {author} {\bibfnamefont {P.}~\bibnamefont
  {Agostini}}, \bibinfo {author} {\bibfnamefont {T.}~\bibnamefont
  {Altinoluk}},\ and\ \bibinfo {author} {\bibfnamefont {N.}~\bibnamefont
  {Armesto}},\ }\href {https://doi.org/10.1007/JHEP07(2024)137} {\bibfield
  {journal} {\bibinfo  {journal} {JHEP}\ }\textbf {\bibinfo {volume} {07}},\
  \bibinfo {pages} {137}},\ \Eprint {https://arxiv.org/abs/2403.04603}
  {arXiv:2403.04603 [hep-ph]} \BibitemShut {NoStop}%
\bibitem [{\citenamefont {Altinoluk}\ \emph
  {et~al.}(2023{\natexlab{b}})\citenamefont {Altinoluk}, \citenamefont
  {Armesto},\ and\ \citenamefont {Beuf}}]{Altinoluk:2023qfr}%
  \BibitemOpen
  \bibfield  {author} {\bibinfo {author} {\bibfnamefont {T.}~\bibnamefont
  {Altinoluk}}, \bibinfo {author} {\bibfnamefont {N.}~\bibnamefont {Armesto}},\
  and\ \bibinfo {author} {\bibfnamefont {G.}~\bibnamefont {Beuf}},\ }\href
  {https://doi.org/10.1103/PhysRevD.108.074023} {\bibfield  {journal} {\bibinfo
   {journal} {Phys. Rev. D}\ }\textbf {\bibinfo {volume} {108}},\ \bibinfo
  {pages} {074023} (\bibinfo {year} {2023}{\natexlab{b}})},\ \Eprint
  {https://arxiv.org/abs/2303.12691} {arXiv:2303.12691 [hep-ph]} \BibitemShut
  {NoStop}%
\bibitem [{\citenamefont {Altinoluk}\ \emph
  {et~al.}(2025{\natexlab{b}})\citenamefont {Altinoluk}, \citenamefont {Beuf},
  \citenamefont {Czajka},\ and\ \citenamefont {Marquet}}]{Altinoluk:2024zom}%
  \BibitemOpen
  \bibfield  {author} {\bibinfo {author} {\bibfnamefont {T.}~\bibnamefont
  {Altinoluk}}, \bibinfo {author} {\bibfnamefont {G.}~\bibnamefont {Beuf}},
  \bibinfo {author} {\bibfnamefont {A.}~\bibnamefont {Czajka}},\ and\ \bibinfo
  {author} {\bibfnamefont {C.}~\bibnamefont {Marquet}},\ }\href
  {https://doi.org/10.1103/PhysRevD.111.014010} {\bibfield  {journal} {\bibinfo
   {journal} {Phys. Rev. D}\ }\textbf {\bibinfo {volume} {111}},\ \bibinfo
  {pages} {014010} (\bibinfo {year} {2025}{\natexlab{b}})},\ \Eprint
  {https://arxiv.org/abs/2410.00612} {arXiv:2410.00612 [hep-ph]} \BibitemShut
  {NoStop}%
\bibitem [{\citenamefont {Altinoluk}\ \emph
  {et~al.}(2025{\natexlab{c}})\citenamefont {Altinoluk}, \citenamefont {Beuf},\
  and\ \citenamefont {Mulani}}]{Altinoluk:2024dba}%
  \BibitemOpen
  \bibfield  {author} {\bibinfo {author} {\bibfnamefont {T.}~\bibnamefont
  {Altinoluk}}, \bibinfo {author} {\bibfnamefont {G.}~\bibnamefont {Beuf}},\
  and\ \bibinfo {author} {\bibfnamefont {S.}~\bibnamefont {Mulani}},\ }\href
  {https://doi.org/10.1103/PhysRevD.111.034028} {\bibfield  {journal} {\bibinfo
   {journal} {Phys. Rev. D}\ }\textbf {\bibinfo {volume} {111}},\ \bibinfo
  {pages} {034028} (\bibinfo {year} {2025}{\natexlab{c}})},\ \Eprint
  {https://arxiv.org/abs/2411.15047} {arXiv:2411.15047 [hep-ph]} \BibitemShut
  {NoStop}%
\bibitem [{\citenamefont {Altinoluk}\ \emph
  {et~al.}(2025{\natexlab{d}})\citenamefont {Altinoluk}, \citenamefont {Beuf},
  \citenamefont {Blanco},\ and\ \citenamefont {Mulani}}]{Altinoluk:2024tyx}%
  \BibitemOpen
  \bibfield  {author} {\bibinfo {author} {\bibfnamefont {T.}~\bibnamefont
  {Altinoluk}}, \bibinfo {author} {\bibfnamefont {G.}~\bibnamefont {Beuf}},
  \bibinfo {author} {\bibfnamefont {E.}~\bibnamefont {Blanco}},\ and\ \bibinfo
  {author} {\bibfnamefont {S.}~\bibnamefont {Mulani}},\ }\href
  {https://doi.org/10.1007/JHEP06(2025)097} {\bibfield  {journal} {\bibinfo
  {journal} {JHEP}\ }\textbf {\bibinfo {volume} {06}},\ \bibinfo {pages}
  {097}},\ \Eprint {https://arxiv.org/abs/2412.08485} {arXiv:2412.08485
  [hep-ph]} \BibitemShut {NoStop}%
\bibitem [{\citenamefont {Kovchegov}\ \emph {et~al.}(2016)\citenamefont
  {Kovchegov}, \citenamefont {Pitonyak},\ and\ \citenamefont
  {Sievert}}]{Kovchegov:2015pbl}%
  \BibitemOpen
  \bibfield  {author} {\bibinfo {author} {\bibfnamefont {Y.~V.}\ \bibnamefont
  {Kovchegov}}, \bibinfo {author} {\bibfnamefont {D.}~\bibnamefont
  {Pitonyak}},\ and\ \bibinfo {author} {\bibfnamefont {M.~D.}\ \bibnamefont
  {Sievert}},\ }\href {https://doi.org/10.1007/JHEP01(2016)072} {\bibfield
  {journal} {\bibinfo  {journal} {JHEP}\ }\textbf {\bibinfo {volume} {01}},\
  \bibinfo {pages} {072}},\ \bibinfo {note} {[Erratum: JHEP 10, 148 (2016)]},\
  \Eprint {https://arxiv.org/abs/1511.06737} {arXiv:1511.06737 [hep-ph]}
  \BibitemShut {NoStop}%
\bibitem [{\citenamefont {Kovchegov}\ \emph
  {et~al.}(2017{\natexlab{a}})\citenamefont {Kovchegov}, \citenamefont
  {Pitonyak},\ and\ \citenamefont {Sievert}}]{Kovchegov:2016zex}%
  \BibitemOpen
  \bibfield  {author} {\bibinfo {author} {\bibfnamefont {Y.~V.}\ \bibnamefont
  {Kovchegov}}, \bibinfo {author} {\bibfnamefont {D.}~\bibnamefont
  {Pitonyak}},\ and\ \bibinfo {author} {\bibfnamefont {M.~D.}\ \bibnamefont
  {Sievert}},\ }\href {https://doi.org/10.1103/PhysRevD.95.014033} {\bibfield
  {journal} {\bibinfo  {journal} {Phys. Rev. D}\ }\textbf {\bibinfo {volume}
  {95}},\ \bibinfo {pages} {014033} (\bibinfo {year} {2017}{\natexlab{a}})},\
  \Eprint {https://arxiv.org/abs/1610.06197} {arXiv:1610.06197 [hep-ph]}
  \BibitemShut {NoStop}%
\bibitem [{\citenamefont {Kovchegov}\ \emph
  {et~al.}(2017{\natexlab{b}})\citenamefont {Kovchegov}, \citenamefont
  {Pitonyak},\ and\ \citenamefont {Sievert}}]{Kovchegov:2016weo}%
  \BibitemOpen
  \bibfield  {author} {\bibinfo {author} {\bibfnamefont {Y.~V.}\ \bibnamefont
  {Kovchegov}}, \bibinfo {author} {\bibfnamefont {D.}~\bibnamefont
  {Pitonyak}},\ and\ \bibinfo {author} {\bibfnamefont {M.~D.}\ \bibnamefont
  {Sievert}},\ }\href {https://doi.org/10.1103/PhysRevLett.118.052001}
  {\bibfield  {journal} {\bibinfo  {journal} {Phys. Rev. Lett.}\ }\textbf
  {\bibinfo {volume} {118}},\ \bibinfo {pages} {052001} (\bibinfo {year}
  {2017}{\natexlab{b}})},\ \Eprint {https://arxiv.org/abs/1610.06188}
  {arXiv:1610.06188 [hep-ph]} \BibitemShut {NoStop}%
\bibitem [{\citenamefont {Kovchegov}\ \emph
  {et~al.}(2017{\natexlab{c}})\citenamefont {Kovchegov}, \citenamefont
  {Pitonyak},\ and\ \citenamefont {Sievert}}]{Kovchegov:2017jxc}%
  \BibitemOpen
  \bibfield  {author} {\bibinfo {author} {\bibfnamefont {Y.~V.}\ \bibnamefont
  {Kovchegov}}, \bibinfo {author} {\bibfnamefont {D.}~\bibnamefont
  {Pitonyak}},\ and\ \bibinfo {author} {\bibfnamefont {M.~D.}\ \bibnamefont
  {Sievert}},\ }\href {https://doi.org/10.1016/j.physletb.2017.06.032}
  {\bibfield  {journal} {\bibinfo  {journal} {Phys. Lett. B}\ }\textbf
  {\bibinfo {volume} {772}},\ \bibinfo {pages} {136} (\bibinfo {year}
  {2017}{\natexlab{c}})},\ \Eprint {https://arxiv.org/abs/1703.05809}
  {arXiv:1703.05809 [hep-ph]} \BibitemShut {NoStop}%
\bibitem [{\citenamefont {Kovchegov}\ \emph
  {et~al.}(2017{\natexlab{d}})\citenamefont {Kovchegov}, \citenamefont
  {Pitonyak},\ and\ \citenamefont {Sievert}}]{Kovchegov:2017lsr}%
  \BibitemOpen
  \bibfield  {author} {\bibinfo {author} {\bibfnamefont {Y.~V.}\ \bibnamefont
  {Kovchegov}}, \bibinfo {author} {\bibfnamefont {D.}~\bibnamefont
  {Pitonyak}},\ and\ \bibinfo {author} {\bibfnamefont {M.~D.}\ \bibnamefont
  {Sievert}},\ }\href {https://doi.org/10.1007/JHEP10(2017)198} {\bibfield
  {journal} {\bibinfo  {journal} {JHEP}\ }\textbf {\bibinfo {volume} {10}},\
  \bibinfo {pages} {198}},\ \Eprint {https://arxiv.org/abs/1706.04236}
  {arXiv:1706.04236 [nucl-th]} \BibitemShut {NoStop}%
\bibitem [{\citenamefont {Kovchegov}\ and\ \citenamefont
  {Sievert}(2019{\natexlab{a}})}]{Kovchegov:2018znm}%
  \BibitemOpen
  \bibfield  {author} {\bibinfo {author} {\bibfnamefont {Y.~V.}\ \bibnamefont
  {Kovchegov}}\ and\ \bibinfo {author} {\bibfnamefont {M.~D.}\ \bibnamefont
  {Sievert}},\ }\href {https://doi.org/10.1103/PhysRevD.99.054032} {\bibfield
  {journal} {\bibinfo  {journal} {Phys. Rev. D}\ }\textbf {\bibinfo {volume}
  {99}},\ \bibinfo {pages} {054032} (\bibinfo {year} {2019}{\natexlab{a}})},\
  \Eprint {https://arxiv.org/abs/1808.09010} {arXiv:1808.09010 [hep-ph]}
  \BibitemShut {NoStop}%
\bibitem [{\citenamefont {Kovchegov}\ and\ \citenamefont
  {Sievert}(2019{\natexlab{b}})}]{Kovchegov:2018zeq}%
  \BibitemOpen
  \bibfield  {author} {\bibinfo {author} {\bibfnamefont {Y.~V.}\ \bibnamefont
  {Kovchegov}}\ and\ \bibinfo {author} {\bibfnamefont {M.~D.}\ \bibnamefont
  {Sievert}},\ }\href {https://doi.org/10.1103/PhysRevD.99.054033} {\bibfield
  {journal} {\bibinfo  {journal} {Phys. Rev. D}\ }\textbf {\bibinfo {volume}
  {99}},\ \bibinfo {pages} {054033} (\bibinfo {year} {2019}{\natexlab{b}})},\
  \Eprint {https://arxiv.org/abs/1808.10354} {arXiv:1808.10354 [hep-ph]}
  \BibitemShut {NoStop}%
\bibitem [{\citenamefont {Kovchegov}\ and\ \citenamefont
  {Santiago}(2020)}]{Kovchegov:2020kxg}%
  \BibitemOpen
  \bibfield  {author} {\bibinfo {author} {\bibfnamefont {Y.~V.}\ \bibnamefont
  {Kovchegov}}\ and\ \bibinfo {author} {\bibfnamefont {M.~G.}\ \bibnamefont
  {Santiago}},\ }\href {https://doi.org/10.1103/PhysRevD.102.014022} {\bibfield
   {journal} {\bibinfo  {journal} {Phys. Rev. D}\ }\textbf {\bibinfo {volume}
  {102}},\ \bibinfo {pages} {014022} (\bibinfo {year} {2020})},\ \Eprint
  {https://arxiv.org/abs/2003.12650} {arXiv:2003.12650 [hep-ph]} \BibitemShut
  {NoStop}%
\bibitem [{\citenamefont {Kovchegov}\ and\ \citenamefont
  {Tawabutr}(2020)}]{Kovchegov:2020hgb}%
  \BibitemOpen
  \bibfield  {author} {\bibinfo {author} {\bibfnamefont {Y.~V.}\ \bibnamefont
  {Kovchegov}}\ and\ \bibinfo {author} {\bibfnamefont {Y.}~\bibnamefont
  {Tawabutr}},\ }\href {https://doi.org/10.1007/JHEP08(2020)014} {\bibfield
  {journal} {\bibinfo  {journal} {JHEP}\ }\textbf {\bibinfo {volume} {08}},\
  \bibinfo {pages} {014}},\ \Eprint {https://arxiv.org/abs/2005.07285}
  {arXiv:2005.07285 [hep-ph]} \BibitemShut {NoStop}%
\bibitem [{\citenamefont {Adamiak}\ \emph {et~al.}(2021)\citenamefont
  {Adamiak}, \citenamefont {Kovchegov}, \citenamefont {Melnitchouk},
  \citenamefont {Pitonyak}, \citenamefont {Sato},\ and\ \citenamefont
  {Sievert}}]{Adamiak:2021ppq}%
  \BibitemOpen
  \bibfield  {author} {\bibinfo {author} {\bibfnamefont {D.}~\bibnamefont
  {Adamiak}}, \bibinfo {author} {\bibfnamefont {Y.~V.}\ \bibnamefont
  {Kovchegov}}, \bibinfo {author} {\bibfnamefont {W.}~\bibnamefont
  {Melnitchouk}}, \bibinfo {author} {\bibfnamefont {D.}~\bibnamefont
  {Pitonyak}}, \bibinfo {author} {\bibfnamefont {N.}~\bibnamefont {Sato}},\
  and\ \bibinfo {author} {\bibfnamefont {M.~D.}\ \bibnamefont {Sievert}}
  (\bibinfo {collaboration} {Jefferson Lab Angular Momentum}),\ }\href
  {https://doi.org/10.1103/PhysRevD.104.L031501} {\bibfield  {journal}
  {\bibinfo  {journal} {Phys. Rev. D}\ }\textbf {\bibinfo {volume} {104}},\
  \bibinfo {pages} {L031501} (\bibinfo {year} {2021})},\ \Eprint
  {https://arxiv.org/abs/2102.06159} {arXiv:2102.06159 [hep-ph]} \BibitemShut
  {NoStop}%
\bibitem [{\citenamefont {Kovchegov}\ \emph {et~al.}(2022)\citenamefont
  {Kovchegov}, \citenamefont {Tarasov},\ and\ \citenamefont
  {Tawabutr}}]{Kovchegov:2021lvz}%
  \BibitemOpen
  \bibfield  {author} {\bibinfo {author} {\bibfnamefont {Y.~V.}\ \bibnamefont
  {Kovchegov}}, \bibinfo {author} {\bibfnamefont {A.}~\bibnamefont {Tarasov}},\
  and\ \bibinfo {author} {\bibfnamefont {Y.}~\bibnamefont {Tawabutr}},\ }\href
  {https://doi.org/10.1007/JHEP03(2022)184} {\bibfield  {journal} {\bibinfo
  {journal} {JHEP}\ }\textbf {\bibinfo {volume} {03}},\ \bibinfo {pages}
  {184}},\ \Eprint {https://arxiv.org/abs/2104.11765} {arXiv:2104.11765
  [hep-ph]} \BibitemShut {NoStop}%
\bibitem [{\citenamefont {Kovchegov}\ and\ \citenamefont
  {Santiago}(2021)}]{Kovchegov:2021iyc}%
  \BibitemOpen
  \bibfield  {author} {\bibinfo {author} {\bibfnamefont {Y.~V.}\ \bibnamefont
  {Kovchegov}}\ and\ \bibinfo {author} {\bibfnamefont {M.~G.}\ \bibnamefont
  {Santiago}},\ }\href {https://doi.org/10.1007/JHEP11(2021)200} {\bibfield
  {journal} {\bibinfo  {journal} {JHEP}\ }\textbf {\bibinfo {volume} {11}},\
  \bibinfo {pages} {200}},\ \bibinfo {note} {[Erratum: JHEP 09, 186 (2022)]},\
  \Eprint {https://arxiv.org/abs/2108.03667} {arXiv:2108.03667 [hep-ph]}
  \BibitemShut {NoStop}%
\bibitem [{\citenamefont {Cougoulic}\ \emph {et~al.}(2022)\citenamefont
  {Cougoulic}, \citenamefont {Kovchegov}, \citenamefont {Tarasov},\ and\
  \citenamefont {Tawabutr}}]{Cougoulic:2022gbk}%
  \BibitemOpen
  \bibfield  {author} {\bibinfo {author} {\bibfnamefont {F.}~\bibnamefont
  {Cougoulic}}, \bibinfo {author} {\bibfnamefont {Y.~V.}\ \bibnamefont
  {Kovchegov}}, \bibinfo {author} {\bibfnamefont {A.}~\bibnamefont {Tarasov}},\
  and\ \bibinfo {author} {\bibfnamefont {Y.}~\bibnamefont {Tawabutr}},\ }\href
  {https://doi.org/10.1007/JHEP07(2022)095} {\bibfield  {journal} {\bibinfo
  {journal} {JHEP}\ }\textbf {\bibinfo {volume} {07}},\ \bibinfo {pages}
  {095}},\ \Eprint {https://arxiv.org/abs/2204.11898} {arXiv:2204.11898
  [hep-ph]} \BibitemShut {NoStop}%
\bibitem [{\citenamefont {Kovchegov}\ and\ \citenamefont
  {Santiago}(2022)}]{Kovchegov:2022kyy}%
  \BibitemOpen
  \bibfield  {author} {\bibinfo {author} {\bibfnamefont {Y.~V.}\ \bibnamefont
  {Kovchegov}}\ and\ \bibinfo {author} {\bibfnamefont {M.~G.}\ \bibnamefont
  {Santiago}},\ }\href {https://doi.org/10.1007/JHEP11(2022)098} {\bibfield
  {journal} {\bibinfo  {journal} {JHEP}\ }\textbf {\bibinfo {volume} {11}},\
  \bibinfo {pages} {098}},\ \Eprint {https://arxiv.org/abs/2209.03538}
  {arXiv:2209.03538 [hep-ph]} \BibitemShut {NoStop}%
\bibitem [{\citenamefont {Borden}\ and\ \citenamefont
  {Kovchegov}(2023)}]{Borden:2023ugd}%
  \BibitemOpen
  \bibfield  {author} {\bibinfo {author} {\bibfnamefont {J.}~\bibnamefont
  {Borden}}\ and\ \bibinfo {author} {\bibfnamefont {Y.~V.}\ \bibnamefont
  {Kovchegov}},\ }\href {https://doi.org/10.1103/PhysRevD.108.014001}
  {\bibfield  {journal} {\bibinfo  {journal} {Phys. Rev. D}\ }\textbf {\bibinfo
  {volume} {108}},\ \bibinfo {pages} {014001} (\bibinfo {year} {2023})},\
  \Eprint {https://arxiv.org/abs/2304.06161} {arXiv:2304.06161 [hep-ph]}
  \BibitemShut {NoStop}%
\bibitem [{\citenamefont {Kovchegov}\ and\ \citenamefont
  {Li}(2024)}]{Kovchegov:2024aus}%
  \BibitemOpen
  \bibfield  {author} {\bibinfo {author} {\bibfnamefont {Y.~V.}\ \bibnamefont
  {Kovchegov}}\ and\ \bibinfo {author} {\bibfnamefont {M.}~\bibnamefont {Li}},\
  }\href {https://doi.org/10.1007/JHEP05(2024)177} {\bibfield  {journal}
  {\bibinfo  {journal} {JHEP}\ }\textbf {\bibinfo {volume} {05}},\ \bibinfo
  {pages} {177}},\ \Eprint {https://arxiv.org/abs/2403.06959} {arXiv:2403.06959
  [hep-ph]} \BibitemShut {NoStop}%
\bibitem [{\citenamefont {Borden}\ \emph {et~al.}(2024)\citenamefont {Borden},
  \citenamefont {Kovchegov},\ and\ \citenamefont {Li}}]{Borden:2024bxa}%
  \BibitemOpen
  \bibfield  {author} {\bibinfo {author} {\bibfnamefont {J.}~\bibnamefont
  {Borden}}, \bibinfo {author} {\bibfnamefont {Y.~V.}\ \bibnamefont
  {Kovchegov}},\ and\ \bibinfo {author} {\bibfnamefont {M.}~\bibnamefont
  {Li}},\ }\href {https://doi.org/10.1007/JHEP09(2024)037} {\bibfield
  {journal} {\bibinfo  {journal} {JHEP}\ }\textbf {\bibinfo {volume} {09}},\
  \bibinfo {pages} {037}},\ \Eprint {https://arxiv.org/abs/2406.11647}
  {arXiv:2406.11647 [hep-ph]} \BibitemShut {NoStop}%
\bibitem [{\citenamefont {Adamiak}\ \emph {et~al.}(2025)\citenamefont
  {Adamiak}, \citenamefont {Baldonado}, \citenamefont {Kovchegov},
  \citenamefont {Li}, \citenamefont {Melnitchouk}, \citenamefont {Pitonyak},
  \citenamefont {Sato}, \citenamefont {Sievert}, \citenamefont {Tarasov},\ and\
  \citenamefont {Tawabutr}}]{Adamiak:2025dpw}%
  \BibitemOpen
  \bibfield  {author} {\bibinfo {author} {\bibfnamefont {D.}~\bibnamefont
  {Adamiak}}, \bibinfo {author} {\bibfnamefont {N.}~\bibnamefont {Baldonado}},
  \bibinfo {author} {\bibfnamefont {Y.~V.}\ \bibnamefont {Kovchegov}}, \bibinfo
  {author} {\bibfnamefont {M.}~\bibnamefont {Li}}, \bibinfo {author}
  {\bibfnamefont {W.}~\bibnamefont {Melnitchouk}}, \bibinfo {author}
  {\bibfnamefont {D.}~\bibnamefont {Pitonyak}}, \bibinfo {author}
  {\bibfnamefont {N.}~\bibnamefont {Sato}}, \bibinfo {author} {\bibfnamefont
  {M.~D.}\ \bibnamefont {Sievert}}, \bibinfo {author} {\bibfnamefont
  {A.}~\bibnamefont {Tarasov}},\ and\ \bibinfo {author} {\bibfnamefont
  {Y.}~\bibnamefont {Tawabutr}},\ }\href@noop {} {\  (\bibinfo {year}
  {2025})},\ \Eprint {https://arxiv.org/abs/2503.21006} {arXiv:2503.21006
  [hep-ph]} \BibitemShut {NoStop}%
\bibitem [{\citenamefont {Kovchegov}\ and\ \citenamefont
  {Li}(2025)}]{Kovchegov:2025gcg}%
  \BibitemOpen
  \bibfield  {author} {\bibinfo {author} {\bibfnamefont {Y.~V.}\ \bibnamefont
  {Kovchegov}}\ and\ \bibinfo {author} {\bibfnamefont {M.}~\bibnamefont {Li}},\
  }\href {https://doi.org/10.1007/JHEP08(2025)206} {\bibfield  {journal}
  {\bibinfo  {journal} {JHEP}\ }\textbf {\bibinfo {volume} {08}},\ \bibinfo
  {pages} {206}},\ \Eprint {https://arxiv.org/abs/2504.12979} {arXiv:2504.12979
  [hep-ph]} \BibitemShut {NoStop}%
\bibitem [{\citenamefont {Borden}\ and\ \citenamefont
  {Kovchegov}(2025)}]{Borden:2025ehe}%
  \BibitemOpen
  \bibfield  {author} {\bibinfo {author} {\bibfnamefont {J.}~\bibnamefont
  {Borden}}\ and\ \bibinfo {author} {\bibfnamefont {Y.~V.}\ \bibnamefont
  {Kovchegov}},\ }\href@noop {} {\  (\bibinfo {year} {2025})},\ \Eprint
  {https://arxiv.org/abs/2508.00195} {arXiv:2508.00195 [hep-ph]} \BibitemShut
  {NoStop}%
\bibitem [{\citenamefont {Cougoulic}\ and\ \citenamefont
  {Kovchegov}(2019)}]{Cougoulic:2019aja}%
  \BibitemOpen
  \bibfield  {author} {\bibinfo {author} {\bibfnamefont {F.}~\bibnamefont
  {Cougoulic}}\ and\ \bibinfo {author} {\bibfnamefont {Y.~V.}\ \bibnamefont
  {Kovchegov}},\ }\href {https://doi.org/10.1103/PhysRevD.100.114020}
  {\bibfield  {journal} {\bibinfo  {journal} {Phys. Rev. D}\ }\textbf {\bibinfo
  {volume} {100}},\ \bibinfo {pages} {114020} (\bibinfo {year} {2019})},\
  \Eprint {https://arxiv.org/abs/1910.04268} {arXiv:1910.04268 [hep-ph]}
  \BibitemShut {NoStop}%
\bibitem [{\citenamefont {Cougoulic}\ and\ \citenamefont
  {Kovchegov}(2020)}]{Cougoulic:2020tbc}%
  \BibitemOpen
  \bibfield  {author} {\bibinfo {author} {\bibfnamefont {F.}~\bibnamefont
  {Cougoulic}}\ and\ \bibinfo {author} {\bibfnamefont {Y.~V.}\ \bibnamefont
  {Kovchegov}},\ }\href {https://doi.org/10.1016/j.nuclphysa.2020.122051}
  {\bibfield  {journal} {\bibinfo  {journal} {Nucl. Phys. A}\ }\textbf
  {\bibinfo {volume} {1004}},\ \bibinfo {pages} {122051} (\bibinfo {year}
  {2020})},\ \Eprint {https://arxiv.org/abs/2005.14688} {arXiv:2005.14688
  [hep-ph]} \BibitemShut {NoStop}%
\bibitem [{\citenamefont {Balitsky}\ and\ \citenamefont
  {Tarasov}(2015)}]{Balitsky:2015qba}%
  \BibitemOpen
  \bibfield  {author} {\bibinfo {author} {\bibfnamefont {I.}~\bibnamefont
  {Balitsky}}\ and\ \bibinfo {author} {\bibfnamefont {A.}~\bibnamefont
  {Tarasov}},\ }\href {https://doi.org/10.1007/JHEP10(2015)017} {\bibfield
  {journal} {\bibinfo  {journal} {JHEP}\ }\textbf {\bibinfo {volume} {10}},\
  \bibinfo {pages} {017}},\ \Eprint {https://arxiv.org/abs/1505.02151}
  {arXiv:1505.02151 [hep-ph]} \BibitemShut {NoStop}%
\bibitem [{\citenamefont {Balitsky}\ and\ \citenamefont
  {Tarasov}(2016)}]{Balitsky:2016dgz}%
  \BibitemOpen
  \bibfield  {author} {\bibinfo {author} {\bibfnamefont {I.}~\bibnamefont
  {Balitsky}}\ and\ \bibinfo {author} {\bibfnamefont {A.}~\bibnamefont
  {Tarasov}},\ }\href {https://doi.org/10.1007/JHEP06(2016)164} {\bibfield
  {journal} {\bibinfo  {journal} {JHEP}\ }\textbf {\bibinfo {volume} {06}},\
  \bibinfo {pages} {164}},\ \Eprint {https://arxiv.org/abs/1603.06548}
  {arXiv:1603.06548 [hep-ph]} \BibitemShut {NoStop}%
\bibitem [{\citenamefont {Balitsky}\ and\ \citenamefont
  {Tarasov}(2017)}]{Balitsky:2017flc}%
  \BibitemOpen
  \bibfield  {author} {\bibinfo {author} {\bibfnamefont {I.}~\bibnamefont
  {Balitsky}}\ and\ \bibinfo {author} {\bibfnamefont {A.}~\bibnamefont
  {Tarasov}},\ }\href {https://doi.org/10.1007/JHEP07(2017)095} {\bibfield
  {journal} {\bibinfo  {journal} {JHEP}\ }\textbf {\bibinfo {volume} {07}},\
  \bibinfo {pages} {095}},\ \Eprint {https://arxiv.org/abs/1706.01415}
  {arXiv:1706.01415 [hep-ph]} \BibitemShut {NoStop}%
\bibitem [{\citenamefont {Boussarie}\ and\ \citenamefont
  {Mehtar-Tani}(2022{\natexlab{a}})}]{Boussarie:2020fpb}%
  \BibitemOpen
  \bibfield  {author} {\bibinfo {author} {\bibfnamefont {R.}~\bibnamefont
  {Boussarie}}\ and\ \bibinfo {author} {\bibfnamefont {Y.}~\bibnamefont
  {Mehtar-Tani}},\ }\href {https://doi.org/10.1016/j.physletb.2022.137125}
  {\bibfield  {journal} {\bibinfo  {journal} {Phys. Lett. B}\ }\textbf
  {\bibinfo {volume} {831}},\ \bibinfo {pages} {137125} (\bibinfo {year}
  {2022}{\natexlab{a}})},\ \Eprint {https://arxiv.org/abs/2006.14569}
  {arXiv:2006.14569 [hep-ph]} \BibitemShut {NoStop}%
\bibitem [{\citenamefont {Boussarie}\ and\ \citenamefont
  {Mehtar-Tani}(2022{\natexlab{b}})}]{Boussarie:2021wkn}%
  \BibitemOpen
  \bibfield  {author} {\bibinfo {author} {\bibfnamefont {R.}~\bibnamefont
  {Boussarie}}\ and\ \bibinfo {author} {\bibfnamefont {Y.}~\bibnamefont
  {Mehtar-Tani}},\ }\href {https://doi.org/10.1007/JHEP07(2022)080} {\bibfield
  {journal} {\bibinfo  {journal} {JHEP}\ }\textbf {\bibinfo {volume} {07}},\
  \bibinfo {pages} {080}},\ \Eprint {https://arxiv.org/abs/2112.01412}
  {arXiv:2112.01412 [hep-ph]} \BibitemShut {NoStop}%
\bibitem [{\citenamefont {Boussarie}\ and\ \citenamefont
  {Mehtar-Tani}(2024)}]{Boussarie:2023xun}%
  \BibitemOpen
  \bibfield  {author} {\bibinfo {author} {\bibfnamefont {R.}~\bibnamefont
  {Boussarie}}\ and\ \bibinfo {author} {\bibfnamefont {Y.}~\bibnamefont
  {Mehtar-Tani}},\ }\href {https://doi.org/10.1007/JHEP10(2024)056} {\bibfield
  {journal} {\bibinfo  {journal} {JHEP}\ }\textbf {\bibinfo {volume} {10}},\
  \bibinfo {pages} {056}},\ \Eprint {https://arxiv.org/abs/2309.16576}
  {arXiv:2309.16576 [hep-ph]} \BibitemShut {NoStop}%
\bibitem [{\citenamefont {Chirilli}(2019)}]{Chirilli:2018kkw}%
  \BibitemOpen
  \bibfield  {author} {\bibinfo {author} {\bibfnamefont {G.~A.}\ \bibnamefont
  {Chirilli}},\ }\href {https://doi.org/10.1007/JHEP01(2019)118} {\bibfield
  {journal} {\bibinfo  {journal} {JHEP}\ }\textbf {\bibinfo {volume} {01}},\
  \bibinfo {pages} {118}},\ \Eprint {https://arxiv.org/abs/1807.11435}
  {arXiv:1807.11435 [hep-ph]} \BibitemShut {NoStop}%
\bibitem [{\citenamefont {Chirilli}(2021)}]{Chirilli:2021lif}%
  \BibitemOpen
  \bibfield  {author} {\bibinfo {author} {\bibfnamefont {G.~A.}\ \bibnamefont
  {Chirilli}},\ }\href {https://doi.org/10.1007/JHEP06(2021)096} {\bibfield
  {journal} {\bibinfo  {journal} {JHEP}\ }\textbf {\bibinfo {volume} {06}},\
  \bibinfo {pages} {096}},\ \Eprint {https://arxiv.org/abs/2101.12744}
  {arXiv:2101.12744 [hep-ph]} \BibitemShut {NoStop}%
\bibitem [{\citenamefont {Li}(2023)}]{Li:2023tlw}%
  \BibitemOpen
  \bibfield  {author} {\bibinfo {author} {\bibfnamefont {M.}~\bibnamefont
  {Li}},\ }\href {https://doi.org/10.1007/JHEP07(2023)158} {\bibfield
  {journal} {\bibinfo  {journal} {JHEP}\ }\textbf {\bibinfo {volume} {07}},\
  \bibinfo {pages} {158}},\ \Eprint {https://arxiv.org/abs/2304.12842}
  {arXiv:2304.12842 [hep-ph]} \BibitemShut {NoStop}%
\bibitem [{\citenamefont {Li}(2024)}]{Li:2024fdb}%
  \BibitemOpen
  \bibfield  {author} {\bibinfo {author} {\bibfnamefont {M.}~\bibnamefont
  {Li}},\ }\href {https://doi.org/10.1103/PhysRevLett.133.021902} {\bibfield
  {journal} {\bibinfo  {journal} {Phys. Rev. Lett.}\ }\textbf {\bibinfo
  {volume} {133}},\ \bibinfo {pages} {021902} (\bibinfo {year} {2024})},\
  \Eprint {https://arxiv.org/abs/2402.17568} {arXiv:2402.17568 [hep-ph]}
  \BibitemShut {NoStop}%
\bibitem [{\citenamefont {Li}(2025)}]{Li:2024xra}%
  \BibitemOpen
  \bibfield  {author} {\bibinfo {author} {\bibfnamefont {M.}~\bibnamefont
  {Li}},\ }\href {https://doi.org/10.1103/PhysRevD.111.034027} {\bibfield
  {journal} {\bibinfo  {journal} {Phys. Rev. D}\ }\textbf {\bibinfo {volume}
  {111}},\ \bibinfo {pages} {034027} (\bibinfo {year} {2025})},\ \Eprint
  {https://arxiv.org/abs/2411.13431} {arXiv:2411.13431 [hep-ph]} \BibitemShut
  {NoStop}%
\bibitem [{\citenamefont {Jalilian-Marian}(2017)}]{Jalilian-Marian:2017ttv}%
  \BibitemOpen
  \bibfield  {author} {\bibinfo {author} {\bibfnamefont {J.}~\bibnamefont
  {Jalilian-Marian}},\ }\href {https://doi.org/10.1103/PhysRevD.96.074020}
  {\bibfield  {journal} {\bibinfo  {journal} {Phys. Rev. D}\ }\textbf {\bibinfo
  {volume} {96}},\ \bibinfo {pages} {074020} (\bibinfo {year} {2017})},\
  \Eprint {https://arxiv.org/abs/1708.07533} {arXiv:1708.07533 [hep-ph]}
  \BibitemShut {NoStop}%
\bibitem [{\citenamefont {Jalilian-Marian}(2019)}]{Jalilian-Marian:2018iui}%
  \BibitemOpen
  \bibfield  {author} {\bibinfo {author} {\bibfnamefont {J.}~\bibnamefont
  {Jalilian-Marian}},\ }\href {https://doi.org/10.1103/PhysRevD.99.014043}
  {\bibfield  {journal} {\bibinfo  {journal} {Phys. Rev. D}\ }\textbf {\bibinfo
  {volume} {99}},\ \bibinfo {pages} {014043} (\bibinfo {year} {2019})},\
  \Eprint {https://arxiv.org/abs/1809.04625} {arXiv:1809.04625 [hep-ph]}
  \BibitemShut {NoStop}%
\bibitem [{\citenamefont {Jalilian-Marian}(2020)}]{Jalilian-Marian:2019kaf}%
  \BibitemOpen
  \bibfield  {author} {\bibinfo {author} {\bibfnamefont {J.}~\bibnamefont
  {Jalilian-Marian}},\ }\href {https://doi.org/10.1103/PhysRevD.102.014008}
  {\bibfield  {journal} {\bibinfo  {journal} {Phys. Rev. D}\ }\textbf {\bibinfo
  {volume} {102}},\ \bibinfo {pages} {014008} (\bibinfo {year} {2020})},\
  \Eprint {https://arxiv.org/abs/1912.08878} {arXiv:1912.08878 [hep-ph]}
  \BibitemShut {NoStop}%
\bibitem [{\citenamefont {Hatta}\ \emph {et~al.}(2017)\citenamefont {Hatta},
  \citenamefont {Nakagawa}, \citenamefont {Yuan}, \citenamefont {Zhao},\ and\
  \citenamefont {Xiao}}]{Hatta:2016aoc}%
  \BibitemOpen
  \bibfield  {author} {\bibinfo {author} {\bibfnamefont {Y.}~\bibnamefont
  {Hatta}}, \bibinfo {author} {\bibfnamefont {Y.}~\bibnamefont {Nakagawa}},
  \bibinfo {author} {\bibfnamefont {F.}~\bibnamefont {Yuan}}, \bibinfo {author}
  {\bibfnamefont {Y.}~\bibnamefont {Zhao}},\ and\ \bibinfo {author}
  {\bibfnamefont {B.}~\bibnamefont {Xiao}},\ }\href
  {https://doi.org/10.1103/PhysRevD.95.114032} {\bibfield  {journal} {\bibinfo
  {journal} {Phys. Rev. D}\ }\textbf {\bibinfo {volume} {95}},\ \bibinfo
  {pages} {114032} (\bibinfo {year} {2017})},\ \Eprint
  {https://arxiv.org/abs/1612.02445} {arXiv:1612.02445 [hep-ph]} \BibitemShut
  {NoStop}%
\bibitem [{\citenamefont {Kovchegov}(2019)}]{Kovchegov:2019rrz}%
  \BibitemOpen
  \bibfield  {author} {\bibinfo {author} {\bibfnamefont {Y.~V.}\ \bibnamefont
  {Kovchegov}},\ }\href {https://doi.org/10.1007/JHEP03(2019)174} {\bibfield
  {journal} {\bibinfo  {journal} {JHEP}\ }\textbf {\bibinfo {volume} {03}},\
  \bibinfo {pages} {174}},\ \Eprint {https://arxiv.org/abs/1901.07453}
  {arXiv:1901.07453 [hep-ph]} \BibitemShut {NoStop}%
\bibitem [{\citenamefont {Boussarie}\ \emph {et~al.}(2019)\citenamefont
  {Boussarie}, \citenamefont {Hatta},\ and\ \citenamefont
  {Yuan}}]{Boussarie:2019icw}%
  \BibitemOpen
  \bibfield  {author} {\bibinfo {author} {\bibfnamefont {R.}~\bibnamefont
  {Boussarie}}, \bibinfo {author} {\bibfnamefont {Y.}~\bibnamefont {Hatta}},\
  and\ \bibinfo {author} {\bibfnamefont {F.}~\bibnamefont {Yuan}},\ }\href
  {https://doi.org/10.1016/j.physletb.2019.134817} {\bibfield  {journal}
  {\bibinfo  {journal} {Phys. Lett. B}\ }\textbf {\bibinfo {volume} {797}},\
  \bibinfo {pages} {134817} (\bibinfo {year} {2019})},\ \Eprint
  {https://arxiv.org/abs/1904.02693} {arXiv:1904.02693 [hep-ph]} \BibitemShut
  {NoStop}%
\bibitem [{\citenamefont {Kovchegov}\ and\ \citenamefont
  {Manley}(2024)}]{Kovchegov:2023yzd}%
  \BibitemOpen
  \bibfield  {author} {\bibinfo {author} {\bibfnamefont {Y.~V.}\ \bibnamefont
  {Kovchegov}}\ and\ \bibinfo {author} {\bibfnamefont {B.}~\bibnamefont
  {Manley}},\ }\href {https://doi.org/10.1007/JHEP02(2024)060} {\bibfield
  {journal} {\bibinfo  {journal} {JHEP}\ }\textbf {\bibinfo {volume} {02}},\
  \bibinfo {pages} {060}},\ \Eprint {https://arxiv.org/abs/2310.18404}
  {arXiv:2310.18404 [hep-ph]} \BibitemShut {NoStop}%
\bibitem [{\citenamefont {Kovchegov}\ and\ \citenamefont
  {Manley}(2025)}]{Kovchegov:2024wjs}%
  \BibitemOpen
  \bibfield  {author} {\bibinfo {author} {\bibfnamefont {Y.~V.}\ \bibnamefont
  {Kovchegov}}\ and\ \bibinfo {author} {\bibfnamefont {B.}~\bibnamefont
  {Manley}},\ }\href {https://doi.org/10.1103/PhysRevD.111.054017} {\bibfield
  {journal} {\bibinfo  {journal} {Phys. Rev. D}\ }\textbf {\bibinfo {volume}
  {111}},\ \bibinfo {pages} {054017} (\bibinfo {year} {2025})},\ \Eprint
  {https://arxiv.org/abs/2410.21260} {arXiv:2410.21260 [hep-ph]} \BibitemShut
  {NoStop}%
\bibitem [{\citenamefont {McLerran}\ and\ \citenamefont
  {Venugopalan}(1994{\natexlab{a}})}]{McLerran:1993ni}%
  \BibitemOpen
  \bibfield  {author} {\bibinfo {author} {\bibfnamefont {L.~D.}\ \bibnamefont
  {McLerran}}\ and\ \bibinfo {author} {\bibfnamefont {R.}~\bibnamefont
  {Venugopalan}},\ }\href {https://doi.org/10.1103/PhysRevD.49.2233} {\bibfield
   {journal} {\bibinfo  {journal} {Phys. Rev. D}\ }\textbf {\bibinfo {volume}
  {49}},\ \bibinfo {pages} {2233} (\bibinfo {year} {1994}{\natexlab{a}})},\
  \Eprint {https://arxiv.org/abs/hep-ph/9309289} {arXiv:hep-ph/9309289}
  \BibitemShut {NoStop}%
\bibitem [{\citenamefont {McLerran}\ and\ \citenamefont
  {Venugopalan}(1994{\natexlab{b}})}]{McLerran:1993ka}%
  \BibitemOpen
  \bibfield  {author} {\bibinfo {author} {\bibfnamefont {L.~D.}\ \bibnamefont
  {McLerran}}\ and\ \bibinfo {author} {\bibfnamefont {R.}~\bibnamefont
  {Venugopalan}},\ }\href {https://doi.org/10.1103/PhysRevD.49.3352} {\bibfield
   {journal} {\bibinfo  {journal} {Phys. Rev. D}\ }\textbf {\bibinfo {volume}
  {49}},\ \bibinfo {pages} {3352} (\bibinfo {year} {1994}{\natexlab{b}})},\
  \Eprint {https://arxiv.org/abs/hep-ph/9311205} {arXiv:hep-ph/9311205}
  \BibitemShut {NoStop}%
\bibitem [{\citenamefont {McLerran}\ and\ \citenamefont
  {Venugopalan}(1994{\natexlab{c}})}]{McLerran:1994vd}%
  \BibitemOpen
  \bibfield  {author} {\bibinfo {author} {\bibfnamefont {L.~D.}\ \bibnamefont
  {McLerran}}\ and\ \bibinfo {author} {\bibfnamefont {R.}~\bibnamefont
  {Venugopalan}},\ }\href {https://doi.org/10.1103/PhysRevD.50.2225} {\bibfield
   {journal} {\bibinfo  {journal} {Phys. Rev. D}\ }\textbf {\bibinfo {volume}
  {50}},\ \bibinfo {pages} {2225} (\bibinfo {year} {1994}{\natexlab{c}})},\
  \Eprint {https://arxiv.org/abs/hep-ph/9402335} {arXiv:hep-ph/9402335}
  \BibitemShut {NoStop}%
\bibitem [{\citenamefont {Belitsky}\ \emph {et~al.}(2003)\citenamefont
  {Belitsky}, \citenamefont {Ji},\ and\ \citenamefont
  {Yuan}}]{Belitsky:2002sm}%
  \BibitemOpen
  \bibfield  {author} {\bibinfo {author} {\bibfnamefont {A.~V.}\ \bibnamefont
  {Belitsky}}, \bibinfo {author} {\bibfnamefont {X.}~\bibnamefont {Ji}},\ and\
  \bibinfo {author} {\bibfnamefont {F.}~\bibnamefont {Yuan}},\ }\href
  {https://doi.org/10.1016/S0550-3213(03)00121-4} {\bibfield  {journal}
  {\bibinfo  {journal} {Nucl. Phys. B}\ }\textbf {\bibinfo {volume} {656}},\
  \bibinfo {pages} {165} (\bibinfo {year} {2003})},\ \Eprint
  {https://arxiv.org/abs/hep-ph/0208038} {arXiv:hep-ph/0208038} \BibitemShut
  {NoStop}%
\bibitem [{\citenamefont {Marquet}\ \emph {et~al.}(2016)\citenamefont
  {Marquet}, \citenamefont {Petreska},\ and\ \citenamefont
  {Roiesnel}}]{Marquet:2016cgx}%
  \BibitemOpen
  \bibfield  {author} {\bibinfo {author} {\bibfnamefont {C.}~\bibnamefont
  {Marquet}}, \bibinfo {author} {\bibfnamefont {E.}~\bibnamefont {Petreska}},\
  and\ \bibinfo {author} {\bibfnamefont {C.}~\bibnamefont {Roiesnel}},\ }\href
  {https://doi.org/10.1007/JHEP10(2016)065} {\bibfield  {journal} {\bibinfo
  {journal} {JHEP}\ }\textbf {\bibinfo {volume} {10}},\ \bibinfo {pages}
  {065}},\ \Eprint {https://arxiv.org/abs/1608.02577} {arXiv:1608.02577
  [hep-ph]} \BibitemShut {NoStop}%
\bibitem [{\citenamefont {Altinoluk}\ and\ \citenamefont
  {Boussarie}(2019)}]{Altinoluk:2019wyu}%
  \BibitemOpen
  \bibfield  {author} {\bibinfo {author} {\bibfnamefont {T.}~\bibnamefont
  {Altinoluk}}\ and\ \bibinfo {author} {\bibfnamefont {R.}~\bibnamefont
  {Boussarie}},\ }\href {https://doi.org/10.1007/JHEP10(2019)208} {\bibfield
  {journal} {\bibinfo  {journal} {JHEP}\ }\textbf {\bibinfo {volume} {10}},\
  \bibinfo {pages} {208}},\ \Eprint {https://arxiv.org/abs/1902.07930}
  {arXiv:1902.07930 [hep-ph]} \BibitemShut {NoStop}%
\bibitem [{\citenamefont {M\"antysaari}\ and\ \citenamefont
  {Zurita}(2018)}]{Mantysaari:2018nng}%
  \BibitemOpen
  \bibfield  {author} {\bibinfo {author} {\bibfnamefont {H.}~\bibnamefont
  {M\"antysaari}}\ and\ \bibinfo {author} {\bibfnamefont {P.}~\bibnamefont
  {Zurita}},\ }\href {https://doi.org/10.1103/PhysRevD.98.036002} {\bibfield
  {journal} {\bibinfo  {journal} {Phys. Rev. D}\ }\textbf {\bibinfo {volume}
  {98}},\ \bibinfo {pages} {036002} (\bibinfo {year} {2018})},\ \Eprint
  {https://arxiv.org/abs/1804.05311} {arXiv:1804.05311 [hep-ph]} \BibitemShut
  {NoStop}%
\bibitem [{\citenamefont {Iancu}\ \emph {et~al.}(2021)\citenamefont {Iancu},
  \citenamefont {Mueller}, \citenamefont {Triantafyllopoulos},\ and\
  \citenamefont {Wei}}]{Iancu:2020jch}%
  \BibitemOpen
  \bibfield  {author} {\bibinfo {author} {\bibfnamefont {E.}~\bibnamefont
  {Iancu}}, \bibinfo {author} {\bibfnamefont {A.~H.}\ \bibnamefont {Mueller}},
  \bibinfo {author} {\bibfnamefont {D.~N.}\ \bibnamefont
  {Triantafyllopoulos}},\ and\ \bibinfo {author} {\bibfnamefont {S.~Y.}\
  \bibnamefont {Wei}},\ }\href {https://doi.org/10.1007/JHEP07(2021)196}
  {\bibfield  {journal} {\bibinfo  {journal} {JHEP}\ }\textbf {\bibinfo
  {volume} {07}},\ \bibinfo {pages} {196}},\ \Eprint
  {https://arxiv.org/abs/2012.08562} {arXiv:2012.08562 [hep-ph]} \BibitemShut
  {NoStop}%
\bibitem [{\citenamefont {Altinoluk}\ \emph {et~al.}(2024)\citenamefont
  {Altinoluk}, \citenamefont {Jalilian-Marian},\ and\ \citenamefont
  {Marquet}}]{Altinoluk:2024vgg}%
  \BibitemOpen
  \bibfield  {author} {\bibinfo {author} {\bibfnamefont {T.}~\bibnamefont
  {Altinoluk}}, \bibinfo {author} {\bibfnamefont {J.}~\bibnamefont
  {Jalilian-Marian}},\ and\ \bibinfo {author} {\bibfnamefont {C.}~\bibnamefont
  {Marquet}},\ }\href@noop {} {\  (\bibinfo {year} {2024})},\ \Eprint
  {https://arxiv.org/abs/2406.08277} {arXiv:2406.08277 [hep-ph]} \BibitemShut
  {NoStop}%
\end{thebibliography}%
\end{document}